\documentclass[aps,prb,twocolumn,superscriptaddress,10pt]{revtex4-2}
\pdfoutput=1 
\usepackage[utf8]{inputenc}
\usepackage[english]{babel}
\usepackage[T1]{fontenc}
\usepackage[pdftex]{graphicx}
\usepackage{amsmath}
\usepackage{cancel}
\usepackage{amsthm}
\usepackage{amssymb}
\usepackage{braket}
\usepackage{xcolor}
\usepackage{dcolumn}
\usepackage{bbm}
\usepackage[normalem]{ulem}
\usepackage{footmisc}
\usepackage{booktabs}
\usepackage{comment}
\usepackage{multirow}
\usepackage{tikz}
\usepackage{mathrsfs} 
\usetikzlibrary{shapes.geometric, arrows, positioning}
\usepackage{tabularx}
\usepackage[colorlinks=true,linkcolor=blue,citecolor=blue,urlcolor=blue,unicode]{hyperref}

\newcommand{\sj}[1]{\vec S_{#1}}
\newcommand{\dt}{\frac{d}{dt}}

\newcommand{\pj}{p_{j}}

\newcommand{\qj}{q_{j}}
\newcommand{\qjp}[1]{q_{j\scalebox{0.6}{+}{#1}}}
\newcommand{\qjm}[1]{q_{j\text{-}{#1}}}

\makeatletter
\newcommand{\doublewidetilde}[1]{{%
  \mathpalette\double@widetilde{#1}%
}}
\newcommand{\double@widetilde}[2]{%
  \sbox\z@{$\m@th#1\widetilde{#2}$}%
  \ht\z@=.9\ht\z@
  \widetilde{\box\z@}%
}
\makeatother

\begin{document}
\title{Weak integrability breaking perturbations in classical integrable models on the lattice}
\author{Sara Vanovac}
\affiliation{Department of Physics and Institute for Quantum Information and Matter,
California Institute of Technology, Pasadena, California 91125, USA}
\author{Catherine McCarthy}
\affiliation{Department of Theoretical Physics, University of Geneva, 24 quai Ernest-Ansermet, 1211 Gen\`eve, Switzerland}
\author{Federica Maria Surace}
\affiliation{Department of Physics and Institute for Quantum Information and Matter,
California Institute of Technology, Pasadena, California 91125, USA}
\affiliation{School of Physics, Trinity College Dublin, Dublin 2, Ireland}
\author{Olexei I. Motrunich}
\affiliation{Department of Physics and Institute for Quantum Information and Matter,
California Institute of Technology, Pasadena, California 91125, USA}

\begin{abstract}
We show how to systematically construct \textit{weak integrability breaking} perturbations (WIBs) for classical integrable models on the lattice. 
These perturbations, which allow quasi-conserved quantities, have mostly been explored in quantum systems, where they are expected to delay the onset of thermalization and diffusive transport to timescales far exceeding those predicted by Fermi's golden rule. 
However, accessing such long-time dynamics in quantum models is computationally challenging. 
Classical integrable lattice models offer a complementary setting for probing transport and long-time dynamics under WIBs. 
In this work, we specialize our general framework to construct several families of WIBs for the Ishimori model, the Toda chain, and the Harmonic Oscillator Chain (HOC).
Such constructions can help quantify how WIBs contribute to anomalous transport and serve as a benchmark for thermalization studies in perturbed integrable models.
An important example is the Fermi–Pasta–Ulam–Tsingou (FPUT) model:
Starting from the HOC, we show that the cubic nonlinearity (the $\alpha$–FPUT interaction) is a genuine WIB perturbation. 
Using the integrals of motion (IoMs) of the Toda lattice, we explicitly construct corrections to the entire hierarchy of the HOC IoMs, thereby obtaining an infinite tower of quasi-conserved quantities for the $\alpha$–FPUT chain. 
We further identify the corresponding adiabatic gauge potential (AGP) as a nontrivial \emph{trilocal} generator in real space, and show that, more generally, any cubic, translationally invariant, momentum-conserving perturbation of the HOC admits such a generator and is therefore a WIB. 
Together with our transport and AGP-variance studies,
our results provide a unified classical framework for weak integrability breaking and for diagnosing anomalous thermalization and transport in nearly integrable Hamiltonian lattice systems.
\end{abstract}
\maketitle

\section*{Introduction}
\label{sec:intro}
Integrable models play an important role in understanding the dynamics of low-dimensional many-body systems: they are analytically tractable and possess an extensive number of conservation laws, which makes them a natural starting point for studying the non-equilibrium behavior. 
A fundamental tenet of statistical mechanics is that a generic interacting many-body system evolving in time will eventually thermalize (even in the case of isolated 1d quantum systems evolving under unitary dynamics)  \cite{Deutsch1991, Srednicki1994}.
In such cases, local observables relax to stationary values determined only by a few macroscopic parameters, as predicted by the Gibbs ensemble \cite{DAlessio2016, Eisert2015, Polkovnikov2011b}.  
Integrable systems, by contrast, retain information about their initial conditions and do not thermalize in the usual sense, mostly due to the presence of extensively many conserved quantities or integrals of motion (IoMs); instead, they relax into non-thermal steady states described by the generalized Gibbs ensemble (GGE)~\cite{Rigol2007, Rigol2009, Cassidy2011, Kollar2011, Gring2012, Ilievski2015, Langen2015, Essler_2016, Vidmar_2016}. 
This includes models that possess quasi-local conserved charges, which must be taken into account to exactly match steady-state predictions to observed values~\cite{Prosen2011, Prosen2013, Ilievski2016}.

In classical systems, the Kolmogorov-Arnold-Moser (KAM) theorem states that under sufficiently small perturbations, integrable systems can remain stable for sufficiently long times \cite{Kolmogorov1954, Arnold1963,Moser1962,Arnol2013}. 
However, the KAM theorem is best understood for systems with a few degrees of freedom; in the many-body or thermodynamic limit, the required non-resonance conditions become increasingly difficult to satisfy.
What happens when one perturbs an integrable \emph{many-body} system is still an active area of research~\cite{Kim2025, Roy_2023, Dhar_2020, Karve_2025, Danieli2024} with long history in classical integrability \cite{Fermi1955, Mazur1969,Toda_1974,Nekhoroshev1977, Lochak1992}.
In the quantum regime, a closely related question about the dynamics of integrable systems under small but integrability-breaking perturbations has seen significant progress in recent years~\cite{Bertini_2021, Bulchandani2021, Gopalakrishnan_2023, Gopalakrishnan2024superdiffusion, Rabson2004, Jung2006, Mazets2008, Marcuzzi2013, Essler2014, Brandino2015, Bertini2015, Mierzejewski2015, Bertini2016, Tang2018, Friedman2020, LeBlond2021, Bastianello2021, Bulchandani, Durnin2021, kuo2023energy, Chen_2026}.
Explorations of ``special'' perturbations that lead to anomalously slow thermalization due to emergent quasi-conservation laws are even more recent~\cite{Pandey2020, Kurlov2022, Orlov_2023, Surace2023, Vanovac2024, Pozsgay_2024}.

While thermalization is expected to take place \textit{eventually} once integrability is broken, its timescale depends on the nature of the perturbation. 
This distinction has concrete consequences, for example, in transport~\cite{Jung2006}.
The fact that many experimental setups and quasi-1D materials are well described by integrable models, despite the presence of noise and imperfections that plague real materials, makes a strong case for developing a more rigorous understanding of integrability breaking in interacting quantum many-body systems and their classical counterparts.

We begin by distinguishing between the initial relaxation to a stationary state of the unperturbed Hamiltonian, called \textit{prethermalization}, and the subsequent thermalization, which occurs later and can typically be modeled using Fermi's golden rule (in quantum systems) or its classical counterpart.
These predict that the 
thermalization time $\tau$ as a function of perturbation strength $\lambda$ typically scales as $\tau \sim \lambda^{-2}$~\cite{Deng_2023, Onorato_2019, Mallayya2019prethermalization}.
While very long thermalization times can happen as a result of a few different physical phenomena, in this work we will not be talking about instances covered by the rigorous theory of prethermalization (see \cite{Kollar_2011, Abanin2017EffectiveHam, Abanin2017Rigorous, Mori_2018, Huveneers_2020}), many body localization (MBL) (see \cite{NandkishoreHuse2015,Abanin2019}), or Hilbert space fragmentation (HSF) that can be observed in kinetically constrained models (see~\cite{Sala2020,Khemani2020,Rakovszky2020,Moudgalya_2022}).
Instead, we focus on systems in which anomalous thermalization---thermalization time parametrically longer than the FGR scaling---is due to a special property of the perturbation: IoMs of the unperturbed model acquire controlled corrections, becoming quasi-conserved in the perturbed model.
In such cases, the perturbed Hamiltonian can be viewed as a finite-order truncation of a long-range deformation of the original Hamiltonian produced by a unitary (quantum) or canonical (classical) transformation \cite{Bargheer2008, Bargheer2009, Beisert2013, Poszgay2020, Surace2023}.
Most widely cited example of a perturbation with this property is the quantum spin-$1/2$ Heisenberg chain with next-to-nearest-neighbor isotropic coupling~\cite{Jung2006, Kurlov2022}.
The special status of this perturbation was first observed in \cite{Jung2006} and was used to explain the unusually large heat conductivity, where it was recognized that the perturbation preserved the energy current up to a higher order in coupling strength~\cite {Jung2006}. 
This was the first example of what we refer to here as weak integrability breaking (WIB), and subsequent works~\cite{Kurlov2022, Surace2023, Orlov_2023} showed that all the IoMs of the original model were also quasi-conserved.
We emphasize that by ``weak integrability breaking'' we mean this precise, qualitative notion—namely, the existence of quasi-conserved quantities—rather than merely a small perturbation strength.

While WIBs have been identified in several quantum systems~\cite{Jung2006,Poszgay2020,Durnin2021,Kurlov2022, Surace2023, Orlov_2023,kuo2023energy,Pozsgay_2024,Yan2025}, this concept has not yet been explored in classical systems. 
Classical models, however, provide a powerful framework for studying integrability breaking, as demonstrated, for example, in tests of superdiffusive transport in SU(2)-invariant chains~\cite{Ljubotina2019,Gopalakrishnan_2023,takeuchi2024,McCarthy2024,Gopalakrishnan2024superdiffusion,Roy_2023,McRoberts2022}.
In particular, they enable simulations of much larger system sizes than are accessible in quantum systems.
This suggests that classical systems may be a useful framework for studying WIBs and their thermalization times.

In this work, we systematically construct a broad family of such special perturbations for \textit{classical} integrable lattice models.
Our approach builds on the analytical framework recently developed in the quantum domain \cite{Surace2023}.
The central idea is that continuous deformations of integrable Hamiltonians, generated by boosted, bilocal, or extensive local operators, can be truncated order-by-order to produce perturbations which preserve quasi-conserved quantities and thus prolong thermalization times to $\tau \sim \lambda^{-2\ell}$ with $\ell>1$ an integer, in contrast to the usual FGR prediction of $\lambda^{-2}$.
We consider several classical integrable models on a lattice, including the Ishimori chain (an integrable classical spin chain), the Toda lattice (a nonlinear integrable model), and a chain of classical harmonic oscillators.
For each case, we explicitly construct WIBs from different classes of generators. 
One of the central results of this work is the existence of a new non-trivial \emph{trilocal} generator of WIBs in the harmonic oscillator chain (HOC), which extends the known classes of WIB generators beyond previously studied cases. 

In addition to the analytical constructions, we present numerical evidence of the prolonged thermalization in the presence of WIBs. 
Inspired by a recent work on transport in the Ishimori model \cite{McCarthy2024}, we look at the energy transport in the presence of WIB perturbations.
Since WIBs break integrability, we expect a truly diffusive regime to emerge eventually; however, we expect the crossover to diffusion to be delayed compared to non-WIB perturbations.
Previous numerical studies have focused on spin transport, which is already known to be superdiffusive in the integrable model with a very long crossover to diffusion in the perturbed models. 
We argue that, in the context of WIBs, examining energy transport (and potentially higher-order IoMs) is more insightful, as energy transport appears to be more sensitive to the type of integrability breaking. 
For example, we find that energy transport in  WIB-perturbed models remains near-ballistic for long times.
While we do not extract the exact crossovers in this study, the advantage of our approach is that, by construction, we can make predictions about thermalization times and transport, which we find are compatible with the numerical results. 
Further numerical studies to determine the precise locations of the crossovers are an interesting direction for future work.

Our study of WIBs in classical systems offers new insights into long-standing questions concerning classical chaos. Classical chaos is generally understood as an exponential sensitivity of the system's trajectories to small perturbations. One of the earliest studies aimed at showing that chaos arises due to the presence of nonlinearities in the system gave a strange result: adding a cubic nonlinearity to a one-dimensional chain of coupled harmonic oscillators initialized in a long-wavelength mode did not lead to the expected ergodic behavior, but instead, unexpected periodic behavior was observed. This now-famous Fermi-Pasta-Ulam-Tsingou (FPUT) numerical experiment from the 1950s, while mostly understood (the FPUT model does thermalize, but over very long times), remains extensively studied \cite{Gallavotti2008, Flach2005, Flach2005b, Henrici2008, Henrici2008b, Ganapa2023, Christodoulidi2014, Christodoulidi2019, Grava2020}.
In recent years, special attention has been given to revisiting the $\alpha$–FPUT chain, which is known to be closely related to the integrable Toda lattice whose integrals of motion have been found to act as adiabatic invariants for the FPUT dynamics \cite{Grava2020}.
Within our WIB framework, we make this relation more precise:
We show that the $\alpha$–FPUT interaction is a WIB perturbation of the harmonic chain, construct explicit corrections to its entire hierarchy of integrals of motion using the Toda charges, and identify the corresponding generator as a trilocal operator.
Our findings provide a structural explanation for the longstanding observation that Toda invariants control the slow relaxation and near recurrences in the $\alpha$–FPUT dynamics~\cite{Christodoulidi2019, Christodoulidi_2025}.

Another useful perspective on weak integrability breaking comes from the adiabatic gauge potential (AGP), which helps unify classical and quantum views of chaos.
The AGP is the generator of adiabatic transformations; in quantum (classical) mechanics, it dictates how eigenstates (trajectories) deform under slow parameter changes~\cite{Demirplak_2003, Berry_2009, Jarzynski2017, Sels2017, Pandey2020}. 
Recent works have proposed using the norm and variance of the AGP as a probe of chaos and integrability in quantum and classical systems, respectively~\cite{Kim2026, Karve_2025}. 
If a system is integrable (or nearly so), its eigenstates (trajectories) respond smoothly to perturbations, yielding a relatively small AGP norm (variance).
In chaotic quantum systems, on the other hand, adiabatic deformations are highly non-trivial, and the AGP norm grows exponentially in system size.
In quantum systems, scaling of the AGP norm with system size is a robust probe of the type of integrability breaking~\cite{Pandey2020, Orlov_2023, Vanovac2024}, and our present study provides additional evidence that the same is true in classical models.

We supplement our analytic constructions of WIBs with numerical studies of the classical AGP (obtained by replacing quantum commutators with Poisson brackets in the semiclassical limit) and its scaling with system size.
Variance of the classical AGP saturates quickly with time if the perturbation is a WIB, while for generic perturbations, it keeps growing.
Additional information about the generator type can be obtained by examining the scaling of the saturation value with system size. 
For each WIB generator, we extract scalings and verify our analytic results. 
Our results support the conclusion that AGP is a very sensitive probe of integrability breaking in both quantum and classical models. 
In fact, one motivation for our work is to use WIBs as a benchmark for various probes of chaos:
If a chaos indicator (such as the AGP norm, out-of-time-order correlators, or operator growth rates) is a truly sensitive probe of the degree of integrability breaking, then the WIB framework provides an excellent testing ground.

To summarize, WIB perturbations break integrability in a controlled manner, thereby revealing hidden structures in the dynamics while prolonging the approach to thermalization.
Our construction can be used to investigate the extent to which WIBs contribute to anomalous transport phenomena (such as superdiffusive or ballistic transport) and can serve as a benchmark for theoretical and numerical studies of thermalization in perturbed integrable models.

The paper is organized as follows.
In Sec.~\ref{sec:prelim}, we lay out the theoretical framework for constructing WIB perturbations and the three main classes of WIB generators: extensive local, boosted, and bi-local operators.
In Secs.~\ref{sec:ishimori},~\ref{sec:toda}, and \ref{sec:choc}, we apply this construction to our three chosen classical lattice models.
For the Ishimori spin chain, the Toda lattice, and the HOC, we explicitly derive the form of WIBs generated by each class of operators. 
For the HOC, we pay particular attention to cubic and quartic perturbations that are related to the FPUT problem. 
In particular, in Sec.~\ref{sec:HOC_VFPUT}, we show that the $\alpha$–FPUT interaction defines a WIB of the HOC model and, using the Toda integrals, we construct explicit corrections to the HOC charges, thereby obtaining an infinite family of quasi-conserved quantities.
We further identify the associated AGP in momentum space.
We find an analytic expression for the AGP in real space and identify it as a new trilocal generator of WIB perturbations.
In Sec.~\ref{sec:numerics} we turn to numerical validation of the above results and constructions. 
We present results on energy transport in the Ishimori model and
study the scaling of the classical AGP of all of the perturbations listed in the three models.
Finally, in Sec.~\ref{sec:conclusions}, we summarize our findings and discuss future directions.

\section{PRELIMINARIES: WEAK INTEGRABILITY BREAKING PERTURBATIONS}
\label{sec:prelim}
We work on a Poisson manifold and define two distinct Poisson structures. 
For the Toda chain and the classical harmonic oscillator chain (HOC), the phase space is 
$$\mathcal{M}= \mathbb{R}^{2N} ~,$$
with canonical coordinates $\vec{q}=(q_1, \dots, q_N)$ and $\vec{p}=(p_1, \dots, p_N)$ and a Hamiltonian $H: \mathcal{M} \rightarrow \mathbb{R}$.
Dynamics of the system are described by a curve in this phase space governed by Hamilton's equations of motion
\begin{equation}
    \frac{d\vec{q}}{dt} = \frac{\partial H}{\partial \vec{p}}, \quad \quad 
    \frac{d\vec{p}}{dt} =- \frac{\partial H}{\partial \vec{q}} ~.
\end{equation}
For functions $F, G$ on such a phase space, the Poisson bracket is 
\begin{equation}
    \{F,G\} = \sum_{j=1}^N \bigg [\frac{\partial F}{\partial q_j} \frac{\partial G}{\partial p_j} - \frac{\partial F}{\partial p_j} \frac{\partial G}{\partial q_j} \bigg],
\end{equation}
with standard relations 
$$\{q_i, q_j\} =0, \quad  \{p_i, p_j\} =0, \quad  \{q_i, p_j\} = \delta_{ij} ~.$$
This Poisson structure is anti-symmetric, bilinear, satisfies the Leibniz rule, and obeys the Jacobi identity.

For the spin chain model, the phase space is 
$$\mathcal{M}_{\mathrm{spin}} = (S^2)^{\otimes N},$$
with spin vectors $\vec{S}_j=(S_j^x, S_j^y, S_j^z)$ at each site and a fixed Casimir constraint $\vec{S}_j^2=1$.
The Poisson structure is the standard  $\mathfrak{su}(2)$ Lie-Poisson algebra,
\begin{equation}
    \label{Pbracket}
    \{S_i^\alpha, S_j^\beta\} = \delta_{ij} \sum_\gamma \epsilon_{\alpha\beta\gamma}S_j^\gamma ~,
\end{equation}
with $\alpha,\beta,\gamma \in \{x,y,z\}$ which preserves the Casimir and constrains each spin to evolve on the sphere $S^2$. Dynamics of a Hamiltonian $H(\{\vec{S}\})$ are described by equations of motion
\begin{equation}
    \frac{d\vec{S}_j}{dt} = \{\vec{S}_j,H\} = 
\frac{\partial H}{\partial \vec{S_j}} \times  \vec{S}_j .
\end{equation}
In general, for any function $F$ on the chosen Poisson manifold, the total time derivative along solutions of the Hamiltonian flow generated by $H$ is 
\begin{equation}
    \frac{d}{dt}F = \frac{\partial F}{\partial t} + \{F, H\}
\end{equation}
The Hamiltonians of interest in this work are sums of local terms, defined on a one-dimensional lattice.
We call a Hamiltonian $H$ integrable if there exists 
$N$ independent functions $Q_\alpha, \alpha=1,\dots N$ that Poisson-commute with the Hamiltonian---they are integrals of motion (IoMs)--- and are in involution ($\{Q_\alpha, Q_\beta\} = 0$) with all other IoMs.
Replacing Poisson brackets with commutators gives the commonly used definition of integrability in quantum systems \cite{Liouville1855, FaddeevTakhtajan1987}. 

In what follows, we consider deformations of such integrable Hamiltonians and characterize perturbations that preserve the existence of a quasi-integrable structure.

\subsection{Quasi-IoMs and weak integrability breaking perturbations (WIBs)}
\label{sec:generators}

We take an integrable classical Hamiltonian on a lattice $H_0$, with conserved quantities $Q_2^{(0)}$, $Q_{3}^{(0)}$, $Q_{4}^{(0)}$, $\dots$, and add a perturbation $V$, where $\lambda$ is a small parameter
\begin{equation}
\label{eq:H}
    H = H_0 + \lambda V ~.
\end{equation}
We call $H$ the perturbed Hamiltonian, which is no longer integrable.
Suppose there exists a generator function $X$ such that the perturbation can be written as a Poisson bracket with $H_0$,
\begin{equation}
    V = \{X, H_0\}~.
    \label{eq:VXH}
\end{equation}
In other words, if the perturbation corresponds to an infinitesimal canonical
transformation generated by $X$, we can define corrections, $Q_\alpha^{(1)}$
\begin{equation}
Q_\alpha^{(1)} = \{X, Q_\alpha^{(0)}\} ~,
\label{eq:ioms_corr}
\end{equation}
of the conserved charges $Q_\alpha^{(0)}$ of $H_0$.
It is easy to show that if Eqs.~(\ref{eq:VXH}) and (\ref{eq:ioms_corr}) are satisfied, the following is also true
\begin{equation}
    \{ H_0 + \lambda V, Q_\alpha^{(0)} + \lambda  Q_\alpha^{(1)}\} = O(\lambda^2) ~.
    \label{eq:corrections}
\end{equation}
This means that $Q_\alpha^{(0)} + \lambda  Q_\alpha^{(1)}$ is conserved with respect to the new Hamiltonian to the leading order in $\lambda$, and we call quantities of this type \textit{quasi-conserved}.
Similarly, $Q_\alpha^{(0)} + \lambda  Q_\alpha^{(1)}$ for different $\alpha$ all Poisson-commute with each other to the same order.
We can hence view them as approximate IoMs or {\it quasi-IoMs} of the perturbed Hamiltonian.

From previous work in quantum systems, we know that there are several classes of 
generators (choices of $X$) that give extensive local $V$'s and $Q_\alpha^{(1)}$'s \cite{Bargheer2009,  Poszgay2020}. 
Before introducing these classes of generators, we here discuss some general properties of the IoMs that will be useful in their definition.

We find it useful to write the original IoMs as sums of local densities:
\begin{equation}
    Q_\alpha^{(0)} \;=\; \sum_j q_{\alpha,j}^{(0)}\,.
\end{equation}
Here \(j\) labels the lattice sites, and each \emph{charge density} \(q_{\alpha,j}^{(0)}\) has finite support around site \(j\).
In the classical setting, the involution condition
$\{Q_\alpha^{(0)}, \,Q_\beta^{(0)}\} = 0$
implies the existence of continuity equations of the form
\begin{equation}
\{q_{\beta,j}^{(0)},\,Q_\alpha^{(0)}\}
= J_{\beta\alpha, j} - J_{\beta\alpha, j+1}\,,
\label{eq:Jdef}
\end{equation}
where $J_{\beta\alpha, j}$ is a \emph{generalized current} with finite support around $j$.
We usually identify the classical Hamiltonian with $Q_2^{(0)} := H_0$, while \(Q_{\alpha\ge3}^{(0)}\) are the nontrivial IoMs. 
Some of the models will also have a conventional conserved quantity, such as the total spin or total momentum; these will be denoted by $Q_1^{(0)}$. 
It is worth noting that while all IoMs are conserved quantities, there is a difference between the higher IoMs $Q_{\alpha\ge3}^{(0)}$ and symmetries $Q_1^{(0)}$ like the total magnetization or momentum in that the latter generate a compact Lie group action. In a quantum setup, this means they can be used to block-diagonalize the Hamiltonian, while the higher IoMs cannot.

\subsection{Generators of WIBs in infinite chains from long-range deformations }
We begin by listing all currently known classes of generators, \(X\), that can yield extensive, local, and translationally invariant perturbations, $V$. 
Whether this list is exhaustive remains an open question (in fact, in Sec.~\ref{sec:HOC_VFPUT} we will introduce a novel type of generator in the HOC system that does not belong to these classes).

\subsubsection{Extensive local generators.}
Let $X$ be an arbitrary translationally-invariant function of the form
\begin{equation}
   X_{\mathrm{ex}} \;=\;\sum_j f(O_j)\,,
\end{equation}
where $O_j$ is a function with finite support around $j$.
Then $V_{\mathrm{ex}}\;=\;\bigl\{X_{\mathrm{ex}},\,H_0\bigr\}$ is a translationally-invariant sum of local terms.

\subsubsection{Boosted generators.} 
Given an IoM \(Q_{\beta}^{(0)}=\sum_j q_{\beta, j}^{(0)}\), we can define the “boosted” generator
\begin{equation}
   X_{\mathrm{bo}} 
   \;=\;
   \,\sum_j j\,q_{\beta,j}^{(0)}\,.
\end{equation}
Note that this object acts locally but inhomogeneously along the chain. Boost is formally only well-defined in infinite chains, and finite-chain expressions have not been found yet (while an attempt has been made in \cite{Vanovac2024}).
From the continuity equation Eq.~(\ref{eq:Jdef}), one obtains
\begin{equation}
\bigl\{X_{\mathrm{bo}},\,Q_{\alpha}^{(0)}\bigr\}
   \;=\;
   \sum_j J_{\beta\alpha, j}
   \; =: \;
   J_{\beta\alpha,\mathrm{tot}}\,.
   \label{eq:Vbo}
\end{equation}
Hence, the Hamiltonian perturbation arising from this generator is again an extensive, local function of the lattice variables:
\begin{equation}
V_{\mathrm{bo}}
\;=\;
\bigl\{X_{\mathrm{bo}},\,H_0\bigr\} = J_{\beta,2; \text{tot}} \,.
\end{equation}
which acts homogeneously along the chain and is translationally invariant and well-defined in systems with periodic boundary conditions. 

\subsubsection{Bi-local generators.}
\label{subsec:bilocal}
From two IoMs, $Q_{\beta}^{(0)}$ and $Q_{\gamma}^{(0)}$, we can define the bilocal generator
\begin{align}
 X_{\mathrm{bi}} &=[\,Q_{\beta}^{(0)} \mid Q_{\gamma}^{(0)}\,] \notag \\
& :=\; 2\sum_{j<k}\!q_{\beta,j}^{(0)}\,q_{\gamma,k}^{(0)} +\,\sum_jq_{\beta,j}^{(0)}\,q_{\gamma,j}^{(0)} ~.
\end{align}
Analogous to the quantum case, we get
\begin{align}
\bigl\{X_{\mathrm{bi}},\,Q_{\alpha}^{(0)}\bigr\} \;=\;&\sum_j \Big[ q_{\beta,j}^{(0)}\Bigl(J_{\gamma\alpha,j} + J_{\gamma\alpha,j+1}\Bigr) \notag \\
& \;-\; q_{\gamma,j}^{(0)}\Bigl(J_{\beta\alpha,j} + J_{\beta\alpha,j+1}\Bigr) \Big] ~.
\label{eq:bilocal}
\end{align}
Compared to the quantum expressions, e.g., Eq.~(12) in Ref.~\cite{Surace2023}, in the classical case, the anticommutator is replaced with usual multiplication. 
This again gives a perturbation,
\begin{equation}
   V_{\mathrm{bi}}
   \;=\;\bigl\{X_{\mathrm{bi}},\,H_0\bigr\},
\end{equation}
that is an extensive, local function.

In the initial papers on the subject (see \cite{Bargheer2008, Bargheer2009, Beisert2013, Poszgay2020} for review), these three types of generators were identified as exhausting the space of operators that satisfy the deforming equation \footnote{Our Eq.~(\ref{eq:ioms_corr}) is formally first order correction of a differential equation relating the family of charges or IoMs 
\begin{align}
    \frac{d}{d\lambda}Q_\alpha(\lambda) = \{ X(\lambda), Q_\alpha(\lambda) \}.
\end{align}
The differential equation guarantees that the algebra of the charges $Q_\alpha$ is independent of $\lambda$, and all of our analysis is done at $\lambda=0$, and the series is cut to first order.
} 
for the spin $1/2$ XYZ family of models by characterizing all of the admissible deformations and classifying their generators into one of the three types or a linear combination of them.
Despite this, there are some indications that other types of generators exist. Recent work on free fermions by Poszgay \textit{et. all}, pointed out that an example that was previously thought to be strongly integrability-breaking (it could not be generated from any of the known generators \cite{Surace2023}) has a polynomially growing AGP norm \cite{Pozsgay2024}. 
In their analysis, corrections to only half of the IoM were identified in the perturbed model, and the exact nature of the generator remained unresolved.
Their main claim, however, that the WIB in question could not be generated by one of the classes of generators known within the framework of long-range deformation, opened the possibility of other types of generators. 
This is one of the main findings of this work, since in Sec.~\ref{sec:HOC_VFPUT} we find WIBs generated by an entirely new type of generator. 

\subsubsection{Generators of strictly local WIBs}
We can also define generators that yield strictly local WIBs, i.e., perturbations with support only in a finite region around a site $j_0$ (akin to a local impurity). 
These were introduced and studied in \cite{Vanovac2024}, where they proved useful in finding finite-size analogs of boosted and bilocal generators. 

We call {\it strictly local generators} all functions
\begin{equation}
X_{\text{loc}} = O_{j_0} ~,
\end{equation}
where $O_{j_0}$ has finite support around the site $j_0$, and all WIBs generated from 
\begin{equation}
V_{\text{loc}}\;=\;\big\{X_{\text{loc}},\,H_0\big\}
\end{equation}
are strictly local WIBs. 
These types of WIBs are related to the extensive local ones generated by $X_\text{ex}$.

More interesting, perhaps, are those generated from {\it step generators}, 
\begin{align}
    X_{\mathrm{step}} = \sum_{j\geq j_0} q_{\beta,j}^{(0)} ~.
\end{align}
From Eq.~(\ref{eq:Jdef}) it follows that perturbations generated from such an $X$ are really local currents, 
\begin{align}
    V_{\mathrm{step}} = \bigl\{X_{\mathrm{step}},\,Q_\alpha^{(0)}\bigr\}= J_{\beta\alpha,j_0} ~.
    \label{eq:vstep}
\end{align}
Perturbations generated in this way are strictly local functions with support around $j_0$. 

\subsection{Adiabatic gauge potential (AGP) as a proxy for generators of WIBs}
To determine whether a generator $X$ satisfying Eq.~(\ref{eq:VXH}) exists (and whether the quasi-conserved quantities in Eq.~(\ref{eq:corrections}) can be constructed), we can recast the problem in terms of adiabatic deformations.
In this language, the question becomes whether the perturbation $V$ can be implemented by a (near-)canonical transformation that preserves the integrable structure to the leading order.
The adiabatic gauge potential (AGP) precisely captures this information~\cite{Demirplak_2003, Berry_2009}.

For a classical system with a Hamiltonian parameterized by a real variable $\lambda$, the AGP $\mathcal{A}_{\lambda}(\vec{q},\vec{p})$ (or $\mathcal{A}_{\lambda}(\{\vec{S}\})$ for a spin model) is a phase-space function whose EOM along a trajectory is given by
\begin{equation}
\frac{d \mathcal{A}_{\lambda}}{dt} = \{\mathcal{A}_{\lambda}, H \} = -\partial_\lambda H + G_\lambda,
\label{eq:classicalAGP}
\end{equation}
where $H$ is the Hamiltonian, $G_\lambda$ is a conserved generalized force, and $\{G_\lambda, H\} = 0$. \cite{Polkovnikov2011, KOLODRUBETZ_2017}

The AGP is understood as describing deformations to trajectories in the
phase space under adiabatic changes.
It is particularly useful for studying integrability-breaking perturbations, where the Hamiltonian is of the form $H=H_0+\lambda V$, as in Eq.~(\ref{eq:H}). Previous works showed that, in the quantum setting, the scaling of the AGP norm with system size serves as a sensitive probe of integrability breaking; the AGP captures the sensitivity of the unperturbed model's eigenstates to the perturbation \cite{Pandey2020,Orlov_2023,Pozsgay_2024, Vanovac2024}.
It was recently argued in~\cite{Kim2026} and showed in~\cite{Karve_2025} that an analogous statement holds in classical models.
One can use the variation of the AGP over the trajectory $\Delta \mathcal{A}_\lambda = \mathcal{A}_\lambda (t) - \mathcal{A}_\lambda (0)$, given by:
\begin{equation}
\label{eq:DAt}
\Delta \mathcal{A}_\lambda (t) =  - \int_0^t d\tau \ \left(\partial_\lambda H(\tau) - \overline{\partial_\lambda H}\right),
\end{equation}
where $\overline{\partial_\lambda H}=\lim_{T\rightarrow \infty} \frac{1}{T}\int_{0}^T d\tau\, \partial_\lambda H(\tau)$ is the time average of $\partial_\lambda H(\tau)$ over the entire trajectory as a probe of integrability-breaking and onset of chaos. This result was obtained by showing that $G_\lambda$ plays the role of the conserved part of $\partial_\lambda H$, such that $\overline{\partial_\lambda H}=G_\lambda$ \cite{Karve_2025}. 

The quantity $\Delta \mathcal{A}_\lambda (t)$
captures the sensitivity of a trajectory in phase-space to the perturbation. 
A large variance of AGP over time along a trajectory serves as an indicator of chaos and strong integrability breaking. 

Returning to our framework, the set of Hamiltonians related by canonical transformations that preserve all IoMs forms an integrable manifold in the space of Hamiltonians.
A perturbation $V$ is a WIB perturbation in the sense of Eq.~(\ref{eq:VXH}) when it is tangent to this manifold, i.e., when it is generated by a near-identity canonical transformation. 
AGP provides a practical way to test if a perturbation $V$ is a WIB.
AGP extracts the component of $V$ that such a canonical transformation can generate and thus acts like a proxy for the generator $X$. 
If $V$ is a WIB, its AGP remains small and well behaved, and the deformation $H_0 \rightarrow H_0 + \lambda V$ can be absorbed into a near-identity canonical transformation [as in Eq.~(\ref{eq:ioms_corr})] that approximately preserves all IoMs. 
In the next sections, we first discuss the models we consider in this work and present examples of WIBs constructed analytically, and then we use AGP to confirm our findings numerically.

\section{Ishimori model}
\label{sec:ishimori}
\label{sec:ishimori_ioms}
A natural starting point for classical spin chains with a well-defined quantum analog is the classical ferromagnetic Heisenberg chain
\begin{equation}
    H_{\text{Heis.}} = - \sum_j \vec{S}_j \cdot \vec{S}_{j+1} ~,
\label{Eq.Heis}
\end{equation}
which arises as the semiclassical limit of the quantum spin $1/2$ Heisenberg model.
Despite the fact that the continuum limit of Eq.~\eqref{Eq.Heis} reduces to the integrable Landau–Lifshitz field theory, the lattice model~\eqref{Eq.Heis} itself is \emph{not} integrable for such simple nearest-neighbor couplings.

The classical nearest-neighbor Heisenberg chain does not admit a Lax pair or zero-curvature representation, and therefore lacks the infinite hierarchy of commuting conserved quantities required for classical integrability~\cite{LakshmananNakamura1984, Takhtajan1977, Wysin1996}. 
In contrast, the integrability of the continuum Landau–Lifshitz equation relies on an underlying classical $r$-matrix structure, which is not preserved under the naive lattice discretization. 
Only specially constructed discretizations retain a Lax formulation and the associated tower of conserved charges.

A minimal integrable lattice modification of Eq.~\eqref{Eq.Heis} that restores the integrability
is known as the Ishimori model~\cite{Ishimori1982},
\begin{equation}
H_{0}
= -2 \sum_{j}
\log \!\left(\frac{1+\vec{S}_j \cdot \vec{S}_{j+1}}{2} \right) ~,
\label{eq:ishimori}
\end{equation}
$H_0$ preserves the same global $\mathrm{SU}(2)$ symmetry and the continuum Landau–Lifshitz limit as the Heisenberg chain, but, unlike Eq.~\eqref{Eq.Heis}, admits a Lax pair and an infinite tower of local integrals of motion.

The connection between the Ishimori and the Heisenberg models has been actively explored in recent years, particularly in the study of solitons, quasi-integrable dynamics, and origins of the KPZ-like superdiffusive spin transport and its crossover to diffusion~\cite{Dhar_2020,McRoberts2022,McRoberts2023,McRoberts2024,McCarthy2024}. 
Recent classical studies reported robust spin superdiffusion in the Ishimori model under nearest- and next-to-nearest-neighbor Heisenberg-like perturbations~ \cite{Das_2019}. 
Recent work on the same model explored the role of local but generic integrability-breaking perturbations (similar to our $V_{\text{loc}}$ and $V_{\text{step}}$ in transport and characterization of the Lyapunov spectrum \cite{Prosen2025}.
Simulations of classical models have been popular in recent years as they allow access to much larger systems compared to the quantum models, which is crucial for studying thermalization and transport. 

Weak perturbations, such as the ones that we systematically construct in the following sections, can be used to further our understanding of transport in weakly perturbed integrable models. 
WIBs that we construct have not been studied in the literature before.

\subsection{Higher IoMs for the Ishimori model}
Integrable models have been around for decades, and for many of them,  expressions for the first few IoMs can be found in the literature, usually constructed directly from the Lax matrix. 
Constructing higher IoMs from the Lax matrix is often tedious, but a significantly easier method can be used if the model possesses a boost operator, a ladder-like generator that can be applied recursively to construct the entire hierarchy of IoMs (thus satisfying the {\it Reshetikhin criterion} \cite{Grabowski1995}).

While the Ishimori model was known to be integrable, to the best of our knowledge, the fact that it has a boost operator has not been appreciated previously \cite{Grabowski1994}. 
This is probably why, beyond the first nontrivial IoM (known as \textit{torsion}), explicit expressions for higher charges of the Ishimori model are not readily available in the literature. 
We find that the Ishimori model satisfies the Reshetikhin criterion and that higher IoMs can be easily obtained using the boost.
We show this explicitly by deriving $Q_4^{(0)}$ (see App.~\ref{app:IoMs_ishimori}).
We list the first IoMs below. We adopt the notation that IoMs are written as sums of strictly local densities.

The first few IoMs of the Ishimori model are
\begin{align}
    & \vec{Q}_1^{(0)} = \sum_j \vec{S}_j, \\
    & Q_2^{(0)} \equiv H_0, \\
    & Q_3^{(0)} = -4 \sum_j \frac{(\sj{j}\times\sj{j+1})\cdot \sj{j+2}}{(1+\sj{j}\cdot\sj{j+1})(1+\sj{j+1}\cdot\sj{j+2})}, \\
    &Q_4^{(0)} = 16\sum_j \bigg [ 
    \frac{ \vec{S}_{j}\cdot\vec{S}_{j+2} + 1}{2(1+\sj{j}\cdot\sj{j+1})(1+\sj{j+1}\cdot\sj{j+2})}\notag \\
    & \quad \quad \quad -\frac{((\sj{j}\times\sj{j+1})\cdot \sj{j+2})^2}{(1+\sj{j}\cdot\sj{j+1})^2(1+\sj{j+1}\cdot\sj{j+2})^2}   \label{eq:ISH_Q4} \\
    & \quad \quad  + \frac{((\sj{j+1}\times\sj{j+2})\cdot \sj{j+3})^2}{(1+\sj{j+1}\cdot\sj{j+2})^2(1+\sj{j+2}\cdot\sj{j+3})^2}  \notag \\
    &+\frac{((\vec{S}_{j}\times \vec{S}_{j+1})\times\vec{S}_{j+2})\cdot\vec{S}_{j+3}}{(1+\vec{S}_{j}\cdot\vec{S}_{j+1})(1+\vec{S}_{j+1}\cdot\vec{S}_{j+2})(1+\vec{S}_{j+2}\cdot\vec{S}_{j+3})} 
    \bigg ]. \notag
\end{align}
By convention, Hamiltonian $H_0$ corresponds to the second IoM, $Q_2^{(0)}$. 
We use $q_{3,j}$ as a shorthand here for the local density of $Q_3^{(0)}$ (without the numerical prefactor). Constructing the IoMs via boost guarantees that the range of the charges grows linearly with their index; indeed, $Q_3^{(0)}$ has range $3$, $Q_4^{(0)}$ has range $4$, etc.

The Ishimori model has the usual time-reversal symmetry $\Theta: \vec{S}_j \to - \vec{S}_j, t \to -t$. 
In addition, the chain is invariant under the lattice inversion $I: \vec{S_j} \rightarrow \vec{S}_{L+1-j}$.
With the densities chosen above, the scalar IoMs $Q_\alpha^{(0)}$ can be organized by symmetry: even charges $(Q_2^{(0)}, Q_4^{(0)}\dots)$ are even under $\Theta$ and even under $I$, while the odd ones like the torsion ($Q_3^{(0)}, Q_5^{(0)},\dots$) are odd under $\Theta$ and odd under $I$.

We compute the associated currents using the continuity equation in Eq.~(\ref{eq:Jdef}) with $\alpha=2$ and list them in terms of their local density (this will matter when constructing the bi-local WIBs) 
\begin{align}
J^z_{12,j} &= -2 \frac{(\sj{j-1}\times \sj{j})^z}{1+\sj{j-1}\cdot\sj{j}} ~,  \label{eq:ISH_J12z} \\
J_{22,j} &= -4 \frac{(\sj{j}\times\sj{j+1})\cdot \sj{j+2}}{(1+\sj{j}\cdot\sj{j+1})(1+\sj{j+1}\cdot\sj{j+2})} ~, \label{eq:ISH_J22} \\
J_{32,j} &= 8 \bigg [\frac{1}{1+\vec{S}_{j}\cdot\vec{S}_{j+1}} -q_{3,j-1} q_{3,j} -  \\
&-\frac{((\vec{S}_{j-1}\times \vec{S}_{j})\times\vec{S}_{j+1})\cdot\vec{S}_{j+2}}{(1+\vec{S}_{j-1}\cdot\vec{S}_{j})(1+\vec{S}_{j}\cdot\vec{S}_{j+1})(1+\vec{S}_{j+1}\cdot\vec{S}_{j+2})}\bigg ]. \notag \label{eq:ISH_J32}
\end{align}
We write the local currents in the exact form required for the bi-local generator equation. 
The spin current $J^z_{12}$ is even under $\Theta$ and odd under $I$.
The Ishimori model belongs to the group of integrable models for which the energy current $J_{2,2;\text{tot}}:= \sum_j J_{22;j} =Q_3^{(0)}$ is conserved, and it is also odd under $\Theta$ and odd under $I$.
$J_{3,2;\text{tot}}$ is not conserved and is even under both $\Theta$ and $I$.

\subsection{Extensive local WIBs}
We begin with the simplest class of generators: extensive local and fully $\mathrm{SU}(2)$-symmetric functions of two spins. Concretely, we take
\begin{equation}
    X_{\text{ex}} = \sum_j f(\vec{S}_j\cdot \vec{S}_{j+1}),
\end{equation}
where $f$ is a smooth function of the nearest-neighbor scalar product.
The perturbation generated from such an $X$ following Eq.~(\ref{eq:VXH}) is 
\begin{align}
\{X_{\text{ex}}, H_0 \} &= \sum_j\big[h'_{j,j+1}f'_{j+1,j+2} - \notag \\& - f'_{j,j+1}h'_{j+1,j+2} \big](\sj{j}\times \sj{j+1}) \cdot \sj{j+2}  ~.
\end{align}
This perturbation inherits symmetries of $X_{\text{ex}}$, apart from the sign change under the time reversal coming from the Poisson bracket.

Since the Hamiltonian is fixed in our case, $h(y)=-2\ln(\frac{1+y}{2})$ and $h'(y)=-2\frac{1}{1+y}$, we can further write this as 
\begin{align}
    \{X_{\text{ex}}, H_0 \} &= -2\sum_j \bigg [(\sj{j}\times \sj{j+1})\cdot \sj{j+2} \bigg ]  \\
    & \times \left(\frac{f'_{j+1,j+2}}{1+\sj{j}\cdot\sj{j+1}} - \frac{f'_{j,j+1}}{1+\sj{j+1}\cdot\sj{j+2}} \right) \notag~.
\end{align}
Any $V_{\text{ex}}$ generated from a function of nearest neighbors will be a range-3 object.
As a concrete example, let $f(y)$ be the Heisenberg interaction. Then $f(y) = y$, $f'(y) = 1$, and 
\begin{align}
V_{\text{ex}} = & 2 \sum_j \bigg[ (\vec{S}_j \times \vec{S}_{j+1})\cdot\vec{S}_{j+2} \bigg ] \, \label{eq:ish_vex1} \\
& \times \left(\frac{1}{1 + \sj{j+1}\cdot\sj{j+2}} -\frac{1}{1 + \sj{j}\cdot\sj{j+1}} \right)
\notag
\end{align}
is an extensive local WIB perturbation of the Ishimori model.
This perturbation is odd under $\Theta$ and even under $I$.
Note that, with the classical model, we have more freedom in the number of perturbations we can construct than in the quantum case, e.g., we obtain non-trivial WIB already from nearest-neighbor extensive local generators.
Thus, in the spin-$1/2$ Heisenberg model, the shortest range SU(2) invariant $V_{\text{ex}}$ we are able to generate is range $4$ (it was also odd under $\Theta$ and even under $I$).
The reason for more freedom in the classical case is that, in some sense, the ``onsite Hilbert space'' is infinite-dimensional.

\subsection{Boosted WIBs}
The boost operator defined in Eq.~(\ref{eq:Vbo}) with $\beta=2$ in the Ishimori model can be used to generate higher order IoMs.
Hence, this $\beta$ does not produce WIBs.
However, using $\beta=3$, i.e., the local density of $Q_3^{(0)}$, instead, lets one construct 
quasi-IoMs. 
This procedure was formalized in \cite{Poszgay2020, Pozsgay2020aug}.

A direct computation using the boost as the generator $X_{\mathrm{bo}} \;\propto\; \mathcal{B}[Q_3^{(0)}] = \sum_j j\, q_{3,j}^{(0)}$ gives
\begin{align}
Q_2^{(1)} &\;\equiv\; \{X_{\mathrm{bo}}, H_0\} \equiv J_{3,2; \text{tot}} ~,
\end{align}
which is just the current of $Q_3^{(0)}$.
This is the direct classical analogue of the idea that the leading perturbing operator of the long–range deformations of spin chains is the current of a conserved charge ~\cite{Poszgay2020,Pozsgay2020aug,Kurlov2022, Surace2023}.
One could already use the $J_{3,2;\text{tot}}$ as a WIB of the boosted type; however, if the model satisfies the Reshetikhin criterion, i.e., a two-site operator $R_{j,j+1}$ exists such that
\begin{align}
\{h_{j,j+1}+h_{j+1,j+2}, g_{j,j+1,j+2}\}=R_{j,j+1} - R_{j+1, j+2} ~,
\end{align}
where \(h_{j,j+1}\) and \(g_{j,j+1,j+2}\) denote, respectively, the local Ishimori Hamiltonian density and the associated local energy current $J_{2,2;j} \equiv q_{3,j}^{(0)}$, then there exists a simplifying structure (See App.~C of \cite{Surace2023}) that lets us reduce the range by combining with other IoMs ($Q_4^{(0)}$ in this case), to construct a shorter range WIB.

In the Ishimori case, just like in the quantum Heisenberg case, we have 
\begin{equation}
    2 Q_2^{(1)} + Q_4^{(0)}
    = \sum_j \{h_{j+1,j+2}, g_{j,j+1,j+2}\}+3R_{j,j+1}
\end{equation}
with 
$$ R_{j,j+1} = \frac{8}{1+\vec{S}_j\vec{S}_{j+1}} ~.$$
This allows us to cancel all range-4 terms and yields an \emph{equivalent} WIB perturbation.
Evaluating the Poisson bracket explicitly for the Ishimori densities gives 
\begin{align}
&V_{\text{bo}} = 4\sum_j  \bigg [\frac{(\vec{S}_{j+2}\cdot(\vec{S}_{j+1}\times \vec{S}_{j}))^2}{(1+\vec{S}_j\cdot\vec{S}_{j+1})^2(1+\vec{S}_{j+1}\cdot\vec{S}_{j+2})^2}  \notag \\
& - \frac{\vec{S}_j\cdot\vec{S}_{j+2} +1}{(1+\vec{S}_j\cdot\vec{S}_{j+1})(1+\vec{S}_{j+1}\cdot\vec{S}_{j+2})} + 2 R_{j,j+1} \bigg] ~,
\label{eq:ish_vbo1} 
\end{align}
where the first term looks like the local density of $Q_3^{(0)}$ squared, which would reduce to a constant or vanish in the quantum spin–\(1/2\) setting, but is nontrivial in the classical model. 
The second term is a genuine range-3 next–nearest–neighbor (NNN) interaction $\vec{S}_j\cdot\vec{S}_{j+2}$ multiplied by Ishimori–type denominators together and a nearest–neighbor part that would be reduced to an IoM in the quantum Heisenberg chain. 
This is why we refer to $V_{\text{bo}}$ as the classical analog of the NNN perturbation in the quantum Heisenberg chain, which sparked interest in WIBs.
In terms of currents, this is 
\begin{align}
V_{\text{bo}} = 2J_{3,2; \text{tot}} + Q_4^{(0)} ~.
\end{align}
This perturbation is even under both $\Theta$ and $I$. 
WIBs generated like this are the shortest range boosted WIBs that do not break SU(2). 
Since the expression in Eq.~(\ref{eq:ish_vbo1}) is not particularly simple, it is unlikely that a perturbation of this type has been studied before (as was the case for some WIBs in the quantum setting). 

In fact, perturbations that previous numerical studies explored, such as 
\begin{align}
V_{\text{s,1}} &= \sum_j \sj{j} \cdot \sj{j+1} \, ,\label{eq:ish_vs1} \\
V_{\text{s,2}} &= \sum_j \sj{j} \cdot \sj{j+2} \, . \label{eq:ish_vs2}
\end{align}
will be our main examples of generic (strong), non-WIB perturbations.
They have the same symmetries as $V_{\text{bo}}$: both are even under $\Theta$ and $I$.

\subsection{Bilocal WIBs}
Given two charges \(Q_\beta\) and \(Q_\gamma\) with local densities \(q_{\beta,j}\) and \(q_{\gamma,j}\) and the corresponding currents \(J_{\beta\alpha,j}\) and \(J_{\gamma\alpha,j}\), where $\alpha=2$ since the Hamiltonian is fixed, we can generate the simplest SU(2) preserving bilocal WIB using 
\begin{equation}
X_{\mathrm{bi}} =[\,Q_{2}^{(0)} \mid Q_{3}^{(0)}\,] 
\end{equation}
which correspond to Hamiltonian density $q_{2,j}$ and the torsion density $q_{3,j}$. Direct computation of Eq.~(\ref{eq:bilocal}) gives
\begin{align} 
     V_{\text{bi,}23} &= \sum_j \bigg[ q_{2,j}(J_{32,j} + J_{32,j+1}) \notag \\
     &\quad \quad \quad  - q_{3,j}(J_{22,j} + J_{22,j+1}) \bigg] .
\label{eq:ish_vbi23}
\end{align}
One of the main reasons we explicitly list both the IoMs and the currents for all models we study is to make the construction of WIBs more transparent.
$V_{\text{bi,}23}$ has interactions of range $4$ and is even under $\Theta$ and $I$.

\subsection{SU(2) breaking WIB perturbations} 
Many of the recent studies of transport in the perturbed Ishimori chain have focused on SU(2)-preserving perturbations, typically adding nearest- and
next-to-nearest-neighbor Heisenberg terms of the form ${\sum_j \vec S_j \!\cdot\! \vec S_{j+r}}$ with $r=1,2$
(see e.g.~Refs.~\cite{Das_2019,McRoberts2023,Roy_2023,McRoberts2024,McCarthy2024}).
In such cases, the integrable Ishimori point is deformed within the SU(2)-symmetric manifold, and the observed persistence of KPZ-type superdiffusion of spin has often been attributed to the isotropic nature of the perturbations.
The exact microscopic origin of the spin superdiffusion is still unknown, but spin-rotation symmetry appears to be the key structural ingredient
(see e.g.~\cite{Gopalakrishnan_2023, Gopalakrishnan2024superdiffusion}). 
This motivates the introduction of SU(2)-breaking perturbations within our WIB framework: by comparing SU(2)-preserving WIBs to SU(2)-breaking WIBs, one could probe the role of spin symmetry in sustaining long-lived anomalous spin transport. 

First, we define an SU(2)-breaking WIB obtained by boosting $Q_1^{(0),z}$ (the $z$-component of the total spin),
\begin{align}
V^z_{\text{bo}} := J^z_{1,2;\text{tot}} = -2 \sum_j \frac{(\sj{j}\times \sj{j+1})^z}{1+\sj{j}\cdot\sj{j+1}}
    \label{eq:ish_vz_weak}
\end{align}
This is an SU(2)-breaking analog of
$V_{\text{bo}}$, and it is equivalent to the $z$-component of the total spin current $J_{1,2;\text{tot}}^{z}$.
As a benchmark for generic SU(2)-breaking perturbations, we include the isotropic spin current of the magnetization in the $z$–direction
\begin{align}
V^z_{\text{s,1}} = 2 \sum_j (\sj{j}\times \sj{j+1})^z ~.
\label{eq:ish_vz_strong}
\end{align}
This term is \emph{not} weak in our sense (it is not generated by a short-range $X$), and will serve as a reference for a generic SU(2)-breaking perturbation.
Both $ V^z_{\text{bo}}$ and $ V^z_{\text{s,1}}$ and  are odd under $\Theta$ and $I$.

Next, we construct an SU(2)-breaking bi-local WIB from  $Q_1^{(0),z}$, $Q_2^{(0)}$, and currents $J^z_{12}$ and $J_{22}$.
The generator $X_{\mathrm{bi}} =[\, Q_{1}^{(0),z} \mid Q_{2}^{(0)}\,]$ gives 
\begin{align}
     V^z_{\text{bi,}12} &= \sum_j \bigg [ - S_j^z \big( J_{22,j}+ J_{22,j+1}  \big)  +\notag \\
     & +2 h_{j,j+1}\cdot \bigg ( \frac{(\sj{j}\times \sj{j-1})^z}{1+\sj{j-1}\cdot\sj{j}} + \frac{(\sj{j+1}\times \sj{j})^z}{1+\sj{j}\cdot \sj{j+1}}  \bigg) \bigg] ~.
     \label{eq:ish_vbi12}
\end{align}
Finally, we consider two extensive local SU(2)-breaking perturbations generated from a staggered field $X_{\text{ex,1}} = \sum_j (-1)^j S_j^{z}$ and a nearest-neighbor Ising term $X_{\text{ex,2}} = \sum_j S_j^{z} S_{j+1}^{z}$
\begin{align}
V^z_{\text{ex},1} &=-4 \sum_j (-1)^j\frac{(\sj{j}\times \sj{j+1})^z}{1+\sj{j}\cdot\sj{j+1}},
\label{eq:ish_vz_stag_weak} \\
V_{\text{ex},2}^z  &= \sum_j J_{12,j}^z (-S_{j-2}^z + S_{j-1}^z - S_j^z + S_{j+1}^z) =\notag \\
&  2 \sum_j \frac{(\vec{S}_j \times \vec{S}_{j+1})^z}{1 + \vec{S}_j \cdot \vec{S}_{j+1}} (S_{j-1}^z - S_j^z + S_{j+1}^z - S_{j+2}^z).
\label{eq:ish_vzz}
\end{align}

We also define another strong SU(2) breaking perturbation
\begin{align}
V_{\text{s},2}^z  &= \sum_j J_{12,j}^z (-S_{j-2}^z + S_{j-1}^z - S_j^z + S_{j+1}^z) =\notag \\
&  2 \sum_j (\vec{S}_j \times \vec{S}_{j+1})^z(S_{j-1}^z - S_j^z + S_{j+1}^z - S_{j+2}^z).
\label{eq:ish_vz_strong2}
\end{align}
Recent studies have often attributed the slow thermalization in the perturbed Ishimori model to the isotropic nature of the perturbations used \cite{McCarthy2024}.
However, SU(2)-breaking WIB perturbation will also, by construction, delay diffusion. 

While we do not pursue the question of the robustness of superdiffusion under SU(2)-breaking perturbations in this work, repeating some of the studies we cited here with the perturbations we constructed above is an interesting future direction. 

\subsection{Strictly local WIB perturbations}
In addition to the extensive generators introduced previously, we can also consider \emph{strictly local} generators, which produce perturbations supported only on a finite region of the chain.
We introduced these types of perturbations in our previous work \cite{Vanovac2024}, where they proved to be useful in finding analytic expressions for finite-size analogs of generators of WIBs \footnote{Boosted and bi-local generators are formally defined only in infinite chains. 
Despite this, the produced perturbations and corrections to IoMs are well-defined within PBC. 
Using strictly local WIBs was particularly insightful to find the relations between formal infinite-size generators and finite-size operators.}.
In this work, we use them to support our later statements on the scaling of the AGP norm with system size. 

For the SU(2)-breaking strictly local perturbations, we obtain a local version of our $V^z_{\text{ex}}$ using $X^z_{\text{loc}} = S_1^z$, which we call $V^z_{\text{loc}}$. 
\begin{align}
V^z_{\text{loc}}=& J_{12,1}^z - J_{12,2}^z = \notag \\
& = -2 \bigg [\frac{(\sj{0}\times \sj{1})^z}{1+\sj{0}\cdot\sj{1}} -\frac{(\sj{1}\times \sj{2})^z}{1+\sj{1}\cdot\sj{2}} \bigg ]~. 
\label{eq:ish_loc1}
\end{align}
We can also obtain a non-trivial step-generated WIB, namely, using $X_{\text{step}}^z = \sum_{j \geq 1} S_j^z$, which gives the local spin current
\begin{align}
&  V^z_{\text{step}} = J_{12,1}^z = -2 \frac{(\sj{0}\times \sj{1})^z}{1+\sj{0}\cdot\sj{1}} 
\label{eq:ish_vstep0}
\end{align}
We define two more step-generated WIBs, which this time do not break SU(2) and correspond to local energy $J_{22,1}$ and torsion $J_{32,1}$ currents.
\begin{align}
&  V_{\text{step,1}} =  -4\frac{(\sj{0}\times\sj{1})\cdot \sj{2}}{(1+\sj{0}\cdot\sj{1})(1+\sj{1}\cdot\sj{2})},
\label{eq:ish_vstep1} \\
&  V_{\text{step,2}} = 8\bigg[ \frac{((\vec{S}_0 \times \vec{S}_1) \times \vec{S}_2) \cdot \vec{S}_3}{(1 + \vec{S}_0 \cdot \vec{S}_1)(1 + \vec{S}_1 \cdot \vec{S}_2)(1 + \vec{S}_2 \cdot \vec{S}_3)} \notag \\
&-\frac{(\sj{0}\cdot(\sj{1}\times \sj{2}))(\sj{1}\cdot(\sj{2}\times \sj{3}))}{(1+\sj{0}\cdot\sj{1})(1+\sj{1}\cdot\sj{2})^2(1+\sj{2}\cdot\sj{3})} 
- R_{1,2} \bigg].
\label{eq:ish_vstep2}
\end{align}
With this, we complete our construction of WIBs for the Ishimori model. 
The special type of integrability breaking we present here applies to all of the limits and regimes of the model.
WIBs can serve as natural benchmarks for numerical studies. 

For example, we show a case in which energy transport can be used to detect the type of integrability breaking, which is useful when the spin transport is robust to both generic and WIB perturbations on numerically observable timescales.
In fact, a recent study by Dhar et al. \cite{Das_2019} reported three regimes of integrability breaking; while superdiffusion persisted in all of them, energy transport displayed different behaviors.
However, both perturbations they studied were generic. 
Computing similar results for WIBs could help contextualize the "anomalous" regime they found. 

\section{Toda chain}
\label{sec:toda}
\subsection{Introduction and connections to the HOC and FPUT models}
The Toda chain (or Toda lattice) is one of the first examples of an integrable nonlinear lattice.
It is a simple model of a 1D crystal and, like the Ishimori chain, it hosts solitons.
It has been used to understand heat conduction in solids \cite{Toda_1974} and soliton dynamics in DNA \cite{Muto1990,Sataric1994}.
Its integrability was proven in \cite{Toda1975, GUTZWILLER1981} for periodic boundary conditions (which we use throughout the paper). 
The Hamiltonian 
\footnote{Note that the Hamiltonian used in~\cite{Danieli2024} differs from the one Toda used originally, but can be recovered by noting that $a = -\frac{1}{2\alpha}$ and $b = -2\alpha$.}
for the $N$-site periodic Toda chain is given by
\begin{equation}
H_0 = \sum_j  \left[\frac{p_j^2}{2} + V_0(q_{j+1} - q_j) \right] 
= \sum_j \left[\frac{p_j^2}{2} + V_0(r_{j}) \right] ~,
\end{equation}
where $q_j$ and $p_j$ are the position and momentum of the $j$-th particle and $r_j := q_{j+1} - q_j$.
The interaction is governed by the exponential potential 
\begin{equation}
    V_0(r) := \frac{a}{b} e^{-br} ~. \label{toda_pot2}
\end{equation}
It is convenient to consider the following potential instead,
\begin{equation}
\widetilde{V}_0(r) := \frac{a}{b} \left(e^{-br} + br - 1 \right)  \label{toda_pot1},
\end{equation}
to make the connection to related models clearer and to set the potential minimum at $r=0$.
Note that $\widetilde{V}_0(r)$ and $V_0(r)$ give the same Hamiltonian (up to an additive constant) for periodic boundary conditions, because $\sum_j r_j=0$; they have the same EOMs and conserve the same IoMs. 
By Taylor expanding $V_0(r) \approx \frac{a}{b}[1 - br + \frac{1}{2!}(-br)^2 + \frac{1}{3!}(-br)^3 +\frac{1}{4!}(-br)^4 +\dots]$, one can see that canceling the first two terms in the expansion by redefining $V_0\rightarrow \tilde V_0$ gives a clear connection to the HOC. 

In numerical studies, an additional step is made to express $a$ and $b$ parameters in terms of a single parameter $\alpha$.
In that case, the potential is usually written as
\begin{equation}
\widetilde{V}_0(r) = \frac{e^{2\alpha r} - 2\alpha r - 1}{4\alpha^2} ~. 
\label{eq:toda_pot3}
\end{equation}
The Toda lattice is known to be closely related to the well-known \textit{non-integrable} Fermi-Pasta-Ulam-Tsingou (FPUT) model and is commonly studied in relation to it. 
This connection is clear in the limit $b\rightarrow 0$ with $ab$ fixed. 
Let us take $ab = 1$ for simplicity, and $b = -2\alpha$.
Taylor-expanding the Toda potential for small $r$ gives
\begin{align*}
\widetilde{V}_0(r) = \frac{1}{2} r^2 + \frac{\alpha}{3} r^3 + \frac{\alpha^2}{6} r^4 + \mathcal{O}(\alpha^3r^5) ~.
\end{align*}
Truncating at third order in $\alpha r$ yields the potential for the $\alpha$-FPUT model directly; truncating at fourth order gives the $\alpha\beta$-FPUT model; and keeping only the fourth-order term is known as the $\beta$-FPUT model.

\begin{figure}[h]
    \includegraphics[width=0.99\linewidth]{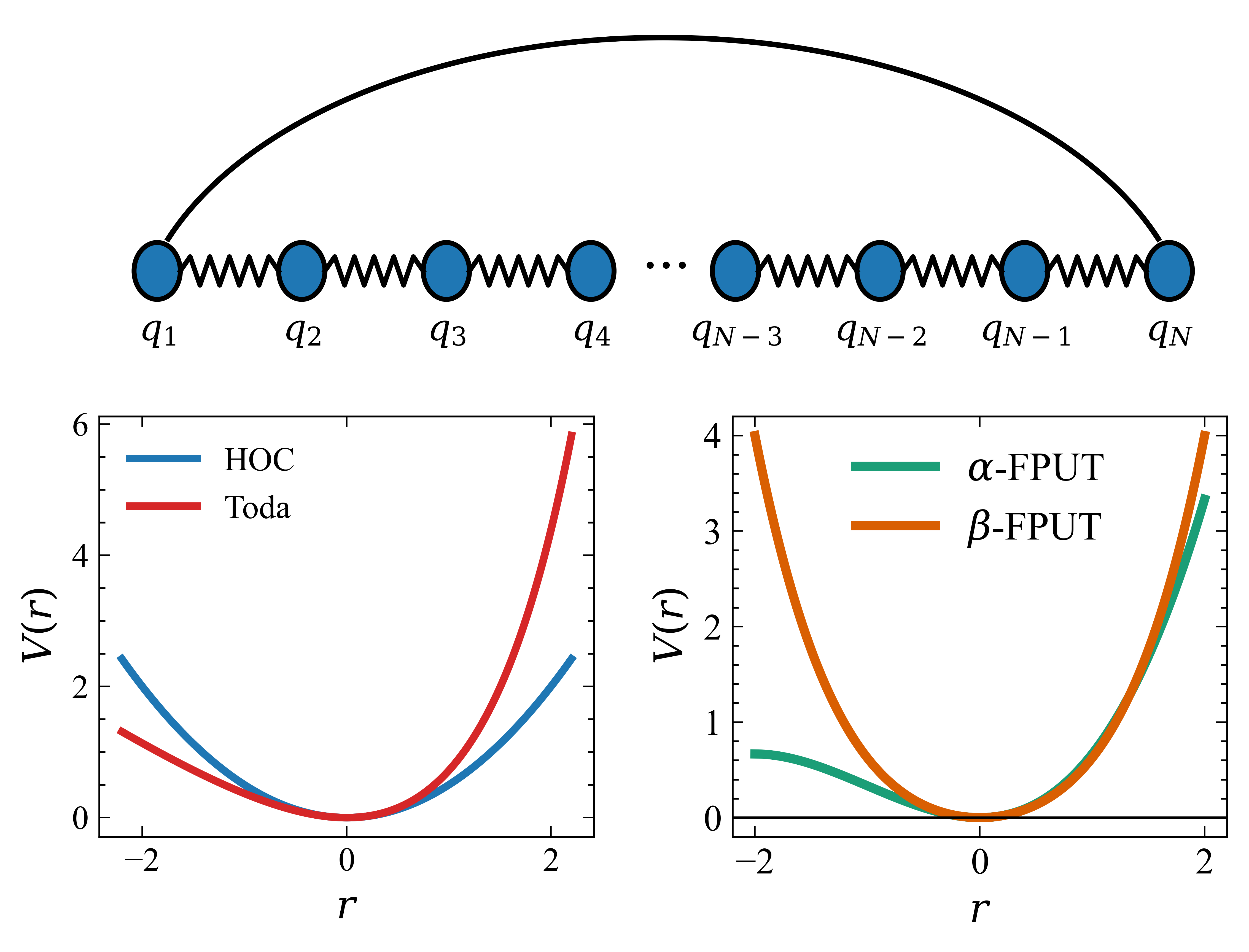}
    \caption{Top: Illustration of a 1D chain of interacting particles (oscillators) with periodic boundary conditions.
    Bottom, left: Toda and HOC potentials. Bottom, right: $\alpha$-FPUT and $\beta$-FPUT potentials. 
    We used Eq.~(\ref{eq:toda_pot3}) with $\alpha = 0.5$ for the left plot and $\frac{1}{2} r^2 + \frac{\alpha}{3} r^3$ and $\frac{1}{2} r^2 + \frac{\alpha^2}{6} r^4$ for the right plot.
    }
    \label{fig:placeholder}
\end{figure}

We define the two main ones here for reference
\begin{align}
&H_{\alpha{\text{-FPUT}}} = \sum_j \left[ \frac{p_j^2}{2} + \frac{(q_{j+1} - q_j)^2}{2} + \frac{\alpha (q_{j+1} - q_j)^3}{3} \right] ~, 
\label{eq:alpha}\\
&H_{\beta{\text{-FPUT}}} = \sum_j \left[ \frac{p_j^2}{2} + \frac{(q_{j+1} - q_j)^2}{2} + \frac{\beta (q_{j+1} - q_j)^4}{4} \right] ~. \label{eq:beta}
\end{align}
We note that a number of numerical studies have looked at these models since the original work by Fermi \textit{et al.}~\cite{Fermi1955} in the '50s, and we only mention some of the reviews here \cite{Weissert_2002,Berman_2005,Gallavotti2008}. 
In the language of perturbations, we define
\begin{align}
V_\alpha & := \frac{1}{3} \sum_j (q_{j+1} - q_j)^3 ~, \label{eq:valpha} \\
V_\beta & := \frac{1}{4} \sum_j (q_{j+1} - q_j)^4 ~. \label{eq:vbeta}
\end{align}
We will later study both of these potentials as perturbations of the integrable Toda and HOC models.
Our findings provide insight into the longstanding problem of anomalous thermalization in FPUT models.

In the next section, we first define Toda's IoMs.
We explicitly list the local densities and currents used to construct WIBs for the Toda model.
We do not find $V_\alpha$ and $V_\beta$ among the WIBs we construct for the Toda chain.
That is, as formal perturbations of the Toda chain at fixed non-zero $a$ and $b$, these perturbations are ``strong'' (non-WIB), which we confirm with the AGP study in Sec.~\ref{sec:agp_toda}.
However, as we will see in the HOC section, the generalized WIB framework can indeed be used to understand the $\alpha$-FPUT model.

\subsection{IoMs and currents of the Toda chain}
The Toda chain does not satisfy the Reshetikhin condition, and one cannot obtain higher IoMs by boosting the energy density. 
Fortunately, Toda IoMs up to $Q_5^{(0)}$ can be found in the literature starting from the original works by Henon \cite{Henon1974, Grava2020}.
They are usually obtained by computing traces of higher powers of the Lax matrix, and several papers detail how to automate this construction using various software tools \cite{Christodoulidi_2025}. 

Here, we state them in their usual form with the definition of the potential given in Eq.~(\ref{toda_pot2}).
\begin{align}
    & Q_1^{(0)} = \sum_{j=1}^{L} p_j, \label{eq:Toda_Q1}\\
    & Q_2^{(0)} = \sum_{j=1}^{L} \left[ \frac{p_j^2}{2} + V_{0}(q_{j+1} - q_j) \right], \label{eq:Toda_Q2} \\
    & Q_3^{(0)} = \sum_{j=1}^{L} \left[ \frac{p_j^3}{3} + (p_{j+1}+p_j)V_0(q_{j+1} - q_j) \right],  \label{eq:Toda_Q3}\\
    & Q_4^{(0)} = \sum_{j=1}^{L} \left[\frac{p_j^4}{4} + (p_{j+1}^2+p_jp_{j+1}+p_j^2)V_0(q_{j+1} - q_j)  \right. \notag\\
    & \left. + \frac{1}{2}V_0(q_{j+1} - q_j)^2 + V_0(q_{j+2} - q_{j+1})V_0(q_{j+1} - q_j) \right] \label{eq:Toda_Q4} ~.
\end{align}
$Q_5^{(0)} $ can be found in App.~\ref{app:toda_ioms_to_corr}.

Toda model is invariant under the time reversal symmetry $\Theta: q_j \to q_j, p_j \to -p_j$.
In addition to lattice translation symmetry $q_j \rightarrow q_{j+1}, p_j \rightarrow p_{j+1}$, it also has the following symmetry involving lattice reflection: $I^{\text{Toda}}: q_j \to -q_{L-j+1}, p_j \to -p_{L-j+1}$ (for concreteness, showing inversion in the chain midpoint; by lattice translation symmetry, however, the inversion point can be chosen at any site or bond center). The even IoMs are even under both $\Theta$ and $I^{\text{Toda}}$, while the odd IoMs are odd under $\Theta$ and $I^{\text{Toda}}$.

For the general Toda model with parameters $a$ and $b$, we write the densities of conserved quantities to have definite symmetry numbers under $I^{\text{Toda}}$:
\begin{align}
\rho_{1,j} \equiv{}& p_j ~, \label{eq:Toda_rho1} \\
\rho_{2,j} \equiv{}& \frac{p_j^2}{2} + \frac{a}{2b} \left( e^{b(q_{j-1} - q_j)} + e^{b(q_j - q_{j+1})} \right) ~, \label{eq:Toda_rho2} \\
\rho_{3,j} \equiv{}& \frac{p_j^3}{3} + \frac{a}{b}  p_j \left( e^{b(q_{j-1} - q_j)} + e^{b(q_j - q_{j+1})} \right) ~, \label{eq:Toda_rho3} \\
\begin{split}\label{eq:Toda_rho4}
\rho_{4,j} \equiv{}& \frac{p_j^4}{4} + \frac{a}{b} p_j^2 \left( e^{b(q_{j-1} - q_j)} + e^{b(q_j - q_{j+1})} \right) \\
& + \frac{a}{2b} p_j \left( p_{j-1} e^{b(q_{j-1} - q_j)} + p_{j+1} e^{b(q_j - q_{j+1})} \right) \\
& + \frac{a^2}{4b^2} \left( e^{2b(q_{j-1} - q_j)} + e^{2b(q_j - q_{j+1})} \right) \\
& + \frac{a^2}{b^2} e^{b(q_{j-1} - q_{j+1})} ~.
\end{split} 
\end{align}
We also write out the currents of conserved quantities with respect to $H_0 = Q_2^{(0)}$, i.e., $J_{\beta,2;j}$ defined in Eq.~(\ref{eq:Jdef}): 
\begin{align}
J_{1,2;j} ={}& a e^{b(q_{j-1} - q_j)} ~, \label{eq:Toda_J12} \\
J_{2,2;j} ={}& \frac{a}{2} e^{b(q_{j-1} - q_j)} (p_{j-1} + p_j) ~, \label{eq:Toda_J22} \\
J_{3,2;j} ={}& a p_{j-1} p_j e^{b(q_{j-1} - q_j)} + \frac{a^2}{b} e^{2b(q_{j-1} - q_{j})} ~, \label{eq:Toda_J32} \\
\begin{split}\label{eq:Toda_J42}
J_{4,2;j} ={}& \frac{a}{2} p_{j-1} p_j (p_{j-1} + p_j) e^{b(q_{j-1} - q_j)} \\
& + \frac{a^2}{b} (p_{j-1} + p_j) e^{2b(q_{j-1} - q_j)} \\
& + \frac{a^2}{2b} \left( p_{j-1} e^{b(q_{j-1} - q_{j+1})} + p_j e^{b(q_{j-2} - q_j)} \right).
\end{split}
\end{align}
Our choice of writing the densities in this way will help in deriving the $V_{\text{bo}}$- and $V_{\text{bi}}$-type WIB perturbations, which will then automatically have definite $I^{\text{Toda}}$ symmetry
and ourexpressions for the currents will be used directly in the construction of the  $V_{\text{bo}}$ and $V_{\text{bi}}$ WIB perturbations.

\subsection{Integrable directions in the parameter space of the Toda model}
We begin by noting that varying $a$ and $b$ moves us within the integrable model parameter space, with IoMs that depend on these parameters.
An infinitesimal variation of $H_0$ can be viewed as producing a ``weak perturbation'' in the sense that we can find the corresponding corrections to all IoMs~\cite{Surace2023}.
For example, we can obtain the following special perturbation:
\begin{equation}
V_a \equiv \frac{\partial H_0}{\partial a} = \sum_j \frac{1}{b} e^{b(q_j - q_{j+1})} ~, \label{eq:vprime}
\end{equation}
with corrections to the IoMs given by $Q_\alpha^{(1)} = \partial Q_\alpha^{(0)}/\partial a$.
It is important to note that this does not guarantee the existence of a simple, extensive local generator $X$ that can produce this perturbation.
Nevertheless, we will see that this perturbation can be generated by a boost-type generator $X_{\text{bo}} = \mathcal{B}[Q_1]$. By combining with $H_0$, we can obtain a perturbation
\begin{equation}
\widetilde{V}_a \equiv V_a -\frac{1}{a} H_0 = -\frac{1}{a} \sum_j \frac{p_j^2}{2} ~.
\label{eq:Toda_Va}
\end{equation}
This perturbation is special in the same way as $V_a$, with the same corrections to all IoMs.
In the same spirit, we obtain another special perturbation
\begin{equation}
V_b \equiv \frac{\partial H_0}{\partial b} = \sum_j \frac{a}{b} e^{b(q_j - q_{j+1})} \left(q_j - q_{j+1} - \frac{1}{b} \right)~. 
\end{equation}
We will see shortly that a particular combination of $V_a$ and $V_b$ can be generated by an extensive local generator $X$ that simply rescales $q$ and $p$ while preserving their Poisson bracket. 

\subsection{WIB perturbations to Toda from extensive local generators}
\label{subsec:Todaex}
Consider the generator $X_{\text{ex,0}} = \sum_j q_j p_j$, which gives
\begin{equation}
\begin{aligned}
V_{\text{ex}, 0} &= \sum_j \left[ p_j^2 - a e^{b(q_j - q_{j+1})} (q_j - q_{j+1}) \right] \\
&= 2H_0 - 3a V_a - b V_b ~.
\label{eq:Toda_Vex0}
\end{aligned}
\end{equation}
The last equality is easy to check directly.
We can trace the underlying reason to the following relation of this generator to the derivatives in the integrable model parameter space.
In fact, we can obtain the exact canonical transformation \footnote{This is, in fact, an example of a well-known canonical transformation called \textit{dilation} which plays a central role in renormalization group; see \cite{KOLODRUBETZ_2017} for review. }
produced by this generator for finite $\lambda$: $q \to e^{-\lambda} q$, $p \to e^{\lambda} p$, i.e., a simple rescaling of the phase space variables that preserves the Poisson bracket.
We can perform the corresponding transformation $H_0 \to H_0(\lambda)$, while the above perturbation is simply $V_{\text{ex},0} = \partial H_0(\lambda)/\partial \lambda$ at $\lambda = 0$.

The above rescaling of the phase space variables can be restated as a particular transformation of the parameters of the Hamiltonian as well as of its overall scale, which we can then use to understand the specific relation between the above $V_{\text{ex},0}$ and $V_a$, $V_b$, and $H_0$.

We note that the above generator $X_{\text{ex},0}$ is not momentum-conserving, but the perturbation $V_{\text{ex},0}$ generated in this way is.
This can be understood by considering the following general argument. 
Starting with the Jacobi identity 
\begin{align}
& \{\{X_{\text{ex}}, H_0\}, P_{\text{tot}}\} + \{\{H_0, P_{\text{tot}}\}, X_{\text{ex}}\} + \notag \\
&~~~ + \{\{P_{\text{tot}}, X_{\text{ex}}\}, H_0\} = 0 ~,
\label{eq:jacobi}
\end{align}
we can use the fact that the total momentum is an IoM ($\{H_0, P_{\text{tot}}\} = 0$) to conclude that the perturbation $V_{\text{ex}}= \{X_{\text{ex}}, H_0\}$ is momentum-conserving, i.e., $\{V_{\text{ex}}, P_{\text{tot}}\} = 0$, if the generator is momentum conserving, $\{P_{\text{tot}}, X_{\text{ex}}\}=0$, or if $\{P_{\text{tot}}, X_{\text{ex}}\}$ is an IoM. 
The perturbation $V_{\text{ex,0}}$ in Eq.~(\ref{eq:Toda_Vex0}) belongs to the second group since the P.B.\ with $P_{\text{tot}}$ gives an IoM.

We now consider several additional generic examples of $X_{\text{ex}}$. For concreteness, it is natural to restrict to lattice-translation-invariant generators, and require that they conserve $P_{\text{tot}} = \sum_j p_j = Q_1^{(0)}$. In other words, generators $X_{\text{ex}}$ are functions of relative coordinate distances $q_j - q_{j'}$ which guarantee that the corresponding perturbations are as well.
We classify the corresponding perturbations $V_{\text{ex}}$ based on their symmetry properties under time reversal $\Theta$ (determined simply by the parity of the powers of momentum $p$ in $V_{\text{ex}}$) and Toda inversion $I^{\text{Toda}}$.
We will label the corresponding symmetry numbers $(\Theta_V, I^{\text{Toda}}_V)$ as superscripts on the perturbations $V$.

First, consider a momentum-conserving generator $X_{\text{ex}} = \sum_j \frac{1}{3} p_j^3$, and the resulting perturbation
\begin{equation}
V_{\text{ex}}^{(+,-)} = \sum_j a e^{b(q_j - q_{j+1})} (p_{j+1}^2 - p_j^2) ~.\label{eq:Toda_Vex1}
\end{equation}
$V_{\text{ex}}^{(+,-)}$ is even under the time reversal and odd under the Toda lattice inversion, $(\Theta_V = +, I^{\text{Toda}}_V = -)$.

Next, consider $X_{\text{ex}} = \sum_j e^{b(q_j - q_{j+1})} (p_j - p_{j+1})$, which gives 
\begin{equation}
\begin{aligned}
V_{\text{ex}}^{(+,+)} = \sum_j \bigg[& b e^{b(q_j - q_{j+1})} (p_j - p_{j+1})^2 \\
& - 2a e^{2b(q_j - q_{j+1})} + 2a e^{b(q_j - q_{j+2})} \bigg] ~. \label{eq:Toda_Vex2}
\end{aligned}
\end{equation}
$V_{\text{ex}}^{(+,+)}$ is an important example with $(\Theta_V = +, I^{\text{Toda}}_V = +)$, 
which we will combine later with $V_{\text{bi}}$- and $V_{\text{bo}}$-type perturbations to construct simpler WIBs.

Finally, consider $X_{\text{ex}} = \sum_j e^{b(q_j - q_{j+1})}$, which gives
\begin{equation}
V_{\text{ex}}^{(-,+)} = \sum_j b e^{b(q_j - q_{j+1})} (p_j - p_{j+1}) ~.
\label{eq:Toda_Vex3}
\end{equation} 
an $V_{\text{ex}}$-type WIB that is odd under time reversal and even under inversion.

\subsection{WIB perturbations to Toda from boosted generators}
Having defined local densities in Eqs.~(\ref{eq:Toda_rho1})-(\ref{eq:Toda_rho4}), and having calculated local currents in Eqs.~(\ref{eq:Toda_J12})-(\ref{eq:Toda_J42}) for all of the IOMs, they immediately give WIB perturbations generated by $X^{[\beta]}_{\text{bo}} = {\mathcal B}[Q_\beta] = \sum_j j \rho_{\beta, j}$, cf.\ Eq.~(\ref{eq:Vbo}):
\begin{align}
V_{\text{bo}}^{[1]} &= \sum_j a e^{b(q_j - q_{j+1})} ~, \label{eq:Toda_Vbo1} \\
V_{\text{bo}}^{[2]} &= \sum_j \frac{a}{2} (p_j + p_{j+1}) e^{b(q_j - q_{j+1})} ~, \label{eq:Toda_Vbo2} \\
V_{\text{bo}}^{[3]} &= \sum_j \bigg [ a p_j p_{j+1} e^{b(q_j - q_{j+1})} + \frac{a^2}{b} e^{2b(q_j - q_{j+1})} \bigg ], \label{eq:Toda_Vbo3} \\
\begin{split}
V_{\text{bo}}^{[4]} &= \sum_j \bigg[ \frac{a}{2} p_j p_{j+1} (p_j + p_{j+1}) e^{b(q_j - q_{j+1})} \\
& \qquad\quad\; + \frac{a^2}{b} (p_j + p_{j+1}) e^{2b(q_j - q_{j+1})} \\
& \qquad\quad\; + \frac{a^2}{2b} (p_j + p_{j+2}) e^{b(q_j - q_{j+2})} \bigg]. \label{eq:Toda_Vbo4}
\end{split}
\end{align}
We organized the terms in all of the expressions above by the leftmost nontrivial degree of freedom.
$V_{\text{bo}}^{[1]}$ and $V_{\text{bo}}^{[3]}$ are even under both the time reversal and the Toda lattice inversion, while $V_{\text{bo}}^{[2]}$ and $V_{\text{bo}}^{[4]}$ are odd under the two symmetries.
As expected, the range and complexity of the perturbations increase as we go to higher $\beta$ in $V_{\text{bo}}^{[\beta]}$, but the initial few are still fairly simple.

We first note that $V_{\text{bo}}^{[1]}$ is a special perturbation that simply moves in the integrable Toda manifold, reproducing $V_a$ discussed in Eq.~(\ref{eq:Toda_Va}) earlier, namely $V_a = \frac{1}{ab} V_{\text{bo}}^{[1]}$.
Thus, $V_a$ perturbation is equivalent to a $V_{\text{bo}}$-type WIB perturbation.
We remark that, in general, finding generators for perturbations that are tangential to the integrability manifold is non-trivial, e.g., we do not know them analytically in the quantum XXZ chain~\cite{Pandey2020, Surace2023}.

Next we consider $V_{\text{bo}}^{[2]}$.
Unlike in the Ishimori model, $V_{\text{bo}}^{[2]}$ is a non-trivial perturbation because the Reshetikhin criterion does not hold (i.e., there is no boost generating conserved quantities).
By combining with $Q_3^{(0)}$, we obtain an equivalent WIB perturbation
\begin{equation}
\widetilde{V}_{\text{bo}}^{[2]} \equiv V_{\text{bo}}^{[2]} - \frac{b}{2} Q_3^{(0)} =
-\frac{b}{6} \sum_j p_j^3 ~,\label{eq:Toda_Vbo2_tilde}
\end{equation}
which has only on-site terms!
We find that a simple $\Theta$-odd and $I^{\text{Toda}}$-odd perturbation---arguably the simplest possible perturbation with such symmetry properties---is a WIB perturbation.

Finally, we turn to $V_{\text{bo}}^{[3]}$ (loosely speaking, this is an analog of $V_{\text{bo}}$ in our Ishimori model study since it was also generated using $\mathcal{B}[Q_3^{(0)}]$, i.e., boost of the first non-trivial IoM).
As generated, it has only nearest-neighbor couplings and is already much simpler than in the Ishimori case, making it particularly appealing for numerical simulations.

We can combine $V_{\text{bo}}^{[3]}$ with $V_{\text{ex}}^{(+,+)}$ and $Q_4^{(0)}$ to obtain an equivalent WIB perturbation that involves only nearest-neighbor couplings and has all terms positive (which could be helpful for more robust simulations):
\begin{equation}
\begin{aligned}
\widetilde{V}_{\text{bo}}^{[3]} & \equiv -V_{\text{bo}}^{[3]} - \frac{a}{2b} V_{\text{ex}}^{(+,+)} + b Q_4^{(0)}=  \\
& = \sum_j \bigg[ \frac{b}{4} p_j^4 + \frac{a}{2} (p_j + p_{j+1})^2 e^{b(q_j - q_{j+1})}  \\
& \qquad\quad\, + \frac{a^2}{2b} e^{2b (q_j - q_{j+1})} \bigg]~. \label{eq:Toda_Vbo3_tilde}
\end{aligned}
\end{equation}
Alternatively, we can find a combination that gives a perturbation where momentum and coordinate variables do not couple directly (one benefit of which is the ability to perform computational studies relying on symplectic integrators such as ABA864, which we used as well \cite{Danieli_2019, Danieli2024}):
\begin{equation}
\begin{aligned}
\doublewidetilde{V}{}_{\!\text{bo}}^{[3]} & \equiv -3 V_{\text{bo}}^{[3]} - \frac{a}{b} V_{\text{ex}}^{(+,+)} + b Q_4^{(0)} \\ 
& = \sum_j \! \bigg[\frac{b}{4} p_j^4 -\frac{a^2}{2b} e^{2b(q_j - q_{j+1})} - \frac{a^2}{b} e^{b(q_j - q_{j+2})} \bigg]. \label{eq:Toda_Vbo3_tilde_tilde}
\end{aligned}
\end{equation}
We will later encounter yet more uses of $V_{\text{bo}}^{[3]}$ to obtain a WIB perturbation that depends only on the $q$ variables and not on the $p$ variables, or that has only nearest-neighbor terms and $q$ and $p$ variables not coupled directly.

$V_{\text{bo}}^{[4]}$ is an analog of $V_{\text{bo},2}$ in our quantum Heisenberg model study in Ref.~\cite{Vanovac2024}.

\subsection{WIB perturbations to Toda from a bilocal generator}
We can obtain the simplest WIB perturbation from a bilocal generator by using $Q_1^{(0)}$ and $Q_2^{(0)}$.
Using $\rho_{1,j}$, $\rho_{2,j}$ from Eqs.~(\ref{eq:Toda_rho1})-(\ref{eq:Toda_rho2}) and $J_{1,2;j}$, $J_{2,2;j}$ from Eqs.~(\ref{eq:Toda_J12})-(\ref{eq:Toda_J22}) and applying Eq.~(\ref{eq:bilocal}), we obtain
\begin{equation}
\begin{aligned}
V_{\text{bi}}^{[1,2]} = \sum_j \bigg[& a p_j p_{j+1} e^{b(q_j - q_{j+1})} \\
& - \frac{a^2}{b} \left(e^{2b(q_j - q_{j+1})} + e^{b(q_j - q_{j+2})} \right) \bigg] ~.
\label{eq:Toda_Vbi12}
\end{aligned}
\end{equation}
We can obtain a simpler WIB perturbation that is a function of only $q$'s by combining with $V_{\text{bo}}^{[3]}$ as follows:
\begin{equation}
\begin{aligned}
\widetilde{V}_{\text{bi}}^{[1,2]} & \equiv - V_{\text{bi}}^{[1,2]} + V_{\text{bo}}^{[3]} \\
& = \frac{a^2}{b} \sum_j \left( 2 e^{2b(q_j - q_{j+1})} + e^{b(q_j - q_{j+2})} \right) ~.
\end{aligned}
\label{eq:Toda_tildeVbi12}
\end{equation}
In the above, we have chosen the overall sign to ensure that the potential is bounded below, so adding this perturbation with a positive coupling would yield a particularly nice model (e.g., with well-defined thermodynamics).

Alternatively, the following combination gives a perturbation with couplings of only up to nearest-neighbors and with the $p$ and $q$ variables not directly coupled:
\begin{equation}
\begin{aligned}
\doublewidetilde{V}{}_{\!\text{bi}}^{[1,2]}
&\equiv - V_{\text{bi}}^{[1,2]}
- 2 V_{\text{bo}}^{[3]} - \frac{a}{b} V_{\text{ex}}^{(+,+)} + b Q_4^{(0)} \\
& = \sum_j \left( \frac{b}{4} p_j^4 +\frac{3 a^2}{2b} e^{2b(q_j - q_{j+1})} \right) ~.
\label{eq:Toda_tildetildeVbi12}
\end{aligned}
\end{equation}

Thus, we see that rather simple perturbations with the same symmetries as the original Toda model are in fact WIB perturbations combining $V_{\text{bi}}$- and $V_{\text{bo}}$-type terms.
In Sec.~\ref{sec:agp_toda}, we will see how the particular WIB character of such perturbations 
[we will study specifically $\widetilde{V}_{\text{bi}}^{[1,2]}$ of Eq.~(\ref{eq:Toda_tildeVbi12})]
manifests itself in the properties of the corresponding AGP in the periodic chain.

We could also construct bilocal generators from other IoMs and currents. For example, $V_{\text{bi}}^{[1,3]}$ is obtained from a bilocal generator constructed from $Q_1^{(0)}$ and $Q_3^{(0)}$. Using $\rho_{1,j}$, $\rho_{3,j}$ from Eqs.~(\ref{eq:Toda_rho1})-(\ref{eq:Toda_rho3}) and $J_{1,2;j}$, $J_{3,2;j}$ from Eqs.~(\ref{eq:Toda_J12})-(\ref{eq:Toda_J32}) and applying Eq.~(\ref{eq:bilocal}), we obtain
\begin{align}
V_{\text{bi}}^{[1,3]} &= \sum_j \bigg[\frac{a}{3}(p_{j-1}^3 - p_j^3)e^{b(q_{j-1} - q_{j})}+\frac{2a^2}{b}p_je^{b(q_{j-1} - q_{j+1})} \notag \\
&-a(p_{j-1}^2p_j + p_{j-1}p_j^2)e^{2b(q_{j-1} - q_{j})}\bigg] 
\label{eq:Vbi13_toda} 
\end{align}
Note that we did not find the $\alpha$-FPUT and $\beta$-FPUT perturbations among the WIBs. 
In fact, the AGP variance calculations in Sec.~\ref{sec:agp_toda} will show that these are strong perturbations and hence cannot be cast as WIBs.

Nevertheless, the WIBs we do find are interesting in their own right.
For example, we can ask if the construction that gave $\doublewidetilde{V}{}_{\!\text{bi}}^{[1,2]}$ can be continued to higher orders in the perturbation while maintaining the nearest-neighbor character of the perturbation, possibly giving an interesting new integrable model.
We leave such explorations for future work.

\section{Harmonic oscillator chain}
\label{sec:choc}
In this section, we consider quasi-integrable deformations of the classical harmonic oscillator chain (HOC) of $L$ particles of unit masses and unit spring constants, whose positions and momenta are denoted by $q_j, p_j$, $j = 1, 2, \dots, L$.
The undeformed Hamiltonian is given by
\begin{equation}
    H_0 = \sum_{j=1}^L \left[ \frac{p_j^2}{2} + \frac{(q_{j+1} - q_j)^2}{2} \right] ~.\label{eq:HOC_hamiltonian}
\end{equation}
We will assume periodic boundary conditions (PBC), $q_{L+1} \equiv q_1$, $p_{L+1} \equiv p_1$. In equations below, we often simply write $\sum_j$ for such PBC sums.
The HOC is an example of an integrable system that is both analytically tractable and has a macroscopic description in terms of hydrodynamics \cite{Doyon_2017,Doyon2017GHD,DeNardis_2018,Bernardin_2019,Lepri_2020}.

The first few integrals of motion of this model are given by
\begin{align}
& Q_1^{(0)} = P_{\text{tot}} = \sum_j p_j ~, \label{eq:HOC_Q1}\\
& Q_2^{(0)} \equiv H_0 ~, \label{eq:HOC_Q2} \\
& Q_3^{(0)} = \frac{1}{2} \sum_j (q_j p_{j+1} - p_j q_{j+1}) ~, \label{eq:HOC_Q3} \\
& Q_4^{(0)} = \frac{1}{2} \sum_j (p_j p_{j+1} + 2 q_j q_{j+1} - q_j^2 - q_j q_{j+2}) ~. \label{eq:HOC_Q4}
\end{align}
Here, $Q_1^{(0)}$ is the conserved total momentum of the particles; $Q_2^{(0)}$ is the Hamiltonian; while $Q_3^{(0)}$ and $Q_4^{(0)}$ are the first nontrivial IoMs. In fact, the nontrivial IoMs generalize to
\begin{align}
& Q_{2n+1}^{(0)} = \frac{1}{2} \sum_j \left(q_j p_{j+n} - p_j q_{j+n} \right) ~, \label{eq:HOC_oddIoMs}\\
& Q_{2n+2}^{(0)} = \frac{1}{2} \sum_j \left [p_j p_{j+n} + q_j (2q_{j+n} - q_{j+n-1} - q_{j+n+1}) \right ] ~. \label{eq:HOC_evenIoMs}
\end{align}
The overall factors $\frac{1}{2}$ are not fundamental; our choice is such that $Q_{2n+2}^{(0)}$ for $n = 0$ matches $H_0$, which is a common definition of $Q_2^{(0)}$, while the expressions are still simple for all other $n$ (see also footnote
\footnote{
For aficionados, the harmonic chain satisfies a Reshetikhin-type criterion~\cite{Grabowski1995}, and the consecutive IoMs can be obtained by using boost of the Hamiltonian, $\mathcal{B}[Q_2^{(0)}] = \sum_j j \rho_{2,j}$, as a ladder operator.
Specifically, using site-inversion-symmetric $\rho_{2,j}$ from Eq.~(\ref{eq:HOC_rho2j}), we have (in a formal infinite system sense):
\begin{align*}
& Q_3 = \left\{ \mathcal{B}[Q_2^{(0)}], Q_2^{(0)} \right\} ~, \\
& Q_4 = \left\{ \mathcal{B}[Q_2^{(0)}], Q_3^{(0)} \right\} ~, \\
& \left\{ \mathcal{B}[Q_2^{(0)}], Q_{2n+1}^{(0)} \right\} = n Q_{2n+2}^{(0)}, \quad n \geq 1 ~, \\
& \left\{ \mathcal{B}[Q_2^{(0)}], Q_{2n+2}^{(0)} \right\} = \left(n + \frac{1}{2} \right) Q_{2n+3}^{(0)} \\
& \qquad\quad -2 n Q_{2n+1}^{(0)} + \left(n - \frac{1}{2} \right) Q_{2n-1}^{(0)}, \quad n \geq 1 ~.
\end{align*}
[Only in the last formula and only for $n = 1$, we use sentinel $Q_1^{(0)} = 0$.]}).

Besides the lattice translation symmetry $q_j \to q_{j+1},~ p_j \to p_{j+1}$, the HOC is invariant under the time reversal $\Theta: q_j \to q_j,~ p_j \to -p_j$, and lattice inversion $I: q_j \to q_{L-j+1},~ p_j \to p_{L-j+1}$ (we took specific inversion point for concreteness, but because of the lattice translation symmetry we can equivalently invert in any site or bond center).
The even IoMs have the same $\Theta$ and $I$ symmetries as $H_0$, while the odd IoMs have the opposite symmetry numbers to $H_0$ (i.e., they are odd under both $\Theta$ and $I$).
Note that when referring to the odd IoMs, we usually exclude $Q_1^{(0)}$, which represents the total momentum and is considered separately, as a conventional conserved quantity.
As we pointed out in Sec.~\ref{sec:toda}, the HOC can be obtained as a particular limit of the Toda chain: $a \to \infty$, $b \to 0$, while keeping $\kappa \equiv a b$ fixed (which then gives the spring constant $\omega^2=\kappa$).
The HOC IoMs can, in principle, be extracted from the Toda chain IoMs.
However, this requires starting from increasingly more complicated Toda IoMs and taking the above limit rather carefully, keeping track of the subleading terms (see App.~\ref{app:toda_ioms_to_corr}), while the listed IoMs for the HOC are simple to derive directly (see, e.g., Ref.~\cite{Pandey2024} for the even IoMs; see also App.~\ref{subapp:HOC_AGP_kspace_IoMcorrs} for expressions in terms of Fourier modes).
Furthermore, while the Toda and HOC share the same time reversal symmetry, the Toda chain does not have the same simple lattice inversion symmetry (the Toda chain has a different symmetry involving lattice inversion $I^{\text{Toda}}$ which is present also for the HOC model but is less convenient).
Since we use the simple lattice inversion as one of the organizing tools, it is more convenient and transparent to work directly with the HOC from scratch.

In our construction of WIBs, we need densities of the first few conserved quantities:
\begin{align}
& \rho_{1,j} \equiv p_j ~, \label{eq:HOC_rho1j} \\
& \rho_{2,j} \equiv \frac{1}{2} p_j^2 + \frac{1}{4} (q_j - q_{j-1})^2 + \frac{1}{4} (q_j - q_{j+1})^2 ~, \label{eq:HOC_rho2j} \\
& \rho_{3,j} \equiv \frac{1}{2} p_j (q_{j-1} - q_{j+1}) ~, \label{eq:HOC_rho3j} \\
& \rho_{4,j} \equiv \frac{1}{4} p_j (p_{j-1} + p_{j+1}) - \frac{1}{2} (q_j - q_{j-1})(q_j - q_{j+1}) ~.  \label{eq:HOC_rho4j}
\end{align}
We have chosen to define densities $\rho_{\alpha,j}$ to commute with $P_{\text{tot}}$ and to have definite transformation properties under inversion in site $j$:
$\rho_{2,j}$ and $\rho_{4,j}$ are even under this inversion, while $\rho_{3,j}$ is odd.
This is convenient later when obtaining $V_{\text{bi}}$-type WIBs, which will then automatically conserve $P_{\text{tot}}$ and have definite inversion symmetry properties.

Using conventions in Eq.~(\ref{eq:Jdef}), the corresponding currents are:
\begin{align}
& J_{1,2;j} = q_{j-1} - q_j ~, \label{eq:HOC_J1j} \\
& J_{2,2;j} = \frac{1}{2}(q_{j-1} - q_j) (p_{j-1} + p_j) ~, \label{eq:HOC_J2j} \\
& J_{3,2;j} = \frac{1}{2} p_{j-1} p_j + \frac{1}{2} (q_{j-1} - q_j)^2 ~, \label{eq:HOC_J3j} \\
& J_{4,2;j} = \frac{1}{4} \big[(q_{j-2} - q_j) p_j + (q_{j-1} - q_{j+1}) p_{j-1} \big] \label{eq:HOC_J4j} ~.
\end{align}
While there is freedom when defining the densities $\rho_{\alpha,j}$, once a choice is made, the expressions for the currents $J_{\alpha,2;j}$ are fixed up to additive constants.

Before proceeding, we can make the following simple observations.
First, on the PBC chain, we have $J_{1,2;\text{tot}} = 0$.
Next, the total energy current $J_{2,2;\text{tot}}$ is equal to $Q_3^{(0)}$ and hence is conserved.
On the other hand, $J_{3,2;\text{tot}}$ is not conserved.
This pattern generalizes to the higher-order IoMs:
For an even IoM, its total current can be expressed in terms of odd IoMs and is hence conserved (e.g., $J_{4,2;\text{tot}} = \frac{1}{2} Q_5^{(0)}$ ).
On the other hand, for an odd IoM, its total current is not conserved.

\subsection{WIB perturbations to HOC from extensive local generators}
There are many WIBs we can construct using $X_{\text{ex}}$ generators.
Here we highlight the most interesting examples and leave the more systematic constructions for  App.~\ref{app:HOC_Vex}.
Furthermore, we only consider lattice-translation-invariant generators.
The total momentum is a particular IoM in the HOC problem that is not related to integrability, and it is natural to focus on WIB perturbations that conserve $P_{\text{tot}}$, i.e., it remains an exact IoM while the other IoMs turn into quasi-IoMs.
In principle, one can consider WIB perturbations that do not conserve $P_{\text{tot}}$, but the manifold of such perturbations is much larger and is beyond this paper. 

In the HOC model, it is natural to consider generators that are polynomial in the phase space variables.
Since the HOC Hamiltonian $H_0$ is quadratic, if $X_{\text{ex}}$ is a polynomial of degree $m$, then $V_{\text{ex}} = \{X_{\text{ex}}, H_0\}$ is also a polynomial of degree $m$.
We start with cubic $X_{\text{ex}}$'s that we use to generate cubic perturbations $V_{\text{ex}}$'s.

Once again, we organize WIBs by the corresponding symmetry numbers.
In this case, the HOC Hamiltonian manifestly has the time reversal $\Theta: q_j \to q_j, p_j \to -p_j$ and lattice inversion $I: q_j \to q_{L-j+1}, p_j \to p_{L-j+1}$ symmetries.
It is easy to see that $V_{\text{ex}}$ gets the opposite time reversal and the same lattice inversion symmetry as its generator $X_{\text{ex}}$.
In our presentation, we denote the symmetry numbers of a perturbation as $\Theta_V$ and $I_V$.

We note also that the Hamiltonian is trivially invariant under $q_j \to -q_j, p_j \to -p_j$, and all degree $m$ polynomials in the phase space variables have the same transformation property $(-1)^m$ under this symmetry, so we do not list it separately.
However, we note that the Toda chain symmetry $I^{\text{Toda}}$ in Sec.~\ref{sec:toda} is a combination of this and the usual lattice inversion $I$.
The latter is not a symmetry of the Toda chain but is a symmetry of the HOC chain and is more natural for us to use here; when comparing $I^{\text{Toda}}$ from Sec.~\ref{sec:toda} and $I$ symmetry numbers here, we only need to remember this difference, e.g., for all cubic polynomials $I^{\text{Toda}} = -I$, while for quartic terms $I^{\text{Toda}} = I$.

We start with the simplest perturbation that is cubic in $p$ (hence odd under time reversal) and odd under the lattice inversion: 
\begin{equation}
V_{\text{ex}}^{(3),(-,-)} = \frac{1}{3} \sum_j (p_j - p_{j+1})^3 ~.
\label{eq:HOC_Vex3_m_m}
\end{equation}
This happens to be a WIB with the generator given in Eq.~(\ref{eqapp:HOC_Vex3_m_m}).

Next, the simplest perturbation that is cubic in $q$ (hence even under the time reversal) and even under the lattice inversion, is
\begin{equation}
\begin{aligned}
V_{\text{ex}}^{(3),(+,+)} = \frac{1}{3} \sum_j & (q_j + q_{j+1} - 2 q_{j+2}) (q_{j+1} + q_{j+2} - 2 q_j) \\
& \times(q_{j+2} + q_j - 2 q_{j+1}) ~.
\label{eq:HOC_Vex3_p_p}
\end{aligned}
\end{equation}
This also happens to be a WIB with the generator given in Eq.~(\ref{eqapp:HOC_Vex3_p_p}).
From just these two examples, we can already see that $V_{\text{ex}}$-type WIB perturbations are important to consider in the corresponding symmetry manifolds of short-range perturbations.

We next consider an example with $(\Theta_V, I_V) = (+,-)$ symmetry numbers:
\begin{equation}
\begin{aligned}
V_{\text{ex}}^{(3),(+,-)} = & \sum_j \!\bigg[ 2 p_j p_{j+1} (q_j - q_{j+1}) + \\
& ~~ + \frac{2}{3} (q_j - q_{j+1})^3 + \frac{1}{3}(q_{j+2} - q_j)^3 \bigg],
\end{aligned}
\label{eq:HOC_Vex3_p_m}
\end{equation}
whose generator is given in Eq.~(\ref{eqapp:HOC_Vex3_p_m}).
This perturbation has the nearest-neighbor cubic term in $q$ that is the same as in the $\alpha$-FPUT problem, so naturally has the same symmetry numbers as the $\alpha$-FPUT perturbation. 
It also has the second-neighbor cubic term in $q$ and the nearest-neighbor $p p q$-type term.
This perturbation will be of interest later where we will get even closer to the $\alpha$-FPUT perturbation by combining with a $V_{\text{bi}}$-type WIB.
Here we also note that there is an important qualitative distinction between this generator and the ones in the previous two examples:
The previous generators in Eq.~(\ref{eqapp:HOC_Vex3_m_m}) and Eq.~(\ref{eqapp:HOC_Vex3_p_p}) are momentum-conserving, while the generator in Eq.~(\ref{eqapp:HOC_Vex3_p_m})
is momentum-nonconserving, and this leads to different scaling of the AGP norms of the corresponding perturbations, which we will discuss in Sec.~\ref{subsubsec:agp_hoc}.

Finally, we consider an example with $(\Theta_V, I_V) = (-,+)$:
\begin{equation}
\begin{aligned}
V_{\text{ex}}^{(3),(-,+)} = \sum_j \bigg[& p_j^3 + \frac{1}{2} p_j p_{j+1} (p_j + p_{j+1}) + \\
& + 2 p_j (q_j - q_{j-1}) (q_j - q_{j+1}) \bigg] ~,
\end{aligned}
\label{eq:HOC_Vex3_m_p}
\end{equation}
which is a more compact and manifestly $P_{\text{tot}}$-conserving rewriting of Eq.~(\ref{eqapp:HOC_Vex3_m_p}).
This has nearest-neighbor $p p p$-type terms and also up to second-neighbor $p q q$-type terms.
We have not found a simpler perturbation within the corresponding symmetry manifold of $V_{\text{ex}}$-type WIB perturbations, but we will see that we can get a simpler one by combining with a $V_{\text{bi}}$-type perturbation.
The generator for this perturbation is given in Eq.~(\ref{eqapp:HOC_Vex3_m_p}) and is also momentum-nonconserving.

\subsection{WIB perturbations to HOC from boosted generators}
According to Eq.~(\ref{eq:Vbo}), a total current of an IoM is a WIB perturbation generated by the corresponding boosted operator, i.e., $V_{\text{bo}}^\beta = J_{\beta,2;\text{tot}}$.
As we have already discussed, for even IoMs, the corresponding total currents happen to be IoMs and are hence trivial non-perturbations.
On the other hand, for odd IoMs, the corresponding total currents are not conserved quantities and hence constitute valid WIB perturbations.
For example, for $\beta = 3$ (indicated as a superscript $[\beta]$ on $V_{\text{bo}}$):
\begin{equation}
V_{\text{bo}}^{[3]} = J_{3,2;\text{tot}} = \frac{1}{2} \sum_j \left[ p_j p_{j+1} + (q_j - q_{j+1})^2 \right] ~.
\label{eq:HOC_Vbo3}
\end{equation}
Note that the WIB character of such perturbations is somewhat trivial, since adding them as perturbations, one obtains new translationally invariant HOC models, which are easily solvable and integrable.
Nevertheless, such a new model will also introduce new IoMs, and the WIB theory can provide leading corrections to the original IoMs to obtain the new IoMs.

Interestingly, even though we first obtained the above perturbation using the boosted generator (first implicitly working in an infinite system and then placing the derived terms on the PBC chain), it can also be reproduced directly in the finite PBC chain---up to additive IoM contributions---using an extensive local generator.
Specifically, we have
\begin{equation}
\begin{aligned}
& X_{\text{ex}} = \frac{1}{8} \sum_j p_j (-2 q_j + q_{j-1} + q_{j+1}) ~, \label{eq:HOC_Xex_for_Vbo3} \\
& V_{\text{bo}}^{[3]} = \{X_{\text{ex}}, H_0 \} + \frac{1}{2} Q_2^{(0)} + \frac{1}{2} Q_4^{(0)} ~.
\end{aligned}
\end{equation}
This turns out to be a special case of a general statement that for the nearest-neighbor unperturbed HOC model $H_0$, for any bilinear perturbation $V$ that is lattice-translation-invariant, is even under the time reversal (i.e., contains only $p p$-type and $q q$-type terms), and that conserves $P_{\text{tot}}$, there exists an extensive local $X_{\text{ex}}$ such that $V = \{X_{\text{ex}}, H_0 \} + \sum_\alpha c_\alpha Q_\alpha^{(0)}$ (with a finite number of additive IoM contributions).

However, we note that while in the above $V_{\text{bo}}^{[3]}$ example the corresponding extensive local generator is momentum-conserving, this is not the case for all such bilinear perturbations.
As an example, consider a generator analogous to Eq.~(\ref{eq:Toda_Vex0})
\begin{equation}
\begin{aligned}
& X_{\text{ex},0} = \sum_j q_j p_j ~, \\
& V_{\text{ex},0} = \sum_j \left[ p_j^2 - (q_j - q_{j+1})^2 \right] ~.
\label{eq:HOC_Vex0} 
\end{aligned}
\end{equation}
The fact that this generator does not conserve momentum leads to different scaling for the AGP norm compared to momentum-conserving generators, as found numerically in Sec.~\ref{sec:numerics} and demonstrated analytically in App.~\ref{app:analytic_scalings}.
As another example, by combining $X_{\text{ex},0}$ with a momentum-conserving generator (hence retaining the same qualitative character as $X_{\text{ex},0}$), we can obtain the second-neighbor quadratic potential as a WIB:
\begin{equation}
\begin{aligned}
& X_{\text{ex},0}' = -X_{\text{ex},0} + \frac{1}{4} \sum_j p_j (2 q_j - q_{j+1} - q_{j-1}) ~, \\
& \sum_j \frac{1}{2} (q_j - q_{j+2})^2 = V_{\text{ex},0}' + 2 Q_2^{(0)} + 2 Q_4^{(0)} ~.
\end{aligned}
\end{equation}

\subsection{WIB perturbations to HOC from bilocal generators}
We now turn to even more interesting WIB perturbations to HOC, namely, those obtained using bilocal generators.
Using the densities and currents from Eqs.~(\ref{eq:HOC_rho1j})-(\ref{eq:HOC_rho4j}) and Eqs.~(\ref{eq:HOC_J1j})-(\ref{eq:HOC_J4j}), we can readily compute $V_{\text{bi}}^{\beta\gamma}$-type WIB perturbations (denoted as $V_{\text{bi}}^{[\beta,\gamma]}$ below).
We first study cases with $\beta = 1$, i.e., using $P_{\text{tot}}$ as one of the IoMs in the bilocal generator.
This gives perturbations that are cubic in the phase space variables:
\begin{align}
\begin{split}
V_{\text{bi}}^{[1,2]} = \sum_j& \bigg[ \frac{1}{3} (q_{j+1} - q_j)^3 + \frac{1}{12} (q_{j+2} - q_j)^3  \\
& + p_j p_{j+1} (q_j - q_{j+1}) \bigg] ~;
\end{split} \label{eq:HOC_Vbi_12}\\
\begin{split}
V_{\text{bi}}^{[1,3]} = \sum_j \bigg[& \frac{1}{2} p_j p_{j+1} (p_j +  p_{j+1}) \\
& + p_j (q_j - q_{j-1}) (q_j - q_{j+1}) \bigg] ~;
\end{split} \\
\begin{split}
V_{\text{bi}}^{[1,4]} = \sum_j \bigg[& \frac{1}{4} p_j^2 (q_{j-2} - q_{j+2}) \\
& - \frac{1}{3} (q_{j+1} - q_j)^3 + \frac{1}{6} (q_{j+2} - q_j)^3 \bigg] ~.
\end{split}
\end{align}
$V_{\text{bi}}^{[1,2]}$ and $V_{\text{bi}}^{[1,4]}$ are even under the time reversal (i.e., have an even number of powers of $p$'s) and odd under the lattice inversion, while $V_{\text{bi}}^{[1,3]}$ has the opposite symmetry numbers.
We can combine these with $V_{\text{ex}}$-type perturbations to obtain somewhat simpler WIB perturbations.
For example, an appropriate combination of $V_{\text{bi}}^{[1,2]}$ with $V_{\text{ex}}$ in Eq.~(\ref{eq:HOC_Vex3_p_m}) gives a WIB perturbation that depends only on coordinates and not on momenta:
\begin{equation}
\begin{aligned}
\widetilde{V}_{\text{bi}}^{[1,2]} & = \frac{1}{2} V_{\text{bi}}^{[1,2]} - \frac{1}{4} V_{\text{ex}}^{(3),(+,-)} \\
&= \frac{1}{3} \sum_j \bigg[(q_{j+1} - q_j)^3 - \frac{1}{8} (q_{j+2} - q_j)^3 \bigg] ~.
\end{aligned}
\label{eq:tVbi12}
\end{equation}
We see that the $\alpha$-FPUT nearest-neighbor cubic perturbation combined with the specific second-neighbor cubic perturbation is a $V_{\text{bi}}$-type WIB perturbation.
We will return to this example in the more general study of AGPs in App.~\ref{app:AGPkspace};
from such AGP studies, we will also learn that the $\alpha$-FPUT perturbation can be viewed as a novel WIB perturbations obtained using a trilocal generator, see Sec.~\ref{sec:HOC_VFPUT}.

We have not found combinations that would significantly simplify $V_{\text{bi}}^{[1,4]}$.
We can obtain, e.g.,
\begin{equation}
\begin{aligned}
\widetilde{V}_{\text{bi}}^{[1,4]} & = V_{\text{bi}}^{[1,4]} - \frac{1}{2} V_{\text{ex}}^{(3),(+,-)} \\
&= \sum_j \bigg[\frac{1}{4} p_j^2 (q_{j-2} - q_{j+2}) - p_j p_{j+1} (q_j - q_{j+1}) \bigg] ~.
\end{aligned}
\end{equation}
Looking through all $V_{\text{ex}}$-type perturbations with $(\Theta_V, I_V) = (+,-)$ symmetry numbers in Appendix~\ref{app:HOC_Vex} and allowing also $V_{\text{bi}}^{[1,2]}$ in the mix, one could also trade $\sum_j p_j p_{j+1} (q_j - q_{j+1})$ in $\widetilde{V}_{\text{bi}}^{[1,4]}$ for $\sum_j p_j^2 (q_{j-1} - q_{j+1})$, which is not a significant simplification.

On the other hand, $V_{\text{bi}}^{[1,3]}$ is odd under the time reversal and even under the lattice inversion.
We can combine it with the $V_{\text{ex}}$ perturbation in Eq.~(\ref{eq:HOC_Vex3_m_p}) to obtain a particularly simple WIB perturbation that contains only $p$'s and only nearest-neighbor couplings:
\begin{equation}
\begin{aligned}
\widetilde{V}_{\text{bi}}^{[1,3]} & =
V_{\text{bi}}^{[1,3]} - \frac{1}{2} V_{\text{ex}}^{(3),(-,+)} \\
& = \sum_j\left[\frac{1}{4} p_j p_{j+1} (p_j + p_{j+1}) - p_j^3 \right] ~.
\end{aligned}
\end{equation}

\subsection{Quartic WIB perturbations to HOC}
We next consider quartic perturbations. 
As our first very simple example of an extensive local generator,
\begin{equation}
\begin{aligned}
& X_{\text{ex}} = \sum_j \pj \qj^3 ~, \\
& V_{\text{ex}} = \sum_j \left[3 \pj^2 \qj^2 + \qjm{1} \qj^3 - 2 \qj^4 + \qj^3 \qjp{1} \right] =: V_{\text{mnc}^2} ~.
\label{eq:HOC_Vex}
\end{aligned}
\end{equation}
Note that both the generator and the perturbation are not momentum-conserving.
The momentum-nonconserving nature of the perturbation differs from all other examples, and we introduced it as an interesting illustration for our later AGP norm studies in Sec.~\ref{sec:numerics}.

We will focus on momentum-conserving perturbations in the rest of the text.
The mechanism that allowed the special momentum-non-conserving cubic $X_{\text{ex}}$ presented earlier that produces momentum-conserving $V_{\text{ex}}$ [discussion around Eq.~(\ref{eq:jacobi}) in Sec.~\ref{subsec:Todaex}] does not operate for quartic generators, since there are no cubic extensive local IoMs in the HOC.

Table~\ref{tab:SHOCquarticVex} in App.~\ref{app:HOC_Vex} lists momentum-conserving $V_{\text{ex}}$ perturbations that are fourth-order polynomials in $q$ and $p$ generated from momentum-conserving quartic $X_{\text{ex}}$'s.
All other aspects of the organization of the generated quartic $V_{\text{ex}}$ perturbations are the same as for the cubic perturbations in the previous subsection.

A simple combination we highlight here is one with $(\Theta_V, I_V) = (+,-)$ from
\begin{align}
X_{\text{ex}} = \sum_j &\big[ (q_j - q_{j+1})^3 (p_j + p_{j+1}) + \notag \\
& + (q_j - q_{j+1}) (p_j^3 + p_{j+1}^3) \big] ~,
\end{align}
which gives 
\begin{align}
V_{\text{ex}}^{(4),(+,-)} &= \sum_j \big[ p_j p_{j+1} (p_{j+1}^2 - p_j^2) + \label{eq:HOC_Vex4_pm}  \\
& + (q_j - q_{j+1})^3 (q_{j-1} - q_j - q_{j+1} + q_{j+2}) \big] ~. \notag
\end{align}
This perturbation has only $pppp$ and $qqqq$-type terms and no direct couplings between the $p$ and $q$ variables, making it a suitable candidate for studies with symplectic integrators; this could be useful in future direct studies of the corresponding perturbed HOC model.

There are no quartic WIB perturbations from the boosted generators. However, one can construct quartic WIBs also from bilocal generators. For example,
\begin{align}
&V_{\text{bi}}^{[2,3]} = \frac{1}{4} \sum_j \bigg\{p_j p_{j+1} (p_j^2 + p_{j+1}^2)+ \notag \\
& + p_j^2 \bigg[ (q_{j-1} - q_j)^2 + (q_j - q_{j+1})^2 - (q_{j-1} - q_{j+1})^2 \bigg] + \notag \\
& + p_j p_{j+1} \bigg[ (q_{j-1} - q_j)^2 + (q_{j+1} - q_{j+2})^2 - \notag \\
& - \frac{1}{2} (q_{j-1} - q_{j+1})^2 - \frac{1}{2} (q_j - q_{j+2})^2 \bigg] +\notag \\
& + (q_j - q_{j+1})^4 + (q_j - q_{j+1})^2 (q_{j+1} - q_{j+2})^2 \bigg\} ~. 
\label{eq:HOC_Vbi_23}
\end{align}
$V_{\text{bi}}^{[2,3]}$ is even under both inversion and time reversal.

Finally, one might wonder how the $V_\beta$ perturbation defined earlier fits into this framework. 
This perturbation on top of the HOC, known as the $\beta$-FPUT model, is nonintegrable for generic $N$, but in the weakly nonlinear regime, it can display long-lived metastable dynamics and very long times to approach equipartition \cite{Benettin_2011, Lvov_2018}, and for small $N = 3$ it has an additional IoM, making it integrable in that case \cite{Arzika_2023}. 
One could ask whether, for general $N$, the $\beta$-FPUT potential $V_\beta$ is a WIB to HOC, and whether the $\beta$-FPUT model has quasi-conserved quantities.

At present, we have not been able to identify a generator in the standard classes (extensive/boosted/bilocal) that yields $V_\beta$.
We also have not found evidence for quasi-conserved quantities associated with this perturbation coming from the Toda chain (see App.~\ref{app:toda_ioms_to_corr_high}).
This observation is consistent with previous studies of the $\beta$-FPUT model using AGP and counterdiabatic driving techniques, which likewise do not reveal a simple local generator structure for the quartic interaction \cite{Gjonbalaj_2022}.
Nevertheless, we do not exclude the possibility that such a structure exists and may emerge from a different construction or perspective (e.g., similar to our analysis of the cubic perturbation $V_\alpha$ in Sec.~\ref{sec:HOC_VFPUT}, we expect that there is a quartic non-local AGP that reproduces $V_\beta$, consistent with our AGP norm study in Sec.~\ref{sec:numerics}; however, unlike the cubic case, we suspect that corrections to some of the IoMs are also non-local).

Finally, we define an additional generic perturbation for which we do not know the generator 
\begin{equation}
\label{eq:HOC_Vs}
\mathrm{V}_{s} = \sum_j p_{j}p_{j+1} (q_j - q_{j+1})^2 ~,
\end{equation}
which we will use for comparison in the later numerical studies with AGP in Sec.~\ref{sec:numerics}.

\section{Quasi-IoMs for the $\alpha$-FPUT perturbation of the HOC}
\label{sec:HOC_VFPUT}
\subsection{The $\alpha$-FPUT problem}
The Fermi-Pasta-Ulam-Tsingou (FPUT) problem \cite{Fermi1955}, posed in the 1950s, is the oldest numerical study of thermalization in integrable systems under small perturbations, yet it remains actively explored more than 70 years later. 
Fermi and collaborators started with an HOC chain, a known non-thermalizing integrable model, and added a weak nonlinear coupling that breaks the integrability of the original chain with an expectation of observing the relaxation of the system towards equilibrium (equipartition of energy among all the modes), even if started from an initial state where all of the energy was concentrated in one mode. 
Instead, they observed quasi-periodic dynamics and near-recurrences of the initial state.
Thermalization, which we now know is achieved~\cite{Benettin_2011}, occurs at much longer times. 
This surprising result, dubbed ``the FPUT paradox,'' led to a vast literature connecting the apparent lack of thermalization in the FPUT problem to integrability (of both HOC and Toda models), the discovery of solitons, the KAM theory, and has played a central role in the development of the nonlinear science~\cite{Weissert1997, Berman_2005, Campbell_2005, Zabusky_2005, Gallavotti2008}. 

A major line of work interprets the unusual thermalization of the $\alpha$-FPUT model through its proximity to the Toda lattice (Sec.~\ref{sec:toda}).
In that picture, the integrals of motion of the Toda chain are said to act as adiabatic invariants for the FPUT dynamics~\cite{Grava2020}.
This Toda-based viewpoint has been used to identify long-lived quasi-integrable regimes, determine energy thresholds associated with the breakdown of invariant tori, and characterize the scaling of equilibration times with system size and nonlinearity strength. 
Most of this literature is formulated either in terms of the dynamics of harmonic normal modes or in terms of Toda invariants themselves, and the resulting explanations of anomalous thermalization are largely supported by numerical studies. 
In contrast, the WIB framework provides a structural mechanism for slow dynamics, deriving anomalous thermalization from the existence of quasi-conserved quantities, which are the original IoMs plus corrections that ensure that their Poisson brackets with the perturbed Hamiltonian vanish to leading order, as in Eq.~(\ref{eq:corrections}).

In this section, we show that the $\alpha$-FPUT interaction is a weak integrability-breaking perturbation of the HOC in this sense. 
To our knowledge, the $\alpha$-FPUT interaction has not been analyzed from the harmonic-chain WIB perspective in a way that (i) constructs quasi-IoMs across the full HOC IoMs hierarchy and (ii) identifies the corresponding generator directly, without assuming an explicit proximity to the Toda chain.
In the remainder of this section, we close this gap, demonstrating that the $\alpha$-FPUT chain fits within the WIB framework and supports an infinite family of quasi-IoMs.
We start by formalizing the connection to Toda's IoMs and use them to derive corrections to IoMs of the HOC. 
More importantly, we identify the generator of the $\alpha$-FPUT as a perturbation of HOC and discuss its implications.
In particular, the latter approach generalizes to arbitrary momentum-conserving cubic perturbations, and we show that they have qualitatively similar WIB character even though they are not related to the nearest-neighbor Toda chain.

\subsection{From Toda IoMs to quasi-IoMs in the HOC under the $\alpha$-FPUT perturbation}

The question of whether the $\alpha$-FPUT is a WIB comes down to finding corrections ($Q_3^{(1)}, Q_4^{(1)}, \dots$) to all of the IoMs ($Q_3^{(0)}, Q_4^{(0)}, \dots$) of the HOC.
This means solving 
\begin{equation}
    \{H_0 + \lambda V, Q_3^{(0)} + \lambda Q_3^{(1)}\} = \mathcal{O}(\lambda^2) ~,
\end{equation}
for $Q_3^{(1)}$ when $V = V_{\alpha{\text{-FPUT}}}$, and similarly for the other IoMs. 
We are essentially looking for an operator $Q_3^{(1)}$ that satisfies
\begin{equation}
\{Q_3^{(1)}, H_0\} \;=\; \{V_{\alpha{\text{-FPUT}}}, Q_3^{(0)}\} ~.
\label{eq:HOC_VPFUT_Q3quasiIoM}
\end{equation}
The right-hand side is easy to evaluate.
However, finding an extensive-local $Q_3^{(1)}$ that satisfies Eq.~(\ref{eq:HOC_VPFUT_Q3quasiIoM}) is, in principle, non-trivial (and not possible for general $V$).
By examining Poisson brackets of various cubic terms with $H_0$, we do find that the following
\begin{equation}
Q_3^{(1)} = -\sum_j \left[ \frac{1}{3} p_j^3 + \frac{1}{2}(p_j + p_{j+1})(q_j - q_{j+1})^2  \right] ~,
\label{eq:Q31aFPUT}
\end{equation}
does satisfy the above equation and is a correction to the case of the $\alpha$-FPUT potential.

One might notice that the expression we find has a structure like Toda's $Q_3^{\text{Toda},(0)}$ in Eq.~(\ref{eq:Toda_Q3}).
And indeed, when $Q_3^{\text{Toda}(0)}$ is Taylor expanded with the HOC limit taken ($b \to 0$, $ab \to \text{const}$) and with our choice of $a$ and $b$, $ab=1$, we see that our correction is proportional to the cubic terms in the expansion:
\begin{equation*}
Q_3^{\text{Toda},(0)} \text{cubic:} \sum_j \left[ \frac{p_j^3}{3} +  \frac{ab}{2} (p_{j+1} + p_j)(q_{j+1} -q_j)^2 \right].
\end{equation*}
This is exactly the correction $Q_3^{(1)}$ we found in Eq.~(\ref{eq:Q31aFPUT}) but with opposite sign, and the reason for the sign is that in our conventions, in this limit the above cubic terms in $Q_3^{\text{Toda}}$ appear as corrections to $-Q_3^{(0)}$ of the HOC problem.
We leave the detailed derivation and discussion from this perspective for Appendix \ref{app:toda_ioms_to_corr}. 
Here, we only briefly note that in the formal expansion, one also recovers IoMs of the HOC; 
in the case of $Q_3^{\text{Toda},(0)}$ we find that the leading term contains $Q_1^{(0)} = P_{\text{tot}}$, the first subleading term contains $Q_3^{(0)}$, and the next subleading term contains $Q_3^{(1)}$

Inspired by this connection, we anticipate that we can also find corrections for higher IoMs.
This turns out to be more subtle, e.g., cubic terms in $Q_4^{\text{Toda},(0)}$ are not corrections to just the HOC $Q_4^{(0)}$.
Yet, a systematic method using Toda's IoMs is possible, and we refer to Appendix~\ref{app:toda_ioms_to_corr} for details.
Here we simply list the corrections for the next two IoMs:
\begin{align}
Q_4^{(1)} = \sum_j & \bigg[ (p_j^2 + p_j p_{j+1} + p_{j+1}^2) (q_{j+1} - q_{j}) + 
\label{eq:Q41aFPUT} \\
&+ \frac{1}{6} (q_{j+2} - q_j)^3 \bigg] ~; \notag \\
Q_5^{(1)} = - \sum_j & \bigg[ \frac{2}{3} p_j^3 + p_j^2 p_{j+1} + p_j p_{j+1}^2 +  \\
&+ \frac{1}{2} (p_j + 2p_{j+1} + p_{j+2}) (q_{j+2} - q_j)^2 \bigg] ~. \notag
\end{align}
These results support the long-standing observation that Toda integrals behave as slow or adiabatic invariants for the $\alpha$–FPUT dynamics~\cite{Grava2020, Christodoulidi_2025}.
More importantly, they demonstrate that the entire hierarchy of harmonic-chain IoMs can be consistently corrected.
In the context of our framework, this establishes that the $\alpha$–FPUT model is a genuine weak integrability-breaking deformation (WIB) of the HOC and is equipped with a tower of quasi-conserved quantities. 
In the following sections, we will demonstrate this further without relying on the proximity to the Toda chain, and we will find the AGP (generator) for the $\alpha-$FPUT both in the real space and the momentum space. 
The object we find is a tri-local operator which does not belong to one of the previously known families of generators and thus represents an entirely new class.

\subsection{$\alpha$-FPUT is a WIB perturbation to the HOC}
In the WIB framework, the notion of “weakness” is defined with respect to a chosen integrable Hamiltonian, in this case, the harmonic chain, HOC.
If a perturbation is weak, then there exists a generator that produces the quasi-local corrections to all IoMs of the perturbed model.
In other words, there must exist a generator $X$ that implements the corresponding near-canonical transformation that corrects all of the IoMs. 
Our analysis shows that such a generator can indeed be constructed, and in fact, that it extends to \emph{any} cubic, translationally invariant, momentum-conserving perturbation of the HOC.
We provide all details of the analysis in Appendix~\ref{app:AGPkspace}, while below we sketch its main steps and summarize key results.

We start by writing the HOC Hamiltonian in momentum space:
\begin{equation}
    H_0=\sum_k\left[ \frac{1}{2}\tilde p_k\tilde p_{-k}+\frac{\omega_k^2}{2}\tilde q_k\tilde q_{-k}\right] ~,
\end{equation}
where $\{\tilde q_k, \tilde p_{k'}\} = \delta_{k,-k'}$ and $\omega_k^2 = 2(1-\cos k) = 4\sin^2\frac{k}{2}$.
We then consider the general expression for a cubic and translationally invariant perturbation $V$
\begin{equation}
\label{eq:Vk}
V = \sum_{k,k'} v(k, k') \tilde q_k \tilde q_{k'} \tilde q_{-k-k'} ~,
\end{equation}
where, without loss of generality, we may assume that 
$v(k,k')$ is fully symmetric under permutations of the three momenta, i.e, $v(k,k')=v(k',k)=v(k,-k-k')=\dots$.
For the moment, we keep the discussion general and allow for an arbitrary function $v(k,k')$, before specializing to the case of the $\alpha$-FPUT model.

We look for an $X$ that satisfies $V=\{X,H_0\}$.
On symmetry grounds, such an $X$ must be i) cubic, ii) translationally invariant, and iii) odd under time reversal.
The most general ansatz satisfying these properties is of the form
\begin{equation}
    X = \sum_{k,k'} \Big[ g(k,k')\tilde p_k \tilde p_{k'}+f(k, k')\tilde q_k \tilde q_{k'}\Big] \tilde p_{-k-k'} ~.
\end{equation}
We can assume, w.l.o.g., that $g(k,k')=g(k',k)=g(k,-k-k')=\dots$ and that $f(k, k')=f(k',k)$. 
We solve for $g$ and $f$ such that $V=\{X, H_0\}$ is satisfied (see Appendix \ref{app:AGPkspace} for details) and we find that
\begin{align}
&g(k,k')=-\frac{v(k, k')}{32 \sin^2\left(\frac{k}{2}\right)\sin^2\left(\frac{k'}{2}\right)\sin^2\left(\frac{-k-k'}{2}\right)} ~,\label{eq:gkk} \\
&f(k,k')=\frac{3}{2}(\omega^2_k+\omega^2_{k'}-\omega^2_{-k-k'})g(k,k') ~.\label{eq:fkk}
\end{align}
Note that for a generic potential $V$, the functions $f$ and $g$ have second-order poles in $k=0$, $k'=0$, $-k-k'=0$. 
Such singularities would generically lead to strongly nonlocal generators $X$.
This is not, a priori, an issue, since generators of WIBs can themselves be non-local (e.g., boosted or bilocal generators) while still producing local IoM corrections. 

Since the locality of the IoM corrections is the key requirement, we now turn to their explicit form and examine whether they are local.
For example, consider corrections to the odd IoMs, Eq.~(\ref{eq:HOC_oddIoMs}), which are, in momentum space, given by 
\begin{equation}
    Q_{2n+1}^{(0)}=-i\sum_k \sin(nk)\tilde q_k \tilde p_{-k} ~.
\end{equation}
To compute corrections, we evaluate the P.B. directly, 
\begin{align}
&\{X, Q_{2n+1}^{(0)}\}=4i\sum_{k,k'}\bigg[g(k,k')\tilde p_k \tilde p_{k'}\tilde p_{-k-k'}+  \\
& + f(k,k')\tilde q_k \tilde q_{k'} \tilde p_{-k-k'} \bigg] \sin\frac{nk}{2}\sin\frac{nk'}{2}\sin\frac{n(-k-k')}{2}.\notag
\end{align}
Comparing this expression with Eqs.~(\ref{eq:gkk}) and (\ref{eq:fkk}), we observe that the triple-sine factor partially cancels the poles in $g$ and $f$.
To completely remove the singularities, we need $v(k,k')$ to have zeros in $k=0$, $k'=0$, and $-k-k'=0$.
Under this condition, the odd IoM corrections are regular in momentum space, and therefore local in real space. 
An analogous calculation for the even IoMs, Eq.~(\ref{eq:HOC_evenIoMs}), leads to the same conclusion (see Appendix \ref{app:AGPkspace}): the corrections are local if $v(k,k')$ vanishes at $k=0$, $k'=0$, and $-k-k'=0$.

Let us know, specialize to the $\alpha$-FPUT case. The $\alpha$-FPUT potential indeed has this special property, since it can be expressed in momentum space as
\begin{align}
    &V_{\text{FPUT}}=\sum_{k,k'}v_{\text{FPUT}}\cdot\tilde q_k \tilde q_{k'} \tilde q_{-k-k'} ~, \\
   &v_{\text{FPUT}}= \frac{(2i)^3}{3\sqrt{L}}\sin\left(\frac{k}{2}\right)\sin\left(\frac{k'}{2}\right)\sin\left(\frac{-k-k'}{2}\right) ~.
\end{align}
It follows then that 
\begin{align}
\label{eq:gfput}
&g_{\text{FPUT}}=\frac{i}{12\sqrt{L} \sin\left(\frac{k}{2}\right)\sin\left(\frac{k'}{2}\right)\sin\left(\frac{-k-k'}{2}\right)} ~,\\
& f_{\text{FPUT}}=\frac{3}{2}\big(\omega_k^2+\omega_{k'}^2-\omega_{-k-k'}^2\big)\, g_{\text{FPUT}} ~,
\end{align}
so  $g_{\text{FPUT}}$ and $f_{\text{FPUT}}$ have only first-order poles (in fact, $f(k,k')$ does not have poles when $k \to 0$ or $k' \to 0$ but has a pole when $-k-k' \to 0$). 
Unlike generic perturbations, the generator of $ V_{\text{FPUT}}$ therefore contains poles of order one rather than two.
As a consequence, although the generator $X_{\text{FPUT}}$ is still non-local, the corresponding IoM corrections are local.
Indeed, we find that the corrections generated from this $X_{\text{FPUT}}$ reproduce the quasi-IOMs in Eqs.~(\ref{eq:Q31aFPUT}) and (\ref{eq:Q41aFPUT}) up to conserved quantities (see Appendix~\ref{app:AGPkspace} for details). 

\subsection{Trilocal $X$: new class of generators of WIB}
This observation leads to an interesting question.
In the case of $V_{\text{FPUT}}$, we were unable to express the corresponding generator in any of the classes we considered (extensive local, boosted, or bilocal operators).
One might wonder whether this reflects a mere lack of creativity on our part, or whether the correct generator for such perturbations truly lies outside all these classes. 
Our analysis shows the second scenario to be correct.
By rewriting the generator from the momentum space to the real space, we obtain the exact expression of $X_{\text{FPUT}}$ at finite size (see Appendix~\ref{app:AGPFPUT}).
Here, we focus on its infinite-size limit, which reads
\begin{align}
    X_{\text{FPUT}}=&\sum_{j_1<j_2<j_3}\frac{2}{3}[(j_2-j_1)-(j_3-j_2)]p_{j_1}p_{j_2}p_{j_3}\nonumber\\
    &+\sum_{j_1<j_2}\frac{1}{3}(j_2-j_1)p_{j_1}p_{j_2}(p_{j_2}-p_{j_1})\nonumber\\
    &+\sum_{j_1<j_2}(p_{j_1}q_{j_2}^2-p_{j_2}q_{j_1}^2) ~.\label{eq:XFPUTr}
\end{align}
This expression has the structure of a {\it trilocal} operator, and correctly reproduces  $V_{\text{FPUT}}=\{X_{\text{FPUT}}, H_0\}$ (see Appendix~\ref{app:AGPFPUT} for details).

The $\alpha$-FPUT model, therefore, provides a concrete example of a WIB perturbation generated by an operator that does not belong to any of the classes identified so far. 
This raises an interesting direction for future work: developing a general and systematic framework for defining trilocal generators.
We note here the possibility of other generators of WIBs and their connection to a recent result in free fermion systems by Pozsgay\textit{et al.}~\cite{Pozsgay2024}. 

\subsection{All cubic, momentum-conserving perturbations to HOC are WIBs}
\label{sec:xagp-all-cubic}
A natural question at this stage is whether the fact that $\alpha$-FPUT is a WIB is a special property that crucially depends on the proximity to the integrable Toda chain, or whether it belongs to a larger class of WIBs, unrelated to the Toda structure.
We find that the latter is the case: this property follows more generally from the invariance of $V$ under the transformation $q_j\rightarrow q_j+c$ (i.e., $V$ is momentum-conserving).
Since the potential only depends on differences between the $q_j$'s, the mode $\tilde q_{k=0}\propto \sum_j q_j$ does not enter the momentum-space expression for the perturbation, and therefore $v(k, k')$ has zeros at $k=0$, $k'=0$, and $-k-k'=0$. 

In other words, momentum-conserving cubic perturbations soften the singularities that would otherwise appear in the construction of the quasi-IOMs. This ensures that the resulting corrections remain regular (i.e., extensive local). 
Several examples of this class are discussed in Appendix~\ref{app:AGPkspace}.

We conclude that any momentum-conserving cubic perturbation is a WIB perturbation in our framework and admits quasi-IOMs with controlled locality properties.
Finally, we note that while such perturbations are generally not related to the integrable Toda chain, one is left to wonder whether they suggest new Toda-like integrable models with further-neighbor interactions; we leave the exploration of this possibility to future work.

\section{NUMERICAL STUDIES} 
\label{sec:numerics}
In this section, we present numerical evidence for our central claim: our systematically constructed WIB perturbations delay the onset of chaos and thermalization on timescales significantly longer than those induced by generic integrability-breaking perturbations. 
We start with a numerical study of AGP, the probe of chaos that proved to be very sensitive to the type of integrability breaking in quantum systems~ \cite{KOLODRUBETZ_2017, Pandey2020}.  
While generic perturbations typically lead to an AGP norm that grows exponentially with system size, WIB perturbations---like the ones we construct---are characterized by a much slower polynomial growth~\cite{Pandey2020, Orlov_2023}.
We compute the variance of AGP, defined for classical systems, and use it to probe the nature of integrability breaking. 
Since the AGP acts as a proxy for the perturbation generator, we show that its scaling behavior with system size can be used to distinguish between generic and WIB perturbations and to distinguish among different classes of WIBs.

We also report transport studies of the Ishimori chain, in which we analyze the crossover from ballistic to diffusive energy transport. 
Spin transport in this family of classical isotropic chains is known to be anomalous (superdiffusive) and exhibits Kardar–Parisi–Zhang (KPZ) scaling \cite{Ljubotina2019,takeuchi2024}.
Related classical spin-chain models, including the lattice Landau–Lifshitz chain and even the non-integrable Heisenberg chain, show similar behavior~\cite{McRoberts2022, McRoberts2023, Das_2019, Das_2020, takeuchi2024}.  
For the Ishimori model, recent numerical studies have shown that, although the crossover to diffusion eventually occurs, its timescale is parametrically large and difficult to access numerically~\cite{McRoberts2024}. 
McCarthy \emph{et.~al.} successfully extracted the crossover by taking advantage of the discrete-time, discrete-space integrable analog of the Ishimori model originally proposed by Krajnik and Prosen \cite{Krajnik2021}. 
Their results were recently confirmed in the quantum setting as well \cite{Moore_2025}.
In the presence of WIBs, we expect the crossover to happen at even longer timescales, which is why we focus on the energy transport instead. 
In the integrable limit, energy transport is ballistic, and its crossover to diffusion upon perturbation signals the onset of conventional thermalizing behavior. 
Analyzing the crossover time enables a direct comparison of the thermalization timescale with the standard prediction from Fermi's Golden Rule. \cite{Bilha_2003}

Our work builds upon recent studies of spin and energy transport in nearly integrable classical systems, particularly in the Ishimori chain under isotropic perturbations~\cite{Das_2019, Das_2020, Roy_2023,  Roy_2023supp,McRoberts2024}.
Due to the substantial computational cost of long-time transport simulations, we restrict the transport analysis to a single WIB perturbation; for the remaining perturbations and models, we rely primarily on the results from our AGP analysis.

\subsection{Numerical Integrators}
The numerical integration of nonlinear differential equations can be performed in several ways, most commonly by discretizing time into short intervals, $\delta t$. 
Standard methods include the Euler and Runge-Kutta algorithms. 
However, for Hamiltonian systems, these methods do not exactly conserve energy \cite{Ruth1983, Yoshida1990, Hairer2006}.
Symplectic integrators, like ABA864, by contrast, preserve the geometric structure of Hamiltonian flows and offer near-exact conservation properties over long times \cite{SanzSerna1994, McLachlan2002}.

In this study, we use two different algorithms: the fourth-order Runge-Kutta (RK4) and ABA864 \cite{Runge1895, Kutta1901, Butcher1996, Butcher2008, HairerNorsettWannerNonstiff_2008,SanzSernaCalvoNHP_1994, LeimkuhlerReichSHD_2004}.
For models in which the Hamiltonian cannot be separated easily, such as the Ishimori chain, we use the RK4 integrator. 
For both the Toda and the HOC, the Hamiltonian can be split into the kinetic and the potential energy parts, $H = A(p) + B(q)$, which is ideal for symplectic methods.
For these systems, we chose the ABA864 algorithm, a highly accurate fourth-order symplectic scheme developed in \cite{Blanes2013}.
This integrator was found to be the most effective overall integration scheme for the $\alpha$-FPUT chain in \cite{Danieli_2019}, making it particularly well-suited for our studies. 
Our implementation of ABA864 for the Toda, HOC, and FPUT models closely follows the detailed procedures outlined in \cite{Danieli2024}, while the remainder of our simulation code was developed independently, enabling cross-validation of previously reported results.

\subsection{Adiabatic gauge potential as a probe of weak integrability breaking}
The classical AGP along a single trajectory up to a final time $T$ is given by Eq.~(\ref{eq:DAt}),
\begin{equation}
    \mathcal{A}(t) = \mathcal{A}(0) - \int_0^{t}dt' \left[V(t') - \overline V\right],
\end{equation}
where
\begin{equation}
    \overline V  = \frac{1}{T}\int_0^T dt\, V(t).
\end{equation}
We are here interested in the AGP at $\lambda=0$, so the system is evolved under the unperturbed integrable Hamiltonian $H_0$. We numerically compute the AGP variance, defined as
\begin{equation}
    \sigma^2(T) = \frac{1}{T}\int_0^T dt \Bigg[ \mathcal{A}(t)^2 - \Bigg( \frac{1}{T}\int_0^T dt' \mathcal{A}(t') \Bigg)^2 \Bigg].
\end{equation}
The average of this quantity over a stationary distribution of initial states is called \textbf{fidelity susceptibility},
\begin{equation}
\chi(T) = \int dq \ dp \ \rho(q,p) \ \sigma^2(T; q,p) ~,
\label{eq_fidSuscClassical}
\end{equation}
and will serve as an analog of the AGP norm in our previous study of WIBs in quantum models~\cite{Vanovac2024}.
As a function of time, the value of $\mathcal{A}(t)$ is analogous to diffusion of a particle in one dimension, with velocity $V - \overline{V} $ where $V$ is the perturbation.
If a given perturbation is a WIB perturbation, we expect $\chi$ to ``quickly'' saturate to a finite value, which we call AGP norm,
\begin{equation}
\text{AGP Norm}\equiv \chi^{\text{sat}}(L)\;=\lim_{T\rightarrow\infty} \chi(T)
\end{equation}
while for a generic perturbation, we expect $\chi$ to grow with time.

We further observe that the scaling of the saturation value of $\chi$ with system size is insightful and can be used to additionally classify not only whether a perturbation is WIB, but also what type of $X$ it was generated from.
For WIBs of the form $V = \{X,H_0\}$, the late-time AGP variance is given by the variance of the generator in an appropriate equilibrium ensemble,
\begin{equation}
\chi^{\text{sat}}\;=\;\big\langle (X-\langle X\rangle)^2 \big\rangle.
\label{eq:AGP-var-as-VarX}
\end{equation}
Indeed, in this case, the numerically integrated $\Delta \mathcal A(t)=\mathcal A(t) -\mathcal A(0)$ 
equals $X(t) - X(0)$ along each trajectory.
We expect the system to be ergodic in time, i.e., time averaging yields ``thermal'' expectation values, subject to the initial-state IoMs.
Furthermore, we average over the ensemble of initial states.
Hence, scaling of $\chi^{\text{sat}}$ with system size follows from equilibrium fluctuations of $X$. 
We refer to Appendix~\ref{app:analytic_scalings} for more details.
We find that scaling with system size provides a way to resolve any doubts about the nature of a given perturbation.

\subsubsection{Initialization Scheme}
\label{subsec:initalization}
To prepare the initial state for the AGP studies of the Toda and HOC models, we first set all positions to zero, $\{q_j=0\}$, and then draw the momenta $\{p_j\}$ from a uniform random distribution on the interval $[-1,1]$. 
The center-of-mass momentum is subtracted to adhere to periodic boundary conditions, and momenta are then uniformly rescaled to fix the energy density at $\epsilon=0.1$.
This initial configuration is then evolved for a long time ($T=10^5$ time units with a time step of $dt=0.01$) using the ABA864 integrator to allow the system to reach a typical state in the microcanonical ensemble.
The resulting phase-space point $\{q_j,p_j\}$ serves as the initial condition for our simulations, which evaluate the AGP variance over much longer runs of order $10^6$. 
We simulate the dynamics up to the final time $T=1.5\times10^6$ with a time step of $dt=0.04$. 

The time step was chosen to be consistent across all system sizes studied here. 
A larger time step, $dt=0.1$, can be used for smaller system sizes to extend the simulation time. 
In the computation of the AGP variance, since the finite time step $dt$ also acts like an integrability-breaking perturbation, we want to choose a sufficiently small time step such that the integrability-breaking due to the integrator happens on a much longer timescale than the integrability-breaking from the Hamiltonian perturbation.
We refer to Ref.~\cite{Danieli2024} for details.  

\begin{figure*}
    \centering
\begin{minipage}[c]{0.33\linewidth}
    {\setlength{\tabcolsep}{2.8pt}
    \renewcommand{\arraystretch}{0.9}
    \begin{tabular}{c c  c c  c}
    \centering
        $\mathbf{V}$ & \textbf{Eq.~ref.} & $\mathbf{\Theta}$ & $\mathbf{I^{\text{Toda}}}$ & ~$\mathbf{\chi^{\text{sat}}}$  \\
        \midrule
        $\mathrm{V_{\text{bo,1}}}$ & Eq.(\ref{eq:Toda_Vbo1}) & even & even & $\mathrm{L}^2$  \\
        $\mathrm{V_{\text{bo,2}}}$ & Eq.(\ref{eq:Toda_Vbo2}) & odd &  odd&  $\mathrm{L}^2$  \\
       $\mathrm{V_{\text{bo,3}}}$ & Eq.(\ref{eq:Toda_Vbo3}) & even & even &  $\mathrm{L}^2$  \\
       $\mathrm{V_{\text{bo,4}}}$ & Eq.(\ref{eq:Toda_Vbo4}) & odd & odd &   $\mathrm{L}^2$  \\
       \midrule
       \midrule
        $\mathrm{V_{\text{bi,1}}}$ & Eq.(\ref{eq:Toda_Vbi12}) & even & even &  $\mathrm{L}^2$  \\
         $\mathrm{V_{\text{bi,2}}}$ & Eq.(\ref{eq:Toda_tildeVbi12}) & even & even & $\mathrm{L}^2$   \\
         \midrule
        $\mathrm{V_{\text{mnc}}}$ & Eq.(\ref{eq:Toda_Vex0})  & even &  even &  $\mathrm{L}^2$   \\
         \midrule
        \midrule
         $\mathrm{V_{\text{ex,1}}}$  & Eq.(\ref{eq:Toda_Vex1}) & even & odd  & $\mathrm{L}$    \\
        $\mathrm{V_{\text{ex,2}}}$  & Eq.(\ref{eq:Toda_Vex2}) & even & even  & $\mathrm{L}$    \\
        $\mathrm{V_{\text{ex,3}}}$  & Eq.(\ref{eq:Toda_Vex3})  & odd & even &  $\mathrm{L}$  \\
        \midrule
        \midrule
        $\mathrm{V_{\alpha}}$ & Eq.(\ref{eq:valpha}) & even  & odd  &   -\\
        $\mathrm{V_{\beta}}$ & Eq.(\ref{eq:vbeta}) & even &  even &     -\\
        $\mathrm{V_{\text{s}}}$& $e^{4b(q_j-q_{j+2})}$ & even & even &   -\\
        \hline
    \end{tabular}}
\end{minipage}
\hfill
\begin{minipage}[c]{0.66\linewidth}
\includegraphics[width=1.01\linewidth]{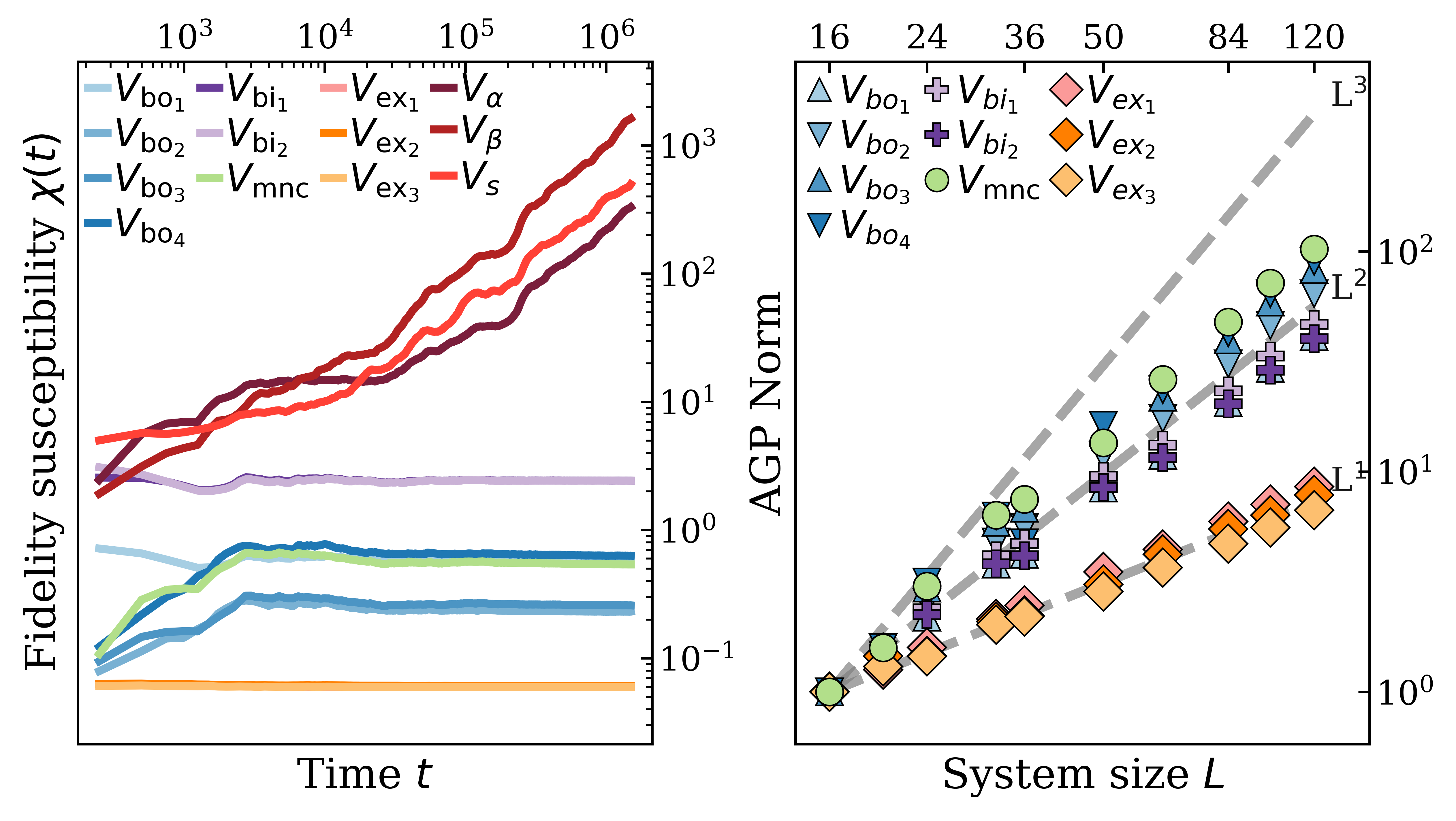}
\end{minipage}
\caption{\textit{Toda chain.} 
\textbf{Left, table:} All perturbations studied with reference to their equation in the main text; time reversal $\Theta$ and inversion $I^{\text{Toda}}$ symmetries; and their observed AGP norm scaling.
\textbf{Middle, plot:} Variance of the AGP averaged over 100 trajectories for the Toda chain under WIBs, and a few generic perturbations.
The system size is $L=100$ with initial energy density $\epsilon=0.1$ and $\alpha=0.5$ (corresponds to setting $b=-1, a=-1$).
After the initialization step, dynamics are simulated up to the final time $T=1.5\times10^6$ with a time step of $dt=0.04$ using the ABA864 symplectic integrator.
For easier visualization, some of the perturbations have been rescaled by a constant ($L$-independent) prefactor.
\textbf{Right, plot:} AGP norm
as a function of system size $L$ for the WIBs studied. Generic perturbations are excluded from this plot as they do not reach saturation.}
\label{fig:agp-toda-global}
\end{figure*}

\begin{figure*}
\begin{minipage}[c]{0.69\linewidth}
\includegraphics[width=0.95\linewidth]{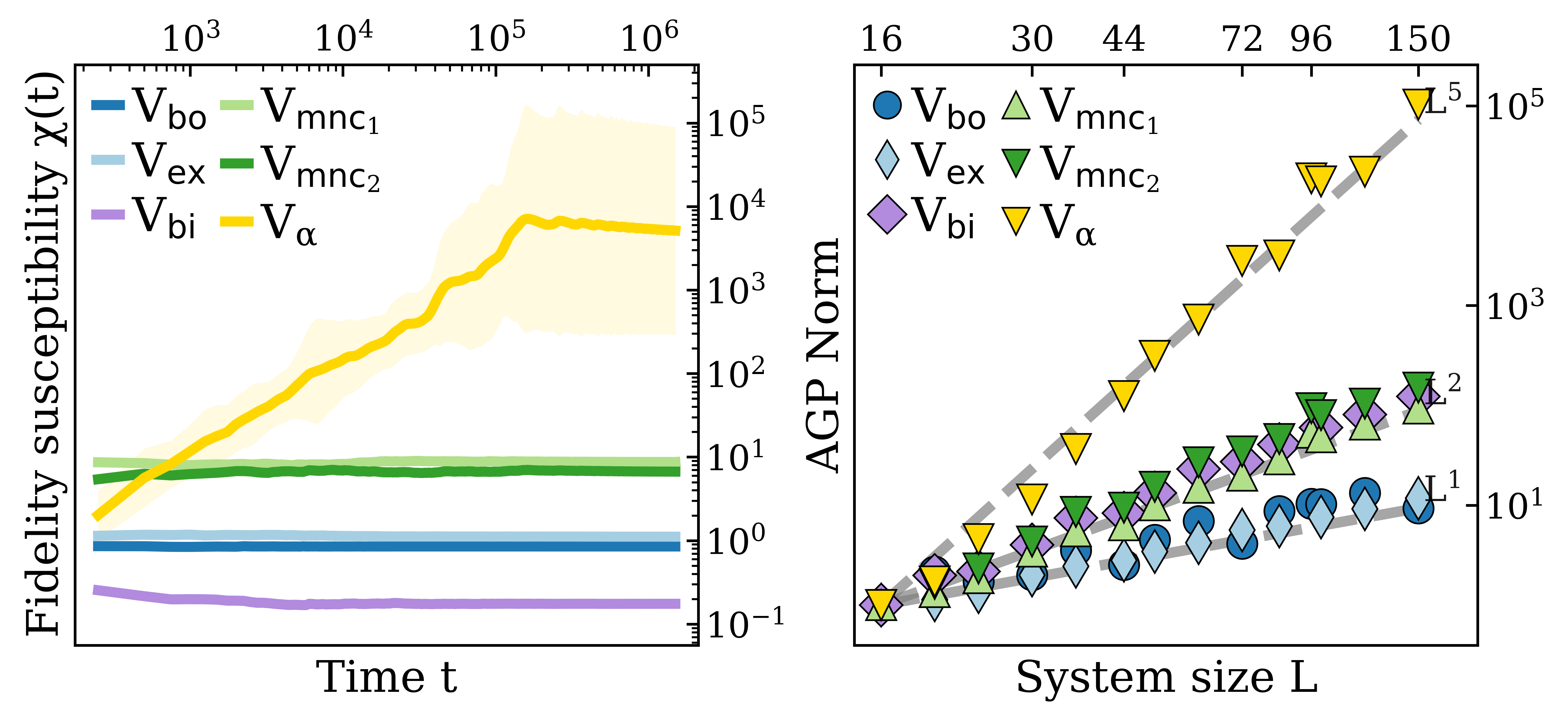} 
    \end{minipage}
    \begin{minipage}[c]{0.30\linewidth}
{\renewcommand{\arraystretch}{1.01}
\centering
    \begin{tabular}{c c  c c  c}
     & \multicolumn{4}{c}{\large Cubic WIBs} \\
    & & & & \\
    $\mathbf{V}$ & \textbf{Eq.~ref.} & $\mathbf{\Theta}$ & $\mathbf{I}$ & ~$\mathbf{\chi^{\text{sat}}}$  \\
    \midrule
    \midrule
    $\mathrm{V}_{\alpha}$ & Eq.(\ref{eq:valpha}) & even & odd & $\mathrm{L}^5$ \\
    \midrule
    $\mathrm{V}_{\text{bo}}$ & Eq.(\ref{eq:HOC_Vbo3}) &  even &  even & $\mathrm{L}$ \\
            $\mathrm{V}_{\text{bi}}$  &Eq.(\ref{eq:HOC_Vbi_12})  & even  & odd &  $\mathrm{L}^2$    \\
            \midrule
            $\mathrm{V}_{\text{ex}}$  &Eq.(\ref{eq:HOC_Vex3_m_m})  & odd  & odd &  $\mathrm{L}$    \\
            \midrule
            $\mathrm{V}_{\text{mnc},1}$ & Eq.(\ref{eq:HOC_Vex3_m_p}) & odd & even  & $\mathrm{L}^2$      \\
            $\mathrm{V}_{\text{mnc},2}$ & Eq.(\ref{eq:HOC_Vex3_p_m}) & even &  odd &  $\mathrm{L}^2$     \\
            \hline
             & & & & \\
        \end{tabular}}
    \end{minipage}
    \begin{minipage}[c]{0.69\linewidth}
    \includegraphics[width=0.95\linewidth]{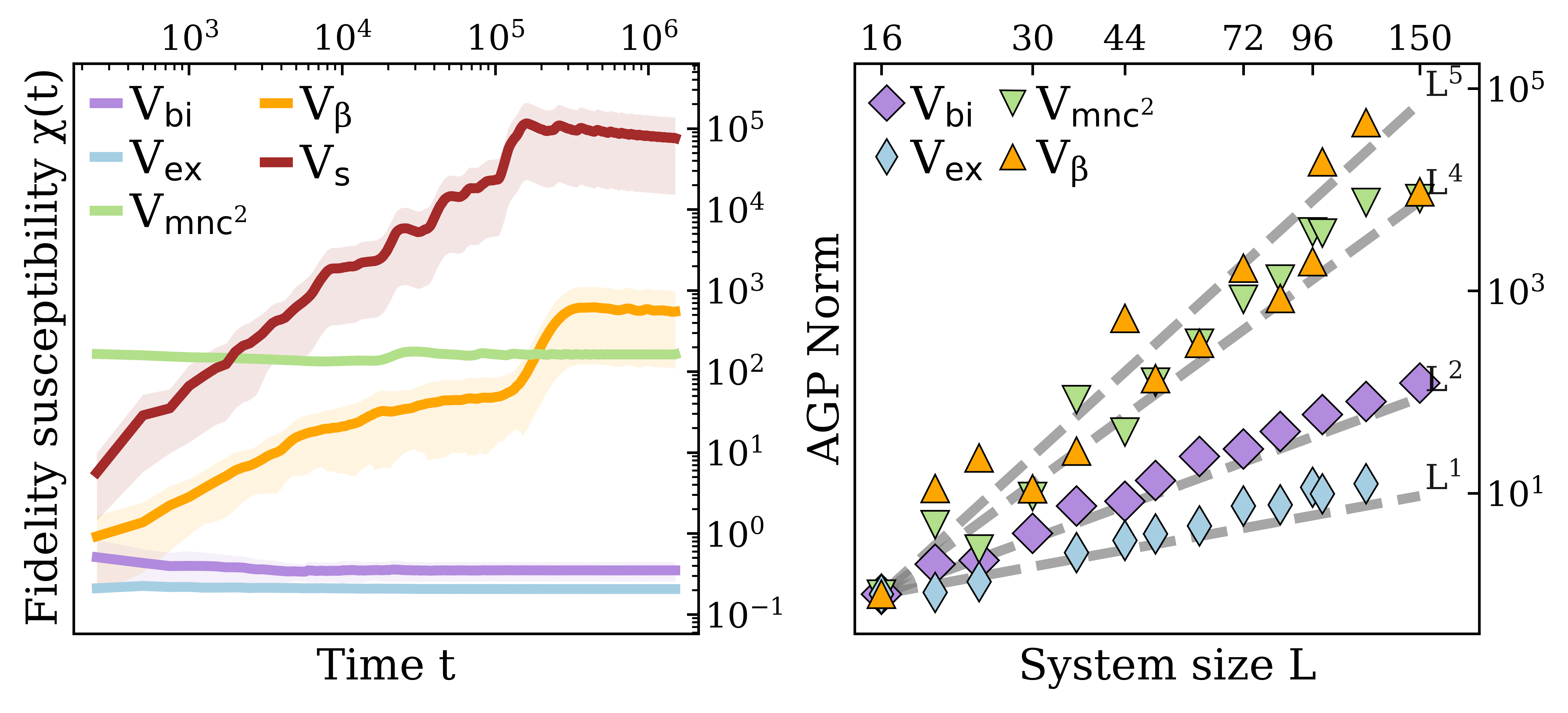}
    \end{minipage}
    \begin{minipage}[c]{0.30\linewidth}{\renewcommand{\arraystretch}{1.01}
  \begin{tabular}{c c  c c  c}
    & \multicolumn{4}{c}{\large Quartic Perturbations} \\
    & & & & \\
    $\mathbf{V}$ & \textbf{Eq.~ref.} & $\mathbf{\Theta}$ & $\mathbf{I}$ & ~$\mathbf{\chi^{\text{sat}}}$  \\
    \midrule
    $\mathrm{V}_{\beta}$ & Eq.(\ref{eq:vbeta}) & even & even &  $\mathrm{L}^4$ \\
    $\mathrm{V}_{s}$ & Eq.(\ref{eq:HOC_Vs}) &  even & even  &  $-$ \\
   \midrule
   \multicolumn{2}{c}{ Quartic WIBs} & &  &\\
   \midrule
    $\mathrm{V}_{\text{bi}} $ & Eq.(\ref{eq:HOC_Vbi_23})&  even  & even   & $\mathrm{L}^2$    \\
   \midrule
    $\mathrm{V}_{\text{ex}}$ & Eq.(\ref{eq:HOC_Vex4_pm})  & even & odd & $\mathrm{L}$     \\
   \midrule
    $\mathrm{V}_{\text{mnc}^2}$ & Eq.(\ref{eq:HOC_Vex}) & even & odd & $\mathrm{L}^4$   \\  
    \hline
     & & & & \\
    \end{tabular}}
    \end{minipage}
    \caption{\textit{HOC chain.} 
    \textbf{Top left, plot:} Fidelity susceptibility, $\chi$ as a function of time $t$, or the variance of AGP for \emph{cubic} perturbations of the HOC at $L=100$ trajectories with initial energy density $\epsilon=0.1$, obtained by averaging over an ensemble of $N=100$ trajectories. Shaded bands indicate $\pm \Delta\chi(t)$, where $ \Delta\chi(t)$ is the run-to-run variance of $\chi$ at time $t$ across all the runs. 
    \textbf{Top middle, plot:} Scaling of the AGP norm (or fidelity susceptibility $\chi(T)$ at final time $T$) with system size $L$ for the WIB perturbations presented in the top right table. The exact fits, as well as theoretically predicted scalings, can be found in the Appendix \ref{app:analytic_scalings}. 
     \textbf{Top right, table:} Perturbations studied (with references to equations in the main text), their time reversal ($\Theta$) and inversion ($I$) symmetries, and the observed AGP norm ($\chi^{\rm sat}(L)$) scaling.
     \textbf{Bottom row:} Everything is the same as in the top row, with the exception that perturbations in the bottom row are quartic perturbations to HOC.
    }
    \label{fig:agp-hoc-a-global}
\end{figure*}

\subsubsection{AGP for the Toda chain}
\label{sec:agp_toda}
We begin by discussing the Toda chain.
We observe that in classical models, the variance of the AGP seems to exhibit two main behaviors: $ \sigma^2(T)$ either grows with time, or it quickly saturates to a stationary value.
In the classical case, the variance of AGP is computed along a given trajectory and must be averaged over the initial distribution, whereas in the quantum case we did not have to do so.
The upside of the classical setup is that we can access much larger system sizes.
Additionally, recent work by Karve {\it et al.}~\cite{Karve_2025} provides a direct baseline for computing the AGP variance. 

In this section, we present results from numerical studies of AGP for all WIBs of the Toda chain discussed in the main text.
Our findings are summarized in Fig.~\ref{fig:agp-toda-global}. 
By looking at the left panel of Fig.~\ref{fig:agp-toda-global}, we can immediately see that there is a clear visual difference between generic and WIB perturbations.
The variance of the AGP for generic perturbations quickly grows, with no visible saturation in time.
This growth is observed for all system sizes considered.
In the case of WIBs, we observe a quick saturation to a stationary value $\chi^\mathrm{sat}$ (which we will simply denote by $\chi$ in the following), which depends on the system size.

We find numerically that the AGP variance for WIB perturbations obtained from extensive local generators saturates to a value that scales linearly with system size, as one would expect from Eq.~(\ref{eq:AGP-var-as-VarX}). WIBs from momentum non-conserving generators, on the other hand, scale as $L^2$. 
Similarly, for boosted and bilocal perturbations,  we find $\chi(L) \sim L^2$ scaling.

Generic perturbations do not seem to saturate even for much longer times (up to $8.5\times10^6$), and therefore we do not extract such scalings here. Among these generic perturbations, we report on the case of the $\alpha$-FPUT and $\beta$-FPUT for comparison with the HOC chain.
Note that the $\alpha$-FPUT perturbation on top of the fixed Toda chain is a generic perturbation, since the cubic potential cannot be generated by any $\{X, H_0\}$.

The Toda chain approaches the HOC in the low energy limit, and we discuss the impact of this in App.~\ref{app:toda_eps}, see Fig.~\ref{fig:agp_with_eps}.
Our simulations presented in this section were done at fixed energy density $\epsilon=0.1$ commonly used in the literature, which is somewhere in the middle of the energy densities in Fig.~\ref{fig:agp_with_eps}. 
One can see and appreciate that for $\alpha$-FPUT and $\beta$-FPUT-like perturbations to Toda, at the energy density used in the present section, the AGP variance grows slower than for generic perturbations due to this, as in Fig.~\ref{fig:agp-toda-global}.
However, looking at Fig.~\ref{fig:agp_with_eps}, we see that with increasing $\epsilon$, this effect goes away. 

\begin{figure*}[ht]
\begin{minipage}[c]{0.33\linewidth}
{\renewcommand{\arraystretch}{0.94}
\centering
    \begin{tabular}{c c  c c  c}
        & \multicolumn{4}{c}{\large SU(2) \textit{conserving}} \\
        \midrule
        $\mathbf{V}$ & \textbf{Eq.~ref.} & $\mathbf{\Theta}$ & $\mathbf{I}$ & ~$\mathbf{\chi^{\text{sat}}}$  \\
        \midrule
        \midrule
        $\mathrm{V}_{\text{bo}}$ & Eq.(\ref{eq:ish_vbo1}) & even & even &  $\mathrm{L}^2$ \\
        $\mathrm{V}_{\text{bi}}$ & Eq.(\ref{eq:ish_vbi23}) & even & even &  $\mathrm{L}^2$ \\
        & & & & \\
        $\mathrm{V}_{\text{ex}}$& Eq.(\ref{eq:ish_vex1}) & odd  & odd & $\mathrm{L}$\\
        \midrule
        $\mathrm{V}_{\text{step}_1}$ & Eq.(\ref{eq:ish_vstep1}) & odd & - &  $\mathrm{L}$\\
        $\mathrm{V}_{\text{step}_2}$ & Eq.(\ref{eq:ish_vstep2}) & even &  -&  $\mathrm{L}$\\
        \midrule
        $\mathrm{V}_{\text{s,1}}$ & Eq.(\ref{eq:ish_vs1}) & even  & even & - \\
        $\mathrm{V}_{\text{s,2}}$ & Eq.(\ref{eq:ish_vs2}) & even  & even & - \\
        \end{tabular}}
    \end{minipage}
    \begin{minipage}[c]{0.655\linewidth}
\includegraphics[width=0.95\linewidth]{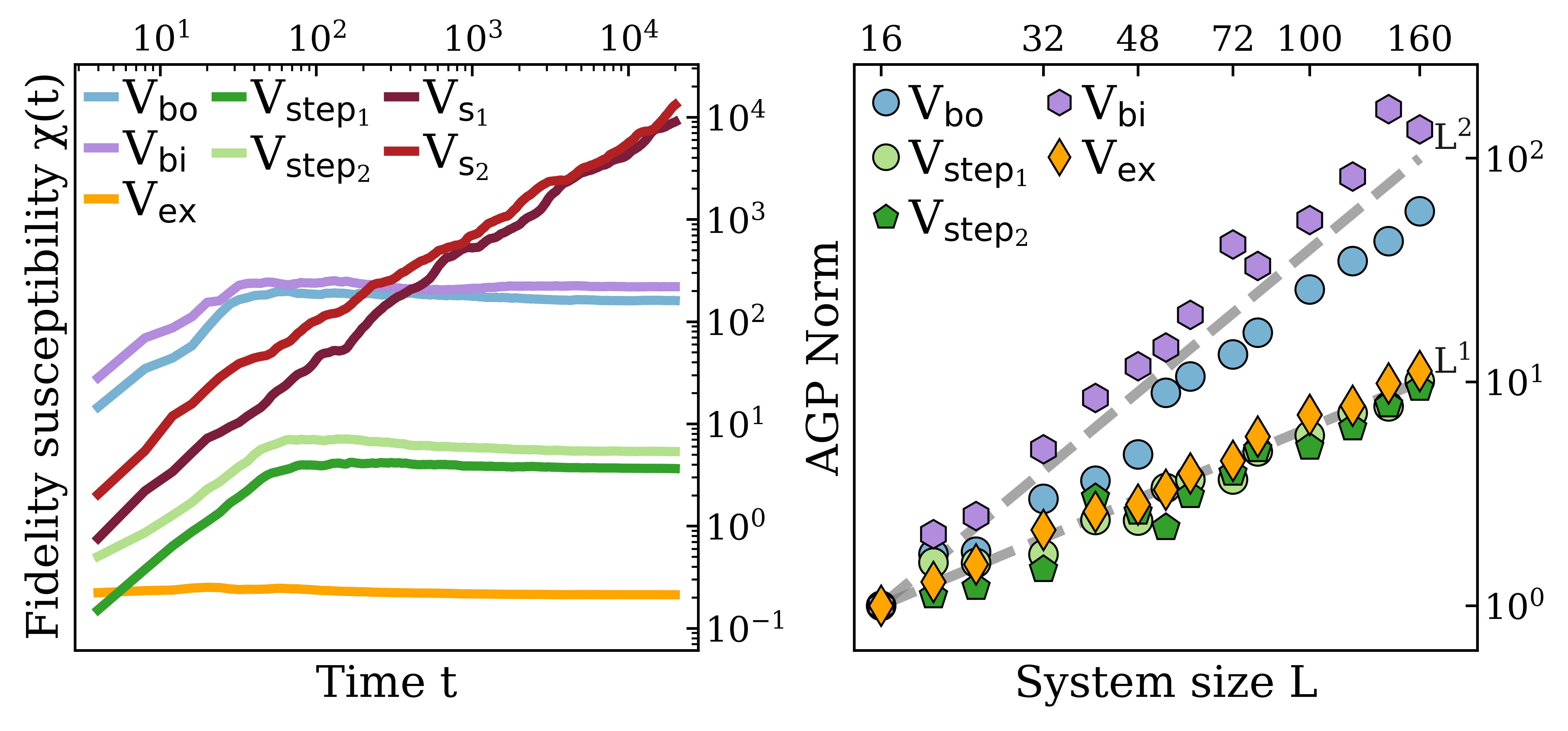} 
    \end{minipage}
    \begin{minipage}[c]{0.34\linewidth}{\renewcommand{\arraystretch}{0.96}
    \begin{tabular}{ccccc}
        &\multicolumn{4}{c}{\large  SU(2) \textit{breaking}} \\
        \midrule
        $\mathbf{V}$ & \textbf{Eq.~ref.} & $\mathbf{\Theta}$ & $\mathbf{I}$ & ~$\mathbf{\chi^{\text{sat}}}$  \\
        \midrule
        \midrule
        $\mathrm{V^z_{\text{bo}}}$ & Eq.(\ref{eq:ish_vz_weak}) & even &  odd & $\mathrm{L}^3$ \\
        $\mathrm{V^z_{\text{bi}}}$ & Eq.(\ref{eq:ish_vbi12}) & even & odd & $\mathrm{L}^3$ \\
        $\mathrm{V^z_{\text{ex}}}$& Eq.(\ref{eq:ish_vz_stag_weak}) &  even & odd & $\mathrm{L}$\\
        \midrule
        $\mathrm{V^z_{\text{step}}}$ & Eq.(\ref{eq:ish_vstep0}) & odd & - & $\mathrm{L}^2$ \\
        $\mathrm{V^z_{\text{loc}}}$ & Eq.(\ref{eq:ish_loc1}) & odd & - & $\mathrm{L}^0$\\
        \hline
        $\mathrm{V^z_{\text{s}_1}}$ & Eq.(\ref{eq:ish_vz_strong}) & even  & odd & - \\
        $\mathrm{V^z_{\text{s}_2}}$ & Eq.(\ref{eq:ish_vz_strong2}) &   &  & - \\
        \end{tabular}}
    \end{minipage}
    \begin{minipage}[c]{0.65\linewidth}
\includegraphics[width=0.97\linewidth]{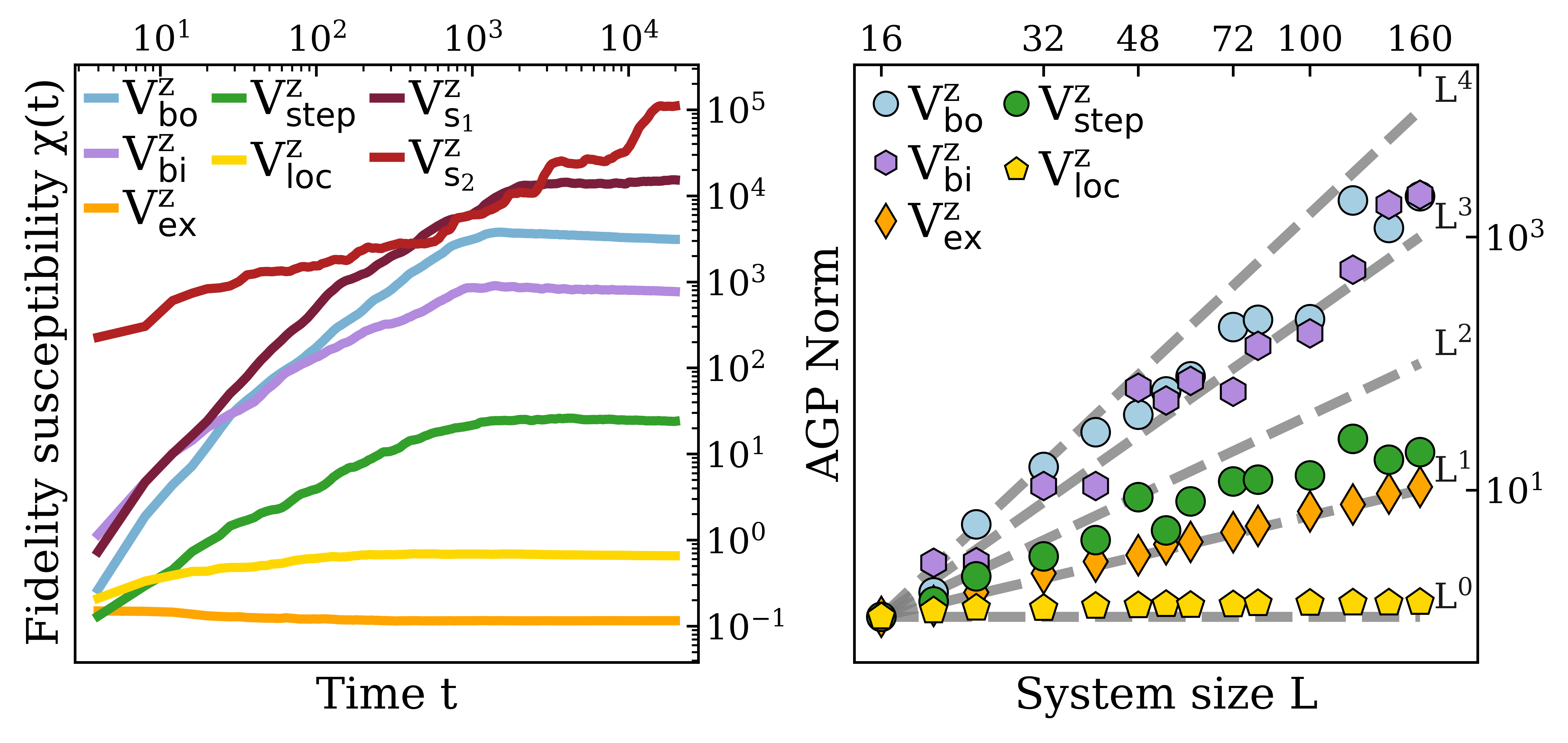}
\label{fig:agp-ishimori-su2}
    \end{minipage}
    \caption{\textit{Ishimori chain.} \textbf{Left, table:} 
    All perturbations studied with reference to their equation in the main text; time reversal $\Theta$ and inversion $I$ symmetries; and their observed AGP norm scaling.
    \textbf{Middle, plot:} 
    Variance of the AGP averaged over 100 trajectories for the Ishimori chain in the presence of WIB perturbations generated from boosted, bilocal, and extensive local generators obtained in Sec.~\ref{sec:ishimori}, for system size $L=72$ with $\beta=1$.
    \textbf{Right, plot:} Scaling of the saturation of AGP variance with system size.}
    \label{fig:agp-ishimori-global}
\end{figure*}

\subsubsection{AGP for the HOC}
\label{subsubsec:agp_hoc}
In contrast to the Toda model, \emph{all} cubic and quartic perturbations to HOC we studied reach a clear late–time plateau of the AGP variance, $\chi(T)$, within our time window ($T\!\lesssim\!1.5\times 10^6$). 
[See Fig.~\ref{fig:ab_longT} in App.~\ref{subapp:HOC_longerT} for additional data demonstrating this on some of the representative perturbations ran for longer timescales ($T\!\lesssim\!8.5\times 10^6$) as well.]
This makes sense in light of our findings in Sec.~\ref{sec:xagp-all-cubic} for cubic perturbations.
The distinction between \emph{weak} versus \emph{generic} perturbations (or in this case WIBs from a known class of $X$'s and WIBs from $X_{\alpha\text{-FPUT}}$ like generators), therefore, appears most evident in the finite–size scaling of the AGP variance saturation (referred to as AGP norm below) (see Fig.~\ref{fig:agp-hoc-a-global}).

The results for both the cubic and quartic perturbations are summarized in Fig.~\ref{fig:agp-hoc-a-global}. 
Our main findings are that for boost-generated and momentum-conserving (MC) extensive local perturbations, the AGP norm scales linearly with $L$, while for the bilocal and the considered momentum-nonconserving (MNC) extensive local perturbations, the AGP norm scales $\sim L^2$.
Thus, for extensive local generators, we identified two types of scaling, depending on whether the generator is momentum-conserving or not.
We provide an analytic argument for this from straightforward calculations of the AGP variance, which, for some of the generators in the HOC, can be performed assuming approximate applicability of equilibrium statistical mechanics ensembles, see App.~\ref{app:analytic_scalings} for details.
Note that the scaling of the AGP variance linearly with $L$ for the boost-generated WIBs supports our earlier findings that boost–generated WIBs to HOC can be expressed as linear combinations of momentum-conserving extensive terms plus IoMs, hence they inherit the $\sim L$ scaling. 
Bilocal generators, containing $O(L^2)$ of local densities, add an extra factor of $L$, yielding $\sim L^2$.

The AGP variance for the $\alpha$–FPUT interaction, on the other hand, scales as $L^5$.  
Although it is not generated by any of the previously known locality classes (extensive, boosted, or bilocal), we showed in Sec.~\ref{sec:xagp-all-cubic} that it \emph{is} a genuine WIB:
We indeed find that its AGP does saturate rapidly, and we know it admits quasi–integrals of motion for the first several HOC charges.
However, its generator is a nontrivial \emph{trilocal} operator in real space, and we find in App.~\ref{app:analytic_scalings} that its AGP norm scales as $L^5$, which is consistent with our numerical findings. 

Quartic perturbations follow the same pattern, with AGP variance for perturbations from MC generators scaling as $\sim L$ and from bilocal generators scaling as $\sim L^2$.
We find that a fairly simple NMC extensive perturbation from an NMC generator has AGP norm that scales as $L^4$.
However, the analytic argument in App.~\ref{app:analytic_scalings} provides an explanation for the scaling with $L$, which might appear to be unusual for a WIB. 
We also provide data for the $\beta$-FPUT perturbation on top of HOC, which also appears to saturate and scale as $L^4$. 
We do not analyze the generator for this perturbation in this work.
Finally, for the quartic perturbation $V_s$ in Fig.~\ref{fig:agp-hoc-a-global}, we see that the fidelity susceptibility also saturationg but to even larger values than for $V_{\beta}$; because of this, we do not attempt to scale the AGP norm in this case (we suspect that it is also polynomial in $L$ but perhaps with higher power than for $V_{\beta}$).

\subsubsection{AGP for the Ishimori model}
\label{sec:agp_ishimori}
Computing the AGP variance for the Ishimori model is more subtle, and a few warnings are in order.
We find that to preserve the spin size (i.e., $\vec{S}_j^2 = 1$) and energy with the RK4 method, we are limited to studying shorter timescales than when using the ABA864 for the Toda and HOC models (for more details on methods see Section \ref{sec:ishimori_techniques}).

Nevertheless, we can still study the AGP variance for WIBs, since they saturate on the timescale we study, in contrast to all generic perturbations we tested, which show continued growth and therefore do not admit a meaningful finite-size scaling analysis of the saturation.
We have checked that our results hold for longer times for selected system sizes and perturbations. 
However, for the systematic study reported in Fig.~\ref{fig:agp-ishimori-global}, we chose a timescale similar to that used in our transport studies.

Our numerical findings support our earlier claims that for all of the perturbations we propose as WIBs, the variance of AGP saturates quickly in time (see Fig.~\ref{fig:agp-ishimori-global}).
This holds for the strictly local and SU(2)- breaking WIB perturbations as well.
The scaling of the saturation value ($\chi$) with system size is particularly informative, because it reflects the locality structure of the underlying generator $X$.

First, let us discuss the scaling we find for SU(2) preserving WIBs. 
We find that $V_{\text{ex}}$ showed linear scaling of $\chi$ with $L$, matching the expectation for $X_{\text{ex}}$. 
For boosted ($V_{\text{bo}}$) and bilocal ($V_{\text{bi}}$) perturbations we find $\chi \propto L^2$.
For the bi-local generator, this matches the naive expectation, obtained by taking the expression Eq.~(\ref{eq:ish_vbi12}) literally on a finite chain of length $L$, which then has $O(L^2)$ terms. 
However, the case of WIBs generated from the boost is more subtle. 
As we mentioned before, $X_{\text{bo}}$ defined in Eq.~(\ref{eq:Vbo}) is an infinite system expression, and the problem of finding an analytic expression for a finite-size version of the boost was previously investigated in the quantum case \cite{Vanovac2024}. 
There, the AGP norm (equivalent to the AGP variance here) was found to scale as $\propto L\log L$, where scaling closer to $L^2$ was expected. 
We note that this might have been due to finite-size effects, as exact diagonalization (ED) in quantum systems allows only small system sizes to be accessed. 
If one looks at Fig.~\ref{fig:agp-ishimori-global}, up to $L=48$, growth of $X^{\text{AGP}}_{\text{bo}}$ seems closer to $L$, and for larger system sizes it jumps to match the $L^2$ scaling. 
Finally, two strictly local perturbations obtained from the step generators we considered, $V_{\text{step},1}$ and $V_{\text{step},2}$, both scale linearly with $L$, as naively expected. 

SU(2)-breaking perturbations seem to display more exotic scalings. 
Starting with the simplest case, $V^z_{\text{loc}}$, which should not depend on $L$, we see in Fig.~\ref{fig:agp-ishimori-global} that it stays constant with system size. 
Next up is $V^z_{\text{ex}},$ which scales linearly with $L$ as expected. 
Both the boosted $V^z_{\text{bo}}$ and bi-local $V^z_{\text{bi}}$ perturbations seem to show stronger scaling with $L$ than for isotropic perturbations, and we find $\chi \propto L^3$ in those cases. 
Additionally, $V^z_{\text{step}}$ (which is a local term of $V^z_{\text{bo}}$) shows stronger than linear scaling with $L$.
We do not offer a conclusive answer regarding the origin of these scalings at this time. 
Given the unusual growth of $\chi$ for $V^z_{\text{bo}}$ and $V^z_{\text{step}}$, it is possible that our initialization scheme or parameter choices could explain this, but we do not rule out the possibility that it reflects a more fundamental effect.
For now, we note that the scaling is polynomial in $L$ and that these perturbations are WIBs.

\subsection{Summary of scalings}
A central tool in our numerical analysis is the classical AGP, and, in particular, the system-size scaling of the long-time saturation value of its variance along trajectories.
In this section, we briefly summarize the findings of our analysis across the three models. 

For all WIB perturbations studied, the AGP variance rapidly saturates, and the resulting plateau value (after averaging over an ensemble of initial conditions) scales polynomially with system size $L$,
In contrast, generic perturbations lead to a variance that continues to grow over the observed simulation times, and thus they do not admit a controlled finite-size scaling analysis of the AGP norm.
We conjecture that if, for generic perturbations, the saturation happens eventually, the corresponding time scales and the AGP norm scale much more strongly with the system size; e.g., in the quantum case, the AGP norm for generic perturbations scales exponentially in the system size, but we do not have at present an analogous understanding in the classical case.
Our results thus reveal a clear difference in the AGP norm scaling directly tied to the type of generating operator.

Across all three models, extensive local WIBs that respect the relevant conservation laws yield the mildest scaling: 
In the Toda and HOC chains, momentum-conserving extensive generators give $\chi^{sat} \sim L$, and in the Ishimori model, the corresponding SU(2)-preserving extensive generators likewise give scaling $\chi^{sat} \sim L$.

The AGP norms for bilocal WIBs in all three models exhibit a robust quadratic scaling with system size $\chi^{sat} \sim L^2$, consistent with the naive estimate. 

Boost-generated WIBs show a difference between interacting and non-interacting models:
In the interacting Toda and Ishimori models, we find $\chi^{sat} \sim L^2$, which is weaker than a naive estimate based on infinite-volume boost (one might guess an additional power of $L$ from the spatial weight $j$), suggesting that cancellations and/or correlations effectively reduce the scaling on a finite ring.
In the noninteracting HOC, on the other hand, boosted operators are equivalent (as generators)  to extensive local ones and thus give scaling $\chi^{sat} \sim L$ instead. 

The HOC chain analysis further reveals that WIBs in this case are not exhausted by the extensive/boosted/bilocal hierarchy as was previously believed.
We find that the $\alpha$-FPUT perturbation, whose generator we identify explicitly in momentum space and show to be trilocal in real space, exhibits the AGP norm scaling $\chi^{sat} \sim L^5$. 
Other higher power polynomial scalings are possible too, as we explained in the case of the quartic momentum-non-conserving perturbation in Eq.~(\ref{eq:HOC_Vex4_pm})

We find that symmetry breaking leads to enhanced scalings.
In the Ishimori chain, the SU(2)-breaking perturbations display enhanced scalings (with $L^3$) for boosted and bilocal cases.
We also find that, in the Toda and HOC chains, WIBs from momentum-nonconserving extensive generators exhibit different scalings from those of momentum-conserving ones.  

Although finite-size effects are stronger in the interacting models, the same structure across the three models emerges:
Linear scaling of the AGP norm with system size signals extensive local generator; quadratic scaling was found for boosted and bilocal generators; and higher powers seem to be related to new types of generators, with the HOC providing concrete examples beyond the standard classes.

\subsection{Energy transport in the perturbed Ishimori chain}
The characterization of the transport of conserved quantities may also serve as a probe of the onset of chaos in nearly-integrable systems.  Diffusive transport ($z=2$) of conserved charges is a hallmark of chaotic dynamics; in contrast, integrable systems host stable, ballistically-propagating quasiparticles with dynamical exponent $z=1$.  Like the quantum Heisenberg spin chain, the Ishimori spin chain features ballistic energy transport ($z=1$) and superdiffusive spin transport with dynamical exponent $z=3/2$---see \cite{Bertini_2021, Bulchandani2021, Gopalakrishnan_2023, Gopalakrishnan2024superdiffusion} for recent reviews on the subject.  Under the effect of an integrability-breaking perturbation, both energy and spin transport will become diffusive at long times as the system's dynamics become chaotic.  Therefore, the timescale of the crossover from faster-than-diffusive transport at short times to diffusive transport at long times is one way to study the onset of chaos in an integrable system subject to a nonintegrable perturbation.  

Here, we numerically probe the timescale $t_\star$ associated with the onset of diffusive energy transport in the perturbed Ishimori spin chain, which provides one way to probe the differences between generic and weak perturbations.  Generically, for quantum systems, $t_\star$ follows from Fermi's Golden Rule (FGR) considerations and scales as $t_\star \sim \lambda^{-2}$ \cite{PhysRevLett_stability}.  On the other hand, as argued in Ref.~\cite{Surace2023}, quasi-conserved charges that commute with the Hamiltonian up to $O(\lambda^\ell)$ are expected to thermalize on an anomalously-long timescale $\tau \sim \lambda^{-2 \ell}$.  Since the weak perturbations constructed in this work have vanishing Poisson brackets with $H$ up to $O(\lambda^2$), we expect a longer crossover timescale for weak perturbations $t_\star \sim \lambda^{-4}$.

\begin{figure}[h!]
\includegraphics[width=0.8\linewidth]{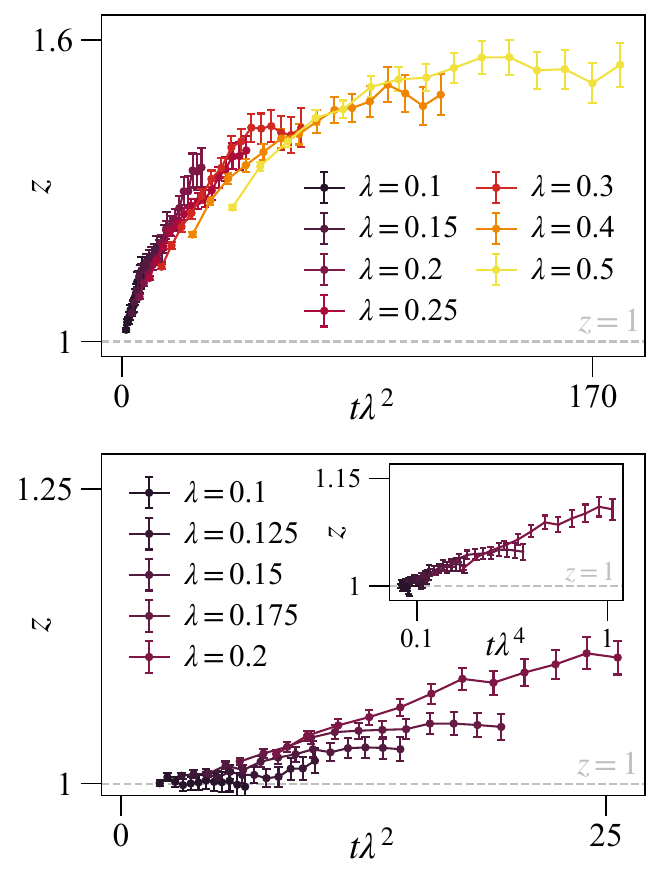}
\caption{Top: Rescaled dynamical exponent $z(t\lambda^2)$ for the Ishimori chain under a generic perturbation $V_g = \sum_j \vec S_j \cdot \vec S_{j+1}$ extracted from fluctuations in energy transfer across a central link.  The curve collapse indicates that the crossover from ballistic to diffusive transport is consistent with the $t_\star \sim \lambda^{-2}$ timescale predicted by Fermi's Golden Rule.  Bottom: The dynamical exponent curves for weak perturbation $V_{\rm ex}$ rescaled by $\lambda^2$ do not neatly collapse, indicating that the crossover timescale $t_\star$ associated with $V_{\rm ex}$ is anomalously long.  Although the data are severely limited by a short numerically accessible timescale, the available data appear to be consistent with the predicted crossover timescale of $t_\star \sim \lambda^{-4}$ (inset).  All data shown is run up to a final time of $t_f=920$ for a system of $L=4000$ sites and averaged over at least $5 \times 10^4$ independent realizations.
}
\label{fig:transport}
\end{figure}

To numerically extract the crossover timescale $t_\star$, we simulate the Ishimori spin chain under the effect of an integrability-breaking perturbation of strength $\lambda$ and extract the dynamical exponent $z$ associated with energy transport; further details on numerical techniques can be found in Appendix~\ref{sec:numerics}.  Figure~\ref{fig:transport} shows the dynamical exponent $z$ for both the WIB perturbation $V_{\rm ex}$, Eq.~(\ref{eq:ish_vex1}), derived in Sec.~\ref{sec:ishimori_ioms} and a generic perturbation $V_g$, Eq.~(\ref{eq:ish_vs1}).
For the generic perturbation $V_g$, the dynamical exponent curves collapse neatly for the rescaled time variable $t \lambda^2$, consistent with the timescale predicted by Fermi's Golden Rule, $t_\star \sim \lambda^{-2}$.  
In contrast, the dynamical exponent curves extracted for the runs with $V = V_{\rm ex,1}$ from Eq.~(\ref{eq:ish_vex1}) clearly do not collapse when rescaling by $t_\star \lambda^2$ and are possibly consistent with a $t\lambda^4$ collapse (see Fig.~\ref{fig:transport}). 

Note that we chose to study $V_{\rm ex}$ simply because the model's EOMs are less prone to blow-ups (some denominators in the EOMs becoming very small), which, for the Ishimori model at high temperatures, are enhanced for some of the other WIBs we construct. Even though the EOMs for the perturbation $V_{\rm ex}$ are more stable than some of the other weak perturbations constructed here, they still contain divergences that pose a significant computational challenge (see Appendix \ref{sec:numerics}).  Therefore, the numerical results are limited to a relatively short timescale in the context of the anticipated $t_\star \sim \lambda^{-4}$ crossover timescale.
Additionally, EOMs for some of the WIBs we construct are more complicated. 
We have nevertheless constructed and tested them. 

As shown in the inset of the bottom panel of Fig.~\ref{fig:transport}, the expected $t_\star \sim \lambda^{-4}$ timescale is consistent with the data; however, the quality of this collapse is limited due to the relatively short numerically-accessible times and the limitation to smaller values of $\lambda$ for weak perturbations (see App.~\ref{sec:ishimori_techniques} for more details).

Our computational focus on the ballistic-to-diffusive crossover in energy transport is motivated by the fact that the superdiffusive-to-diffusive crossover in spin transport for such integrable spin chains under the effect of isotropic perturbations (i.e., perturbations invariant under spin-rotation symmetry) is known to be anomalously long.  A crossover timescale of $t_\star \sim \lambda^{-6}$ has been observed in both classical \cite{McCarthy2024}, and quantum \cite{Wang2025} integrable spin chains, while Ref.~\cite{McRoberts2024} reported a crossover timescale of $t_\star \sim \lambda^{-3}$ numerically observed in a classical Ishimori chain.  Based on the argument presented above, we then expect crossover to spin diffusion due to a weak perturbation to occur on a timescale of either $t_\star \sim \lambda^{-12}$ or $t_\star \sim \lambda^{-6}$.

\subsection{Other probes of chaos: Lyapunov Exponent} 
In classical systems, a commonly used measure of chaos is the largest Lyapunov exponent $\lambda_{\text{max}}$, which is expected to go to zero for integrable systems and to some finite value for all others. 
The challenge with this is that for models such as the $\alpha$-FPUT, the Lyapunov exponent may look almost zero for long times when parameters are chosen so that the model is close to the integrable Toda lattice \cite{Danieli2024}. 
We note that computing $\lambda_{\text{max}}$ is more challenging than simply computing the AGP, since, for example, one needs to compute the time evolution of the perturbed model, and not only the Jacobian but also the Hessian of the Hamiltonian. Additionally, sensitivity to perturbation strength needs to be taken into account.
Furthermore, EOMs for the perturbed model tend to be more complicated and even mix terms, posing an added challenge to the symplectic methods such as the one we used here \cite{Blanes2013}. 
Still, recent works have revisited the usefulness of the Lyapunov spectrum and the maximal Lyapunov exponent \cite{Prosen2025,Kim2025}.

While we do not pursue this question in detail in the present work, our WIB construction produces several nearest-neighbor perturbations with simple structure and no mixed terms, which are particularly amenable to symplectic integration and Lyapunov analysis. 
It would be interesting to revisit the relationship between Lyapunov time, integrability-breaking time, and thermalization times in light of WIBs, as previous numerical studies have focused exclusively on generic perturbations. 
Our new classification of the $\alpha$–FPUT interaction as a WIB with respect to the HOC may also help reinterpret earlier observations in which the Lyapunov exponent remained anomalously small over extended time intervals before saturating to a nonzero value.
A better understanding of the hierarchy of timescales associated with different dynamical regimes on the way to diffusion remains an important open problem.

\section{Conclusions and Future Directions }
\label{sec:conclusions}
In this work, we establish a systematic framework for engineering weak integrability-breaking (WIB) perturbations in \emph{classical} integrable lattice models, extending a construction recently developed for quantum spin chains. \cite{Surace2023}
We show how different families of generators (extensive local, boosted, and bilocal) can be used to produce local, extensive perturbations that endow the perturbed model with an extensive set of quasi-integrals of motion, which are conserved up to order $\mathcal{O}(\lambda^2)$ in the perturbation strength $\lambda$.
We explicitly derive WIBs for three classical systems: the Ishimori spin chain, the nonlinear Toda lattice, and the harmonic oscillator chain.

A central tool in our numerical analysis is the classical adiabatic gauge potential (AGP). \cite{KOLODRUBETZ_2017}
We find that the AGP serves as a powerful probe of integrability breaking in classical integrable systems as well. 
For all WIB perturbations studied, the AGP variance, $\chi$, rapidly saturates with time to a value $\chi^{\text{sat}}$ that exhibits a polynomial scaling with system size $L$.
In contrast, generic perturbations lead to variance that continues to grow over time and/or exhibits much stronger scaling with system size.
Across all three models considered, clear universality classes emerge, where the type of polynomial scaling with $L$ is determined by the structure of the generating operator.
The harmonic chain further reveals that WIBs extend beyond the standard extensive/boosted/bilocal generator hierarchy.
We also find that reducing symmetry (of perturbations or generators compared to the unperturbed model) leads to stronger scaling of the AGP norm: 
In the Ishimori chain, SU(2)-breaking perturbations display stronger growth (e.g., $\chi^{\text{sat}} \sim L^3$ for boosted and bilocal cases), resembling the enhanced scaling behavior of extensive momentum-nonconserving generators in the Toda and HOC chains.
Overall, the system-size scaling of the AGP norm provides a unifying and model-independent classification of weak integrability breaking: 
Linear scaling signals extensive local structure of generators; quadratic scaling corresponds to bilocal or boost-type generators; and higher powers indicate more nonlocal or reduced-symmetry mechanisms.

Within the HOC, we further analyzed the AGP in terms of Fourier modes.
For general cubic, translationally invariant perturbations, we obtained an explicit expression for the AGP and showed that momentum-conserving such perturbations soften the infrared singularities in the AGP kernel, ensuring that the corresponding quasi-IoMs remain extensive local. 
We then specialized to the $\alpha$-FPUT problem.
We found that the $\alpha$–FPUT interaction defines a genuine WIB of the HOC: 
First, using the Toda IoMs, we explicitly construct corrections to the HOC charges.
We showed that the same quasi-IoMs can be obtained using our analytical result for the AGP, thereby confirming our previous finding that the $\alpha$-FPUT perturbation is a WIB to the HOC.

We further analyzed the generator in real space and found that it takes the form of a trilocal operator.
This provides a concrete example of a WIB generated by an operator that does not fall into the previously known extensive local, boosted, or bilocal classes, and suggests the existence of a richer “higher-order” locality structure in the space of WIB generators.
Similar polynomial scaling of the AGP norm was also predicted in free-fermion systems with quartic perturbations \cite{Pozsgay_2024}. 
Interestingly, in the free-fermion examples beyond the standard classes, only a subset of IoMs was corrected~\cite{Pozsgay_2024}, whereas in the HOC with cubic perturbations, all IoMs were corrected. 
It will be interesting to do a future study with the $\beta$-FPUT perturbation, as it may be more similar to the perturbations in the free-fermion case, where not all IoMs got proper extensive local correction. 
While the $\beta$-FPUT perturbation exhibits polynomial AGP norm growth, we conjecture that it may represent a more generic form of integrability breaking of the HOC model compared to the $\alpha$-FPUT perturbation.

More generally, our Fourier-space analysis shows that any cubic, translationally invariant, momentum-conserving perturbation of the harmonic oscillator chain admits a trilocal AGP generator where all IoMs get extensive local corrections and is therefore a WIB.
In this sense, the $\alpha$–FPUT interaction is not an isolated example but rather a representative member of a broader class of cubic perturbations that share the same structural property. 
This observation raises an interesting open question: whether there exist longer-range integrable models, analogous to the Toda lattice, whose hierarchy of conserved quantities naturally connects to these more general cubic perturbations and could provide a systematic integrable framework for understanding their quasi-conserved structures.

The delayed crossover to diffusion observed in our work studying the perturbed Ishimori model invites a more detailed investigation of transport. 
An important open question is how different classes of WIB perturbations (e.g., boosted vs.~bilocal) quantitatively affect the scattering of quasi-particles within the framework of generalized hydrodynamics.
Understanding this could help explain the remarkable persistence of anomalous transport phenomena, such as superdiffusion, observed in numerical and experimental studies of nearly integrable systems. \cite{Das_2019, Roy_2023, McRoberts2024}
It would be interesting to perform similar studies of transport (of both energy and momentum) in the perturbed Toda and HOC models as well.

An outstanding, closely related question concerns the precise characterization of thermalization times in perturbed integrable systems.
In generic settings, thermalization in such models is often described by Fermi’s golden rule (FGR), which predicts a relaxation time $\tau \sim \lambda^{-2}$ in terms of the perturbation strength $\lambda$. 
In classical systems, an analogue of FGR can be formulated in terms of phase-space (or Wigner-function) correlators~\cite{Bilha_2003}.
Our WIB framework suggests a systematic hierarchy of perturbations for which the effective relaxation timescale is parametrically longer, $\tau \sim \lambda^{-2\ell}$ with $\ell > 1$, due to the presence of quasi-IoMs.
More precisely, by extending the derivation in the quantum case (see App.~F of~\cite{Surace2023}) to the classical domain, one can prove that the quasi-IoMs remain approximately conserved up to timescales of $O(\lambda^{-2})$, significantly longer than the timescale $O(\lambda^{-1})$ expected for generic quantities with no such corrections.
These long timescales control the initial relaxation toward a prethermal plateau. 
Furthermore, if we conjecture the existence of a nearby integrable deformed Hamiltonian that approximates the perturbed system to $O(\lambda^2)$, an FGR argument as in~\cite{Surace2023} suggests that subsequent decay from this prethermal state to full thermalization occurs only over timescales $\lambda^{-4}$.
Directly testing this prediction and establishing a rigorous framework for such hierarchies in classical systems remains an intriguing open challenge.

Our results support the physical picture that WIBs delay chaos and thermalization far beyond FGR expectations, and that AGP offers a sensitive classifier of ``weak'' (i.e., special) directions inside the full perturbation space in both the classical and quantum models.

In this work, we have primarily inferred prolonged thermalization indirectly, through the scaling of the AGP variance and the delayed crossover to diffusive transport, rather than by directly extracting $\tau$.
A natural future direction is to combine AGP-based diagnostics with explicit measurements of relaxation times for suitably chosen observables, and to compare these to kinetic or FGR estimates.
Such a comparison could clarify how different notions of ``integrability-breaking strength'' (AGP scaling, Lyapunov spectrum, transport coefficients, and spectral measures) relate to one another and how they organize the space of perturbations into distinct classes.
While we have focused on the AGP, a systematic comparison with other probes, such as the maximal Lyapunov exponent and classical out-of-time-order correlators, would be highly valuable.

The WIB framework can be readily applied to other classical integrable systems, such as the Calogero–Moser or classical Gaudin models, where long-range interactions and nontrivial phase-space geometries may give rise to different classes of generators and quasi-IoMs. 
Looking further ahead, extending these concepts to Floquet-integrable systems and circuits is an interesting yet challenging direction, both in classical and quantum models.
While extensive local generators can be readily adapted, adapting the boost and bilocal formalisms to a discrete-time evolution setting remains an open, nontrivial problem. 
Additionally, advances in control of experimental platforms, such as trapped-ion arrays and cold-atom setups, may soon enable the direct engineering of Hamiltonians tuned to realize WIB perturbations and test the predictions of our framework in controlled laboratory conditions.
In fact, a recent study in this direction was reported in \cite{Chen_2026}.

Overall, our results indicate that WIBs provide a controlled setting in which to study the various timescales and mechanisms of integrability breaking, and offer a fertile ground for future analytical and numerical studies of classical and quantum many-body dynamics.

\section{Acknowledgments}
We thank Nachiket Karve, Hyeongjin Kim, Jorge Kurchan, Anatoli Polkovnikov, and Balázs Pozsgay for useful discussions.
SV and OIM acknowledge support by the National Science Foundation through grant DMR-2001186.
CM acknowledges support from the National Science Foundation through grant GRFP-1938059.
FMS acknowledges support provided by the U.S.\ Department of Energy Office of Science, Office of Advanced Scientific Computing Research (DE-SC0020290); DOE National Quantum Information Science Research Centers, Quantum Systems Accelerator; and by Amazon Web Services, AWS Quantum Program. A part
of this work was done at Les Houches School of Physics.  This research was supported in part by a grant NSF PHY-2309135 to the Kavli Institute for Theoretical Physics (KITP).

\bibliography{bib}
\appendix
\onecolumngrid

\section{Equations of motion for the Runge-Kutta (RK4) algorithm} \label{sec:eom}
To study integrability-breaking in the classical Ishimori spin chain, we consider 
\begin{equation}
    H = (1-\lambda)H_0 + \lambda V \label{eq:ham}
\end{equation}
where $H_0$ is the integrable Ishimori Hamiltonian given in Eq.~(\ref{eq:ishimori}), and $V$ is an integrability-breaking perturbation (either weak or generic).
By tuning the parameter $0 \leq \lambda \leq 1$, the strength of the perturbation is smoothly increased with respect to the strength of the integrable Hamiltonian; for $\lambda=0$, Eq.~(\ref{eq:ham}) reduces to the integrable Ishimori Hamiltonian, while choosing $\lambda=1$ gives an entirely non-integrable Hamiltonian.

For the Ishimori Hamiltonian, EOMs are given by
\begin{align}
\dt \sj{j} = \{\sj{j}, H_0\} & = 
    2\sj{j}\times \bigg[ \frac{\sj{j-1}}{1 + \sj{j-1} \cdot \sj{j}} + \frac{\sj{j+1}}{1 + \sj{j} \cdot \sj{j+1}}  \bigg] ~.
\end{align}
For the generic nearest-neighbor perturbation $V_g = \sum_j \vec{S}_j \cdot \vec{S}_{j+1}$, we just add to the EOMs $\lambda \vec S_j \times \big(\vec S_{j-1} + \vec S_{j+1} \big)$. 
For the extensive local WIB, $V_{\text{ex}}$ in Eq.~(\ref{eq:ish_vex1}),
the additional terms to EOMs are given by
\begin{align}
\vec{S}_j\times\frac{\partial V_{\rm ex}}{\partial \vec{S}_j}  
&= -2 \Biggl[
  \bigl((\sj{j+1}\times \sj{j+2})\times \sj{j}\bigr)\,
    \Bigl(\frac{1}{1 + \sj{j}\cdot\sj{j+1}}
          \;-\;
          \frac{1}{1 + \sj{j+1}\cdot\sj{j+2}}\Bigr)
    \;-\;
  \frac{\bigl((\sj{j}\times \sj{j+1})\cdot \sj{j+2}\bigr)\,
        \bigl(\sj{j+1}\times \sj{j}\bigr)}
       {\bigl(1 + \sj{j}\cdot\sj{j+1}\bigr)^2}
\nonumber\\ 
&\qquad
  +\;
  \bigl((\sj{j+1}\times \sj{j-1})\times \sj{j}\bigr)\,
    \Bigl(\frac{1}{1 + \sj{j-1}\cdot\sj{j}}
          \;-\;
          \frac{1}{1 + \sj{j}\cdot\sj{j+1}}\Bigr)
  \;-\;
  \frac{\bigl((\sj{j-1}\times \sj{j})\cdot \sj{j+1}\bigr)\,
        \bigl(\sj{j-1}\times \sj{j}\bigr)}
       {\bigl(1 + \sj{j-1}\cdot\sj{j}\bigr)^2}
\nonumber\\
&\qquad
+\;\frac{\bigl((\sj{j-1}\times \sj{j})\cdot \sj{j+1}\bigr)\,
        \bigl(\sj{j+1}\times \sj{j}\bigr)}
       {\bigl(1 + \sj{j}\cdot\sj{j+1}\bigr)^2}
  \;+\;
  \bigl((\sj{j-2}\times \sj{j-1})\times \sj{j}\bigr)\,
    \Bigl(\frac{1}{1 + \sj{j-2}\cdot\sj{j-1}}
          \;-\;
          \frac{1}{1 + \sj{j-1}\cdot\sj{j}}\Bigr)
\nonumber\\ 
&\qquad
  +\;
  \frac{\bigl((\sj{j-2}\times \sj{j-1})\cdot \sj{j}\bigr)\,
        \bigl(\sj{j-1}\times \sj{j}\bigr)}
       {\bigl(1 + \sj{j-1}\cdot\sj{j}\bigr)^2}
  \Biggr].
\end{align}
Note that the equations of motion for both the Ishimori Hamiltonian $H_0$ and the WIB perturbation $V_{\rm ex}$ diverge when neighboring spins become anti-aligned. To circumvent this challenge in transport studies, we prepare the system in a lower-temperature state in which the spins are mostly aligned (see App.~\ref{sec:ishimori_techniques}). 
Following this procedure allows the simulation of the Ishimori Hamiltonian for all values of $\lambda$; however, due to stronger divergences in the equations of motion for $V_{\rm ex}$, numerical results were only obtained up to $\lambda = 0.2$.

For our AGP studies, we only need the unperturbed Ishimori EOMs; in this case, we keep the high-temperature regime with inverse temperature $\beta=1$.

We also present the EOMs for the implementation of $V_{\rm bo}$, for which we do not report results at this time, but which might be of interest to readers.
For 
\begin{align}
V_{\text{bo},1} =
-4J^3\sum_j 
\frac{1+\vec{S}_j \cdot \vec{S}_{j+2}}
{(1+\vec{S}_j \cdot \vec{S}_{j+1})(1+\vec{S}_{j+1}\cdot \vec{S}_{j+2})}
-\frac{(\vec{S}_{j+2}\cdot(\vec{S}_{j+1}\times \vec{S}_{j}))^2}
{(1+\vec{S}_j\cdot\vec{S}_{j+1})^2(1+\vec{S}_{j+1}\cdot\vec{S}_{j+2})^2}
+ \frac{2}{1+\vec{S}_j\cdot\vec{S}_{j+1}}
\end{align}

\begin{align}
\dot{\vec{S}}_{j}
&=
\frac{\partial V_{\text{bo},1}}{\partial \vec{S}_j}\times \vec{S}_{j}
=
\{\sj{j}, H_0\}
=
\{\sj{j}, g_{j-2,j+1,j} + g_{j-1,j,j+1} + g_{j,j+1,j+2} \}=
\notag\\
&=
(-4J^3)\Biggl[
\frac{-2\vec{S}_{j-1}}{(1+\vec{S}_{j-1}\cdot\vec{S}_{j})^2}
+
\frac{-2\vec{S}_{j+1}}{(1+\vec{S}_{j}\cdot\vec{S}_{j+1})^2}
+
\frac{-\vec{S}_{j-1}}
{(1+\vec{S}_{j-1}\cdot\vec{S}_{j})^2(1+\vec{S}_{j}\cdot\vec{S}_{j+1})}
\notag\\
&\qquad
+
\frac{-\vec{S}_{j+1}}
{(1+\vec{S}_{j-1}\cdot\vec{S}_{j})(1+\vec{S}_{j}\cdot\vec{S}_{j+1})^2}
+
\frac{-\vec{S}_{j-1}}
{(1+\vec{S}_{j-2}\cdot\vec{S}_{j-1})(1+\vec{S}_{j-1}\cdot\vec{S}_{j})^2}
+
\frac{-\vec{S}_{j+1}}
{(1+\vec{S}_{j}\cdot\vec{S}_{j+1})(1+\vec{S}_{j+1}\cdot\vec{S}_{j+2})^2}
\notag\\
&\qquad
+
\frac{-(\vec{S}_{j-1}\cdot\vec{S}_{j+1})\vec{S}_{j-1}}
{(1+\vec{S}_{j-1}\cdot\vec{S}_{j})^2(1+\vec{S}_{j}\cdot\vec{S}_{j+1})}
+
\frac{-(\vec{S}_{j-1}\cdot\vec{S}_{j+1})\vec{S}_{j+1}}
{(1+\vec{S}_{j-1}\cdot\vec{S}_{j})(1+\vec{S}_{j}\cdot\vec{S}_{j+1})^2}
+
\frac{-(\vec{S}_{j}\cdot\vec{S}_{j+2})\vec{S}_{j-1}}
{(1+\vec{S}_{j-2}\cdot\vec{S}_{j-1})(1+\vec{S}_{j-1}\cdot\vec{S}_{j})^2}+
\notag\\
&\qquad
+
\frac{-\vec{S}_{j+1}}
{(1+\vec{S}_{j}\cdot\vec{S}_{j+1})(1+\vec{S}_{j+1}\cdot\vec{S}_{j+2})^2}
+
\frac{-\vec{S}_{j-1}}
{(1+\vec{S}_{j-1}\cdot\vec{S}_{j})(1+\vec{S}_{j}\cdot\vec{S}_{j+1})}
+
\frac{-\vec{S}_{j+1}}
{(1+\vec{S}_{j}\cdot\vec{S}_{j+1})(1+\vec{S}_{j+1}\cdot\vec{S}_{j+2})}
\Biggr]\times \vec{S}_j+
\notag\\
&\quad
+8J^3 \Biggl[
\frac{\bigl((\vec{S}_{j}\times \vec{S}_{j+1})\cdot \vec{S}_{j+2}\bigr)}
{(1+\vec{S}_{j}\cdot\vec{S}_{j+1})(1+\vec{S}_{j+1}\cdot\vec{S}_{j+2})}
\frac{1}{1+\vec{S}_{j+1}\cdot\vec{S}_{j+2}}
\Biggl(
\frac{-\bigl((\vec{S}_{j}\times \vec{S}_{j+1})\cdot\vec{S}_{j+2}\bigr)\vec{S}_{j+1}}
{(1+\vec{S}_{j}\cdot\vec{S}_{j+1})^2}
+
\frac{\vec{S}_{j+1}\times\vec{S}_{j+2}}
{1+\vec{S}_{j}\cdot\vec{S}_{j+1}}
\Biggr)
\Biggr]\times \vec{S}_{j}+
\notag\\
&\quad
+8J^3 \Biggl[
\frac{\bigl((\vec{S}_{j-2}\times \vec{S}_{j-1})\cdot \vec{S}_{j}\bigr)}
{(1+\vec{S}_{j-2}\cdot\vec{S}_{j-1})(1+\vec{S}_{j-1}\cdot\vec{S}_{j})}
\frac{1}{1+\vec{S}_{j-2}\cdot\vec{S}_{j-1}}
\Biggl(
\frac{-\bigl((\vec{S}_{j-2}\times \vec{S}_{j-1})\cdot\vec{S}_{j}\bigr)\vec{S}_{j-1}}
{(1+\vec{S}_{j-1}\cdot\vec{S}_{j})^2}
+
\frac{\vec{S}_{j-2}\times\vec{S}_{j-1}}
{1+\vec{S}_{j-1}\cdot\vec{S}_{j}}
\Biggr)
\Biggr]\times \vec{S}_{j}+
\notag\\
&\quad
+8J^3 \Biggl[
\frac{\bigl((\vec{S}_{j-1}\times \vec{S}_{j})\cdot \vec{S}_{j+1}\bigr)}
{(1+\vec{S}_{j-1}\cdot\vec{S}_{j})(1+\vec{S}_{j}\cdot\vec{S}_{j+1})}
\Biggl(
\frac{-\bigl((\vec{S}_{j-1}\times \vec{S}_{j})\cdot\vec{S}_{j+1}\bigr)\vec{S}_{j-1}}
{(1+\vec{S}_{j-1}\cdot\vec{S}_{j})^2(1+\vec{S}_{j}\cdot\vec{S}_{j+1})}
\notag\\
&\qquad\qquad\qquad\qquad\qquad
+
\frac{-\bigl((\vec{S}_{j-1}\times \vec{S}_{j})\cdot\vec{S}_{j+1}\bigr)\vec{S}_{j+1}}
{(1+\vec{S}_{j-1}\cdot\vec{S}_{j})(1+\vec{S}_{j}\cdot\vec{S}_{j+1})^2}
+
\frac{\vec{S}_{j+1}\times\vec{S}_{j-1}}
{(1+\vec{S}_{j-1}\cdot\vec{S}_{j})(1+\vec{S}_{j}\cdot\vec{S}_{j+1})}
\Biggr)
\Biggr]\times \vec{S}_{j}.
\end{align}

\section{Numerical techniques for transport studies} 
\label{sec:ishimori_techniques}
To simulate the time evolution of the Ishimori chain, we follow the numerical approach of Ref.~\cite{Roy_2023}.
The system is prepared in a thermal state using a Monte Carlo scheme, after which its time evolution is performed using a fourth-order Runge-Kutta scheme.  
The initial state is prepared using the Monte Carlo scheme proposed in Ref.~\cite{Alzate-Cardona_2019}.
In this scheme, the system evolves towards an equilibrium state for some temperature $\beta$ by sequentially applying an ``adaptive Gaussian move'' to each site, which constitutes one Monte Carlo step.  Here, beginning from a state in which all classical spins are aligned along $\hat z$, an equilibrium state is generated by applying 5000 Monte Carlo steps to the system. After equilibrium is reached, additional initial states are generated by saving the system state every 500 steps.

To time-evolve the system, we employ an adaptive fourth-order Runge-Kutta integration scheme as in Ref.~\cite{Roy_2023}.
In this adaptive scheme, the timestep is adjusted during the run so that the accumulated error at each timestep falls below a set error tolerance, $\delta_0$.
In order to estimate the error $\delta$ associated with updating the system $\{\vec{S}_j(t)\}$ by a timestep $\Delta t$, we generate two copies of the system at time $t + \Delta t$ by performing either one Runge-Kutta step of size $\Delta t$ to obtain $\{ \vec{S_j}(t)\} \rightarrow \{\vec{S}_j'(t+\Delta t)\}$ or two Runge-Kutta steps of size $\Delta t/2$ to obtain $\{ \vec{S_j}(t)\} \rightarrow \{\vec{S}_{j, \mathrm{temp}}''(t+\Delta t/2)\} \rightarrow \{\vec{S}_j''(t+\Delta t)\}$.
The error $\delta$ is defined as the largest component-wise difference between any of the $L$ classical spins in the two copies of the system, or $\delta = {\rm max}\big(|S_{j,\alpha}'-S_{j,\alpha}''|\big)$ for $i=1 \dots L$ and $\alpha=x,y,z$.
A timestep is accepted if the error associated with the timestep $\delta$ is less than the error tolerance $\delta_0$; if this condition is not met, the timestep is reduced, and the process is repeated. After a timestep is accepted, the system is updated, and the norm of each classical spin is manually enforced to be exactly $|\vec{S}_j| = 1$.

In the present work, the system was initialized at a temperature $\beta=5$ with $L=4000$ classical spins.  The error tolerance for the adaptive Runge-Kutta scheme was set to $\delta_0 = 10^{-4}$, which was numerically verified to yield converged results for all simulated times.  All simulations were run up to a final time of $t_f = 920$ steps and averaged over at least $5 \times 10^5$ independent realizations.

To extract the crossover, we simulate the Ishimori spin chain under the effect of an integrability-breaking perturbation as parameterized in Eq.~(\ref{eq:ham}).  
To characterize energy transport, we consider the energy transfer across a central link in the chain, $ \Delta E(t) = \sum_{n<L/2} [h_{j,j+1}(t) - h_{j,j+1}(0)] - \sum_{j\geq L/2} [h_{j,j+1}(t) - h_{j,j+1}]$.
At long times, fluctuations in energy transfer across the link are expected to scale as $\langle \Delta E^2 \rangle \sim t^{1/z}$.  The dynamical exponent $z$ may be extracted from the energy transfer data through taking a logarithmic derivative $z(t) = \Big(\frac{\mathrm{d} \ln(\Delta E^2(t))}{\mathrm{d} (\ln t)}\Big)^{-1}$.

\section{Reshetikhin criterion, boost operator, and construction of higher order IoMs for the Ishimori chain} 
\label{app:IoMs_ishimori}
Here we provide some details supporting our claims in Section \ref{sec:ishimori_ioms} and show that the Ishimori model admits a boost operator. 
We then use the boost to derive $Q_3^{(0)}$ and $Q_4^{(0)}$ charges. 
We do not reproduce all the details and instead refer the interested reader to the App. C. of \cite{Surace2023}.

Consider a nearest-neighbor Hamiltonian with local energy density given by $H=\sum_j h_{j,j+1}$. The boost of the Hamiltonian local density $H= \sum_j h_{j,j+1}$ with the Hamiltonian $H$ gives an extensive local range 3 operator 
\begin{align}
    \bigg\{\sum_j jh_{j,j+1},H \bigg\} = - \sum_j g_{j,j+1,j+2}
\end{align}
Direct computation shows that local terms are given by
\begin{align}
&g_{j,j+1,j+2} = \{h_{j,j+1},h_{j+1,j+2} \}
= -4 \frac{(\sj{j}\times\sj{j+1})\cdot \sj{j+2}}{(1+\sj{j}\sj{j+1})(1+\sj{j+1}\sj{j+2})} \equiv Q_3^{(0)},
\end{align}
which is exactly the first non-trivial IoM of the Ishimori mode---\textit{torsion}---as expected.
The Ishimori model is then one of the integrable models whose local energy current equals its first non-trivial conserved charge. 

Next, we construct two higher-order IoMs. In order to use the simplifying expression from \cite{Surace2023} to construct $Q_4^{(0)}$, we first need to check if the Reshetikhin criterion is satisfied.
We need to check if a two-site operator $R_{j,j+1}$ exist such that
\begin{align}
    &\{ h_{j,j+1} + h_{j+1,j+2}, g_{j,j+1,j+2} \}  = R_{j,j+1} - R_{j+1, j+2}
\end{align}
where \(h_{j,j+1}\) and \(g_{j,j+1,j+2}\) denote, respectively, the local Ishimori
Hamiltonian density, the local torsion density. (Note that only recently was it proven that if a model satisfies the Reshetikhin criterion, it is guaranteed to be integrable \cite{Shiraishi_2025, Hokkyo_2026, Zhang_2026}, and a generalized Reshetikhin criterion was formulated. The Reshetikhin criterion has been used in AI-driven searches for new integrable models. \cite{Lal_2025,Lal_2023})
\begin{align}
    &\{ h_{j,j+1} + h_{j+1,j+2}, g_{j,j+1,j+2} \} =
    8 \bigg[\frac{1}{1+\sj{j}\sj{j+1}} - \frac{1}{1+\sj{j+1}\sj{j+2}}  \bigg] = R_{j,j+1} - R_{j+1, j+2}
\end{align}
This helps greatly in constructing $Q_4^{(0)}$, which we find by boosting $Q_3^{(0)}$ and re-using expressions from \cite{Surace2023}
\begin{align}
 Q_4^{(0)} &= \Big\{ \sum_j j h_{j,j+1}, \sum_{j'} g_{j',j'+1,j'+2} \Big\} =
 \sum_j \Big( -2 \{h_{j,j+1}, g_{j+1,j+2,j+3}\} + \{h_{j+1,j+2}, g_{j,j+1,j+2}\} + R_{j,j+1} \Big)
 \end{align}
All we need to compute now is $f_{j,j+1,j+2,j+3}=\{h_{j,j+1}, g_{j+1,j+2,j+3}\}$ which will give us the range 4 terms. We find 
\begin{align}
  f_{j\dots j+3}= 8\bigg [ \frac{((\vec{S}_{j}\times \vec{S}_{j+1})\times\vec{S}_{j+2})\cdot\vec{S}_{j+3}}{(1+\vec{S}_{j}\vec{S}_{j+1})(1+\vec{S}_{j+1}\vec{S}_{j+2})(1+\vec{S}_{j+2}\vec{S}_{j+3})} - \frac{(\sj{j}\cdot(\sj{j+1}\times \sj{j+2}))(\sj{j+1}\cdot(\sj{j+2}\times \sj{j+3}))}{(1+\sj{j}\sj{j+1})(1+\sj{j+1}\sj{j+2})^2(1+\sj{j+2}\sj{j+3})} \bigg].
\end{align}
Combining the individual terms and taking the sum gives exactly the expression for $Q_4^{(0)}$ listed in \ref{eq:ISH_Q4}.

\section{Equations of motion for ABA864 algorithm for the Toda and HOC chain}
\label{app:toda-ioms}
We implement the symplectic integrator scheme described in \cite{Danieli2024}, and our equations of motion for the A and B steps are identical, except that we use periodic boundary conditions. 
The Hamiltonian of the Toda chain with periodic boundary conditions in terms of coordinates $q_n$ and canonically conjugate momenta $p_n$ reads
\begin{align*}
\label{eq:toda_ham}
H_\mathrm{T} = &\Bigg[\sum_{n=1}^{N} \frac{p_n^2}{2}\Bigg] + \Bigg[ \sum_{n=1}^{N} \frac{e^{2\alpha(q_{n+1}-q_n)} - 2\alpha(q_{n+1}-q_n) -1 }{4\alpha^2} \Bigg] = A+B
\end{align*}
For periodic boundary condition ($q_{1} = q_{N+1},  p_{1} = p_{N+1}$), Hamilton's  equations of motion and the variational equations for the deviations  $\{ \delta q_n,\delta p_n\}$ become
\begin{equation}
\label{eq:toda_eq}
\begin{cases}
\dot{q}_n = p_n, \quad \text{for } n = 1, \ldots, N, \\
\dot{p}_n = \frac{1}{2\alpha}\left[ e^{2\alpha(q_{n+1}-q_n)} - e^{2\alpha(q_{n}-q_{n-1})}  \right], \text{for } n = 1, \ldots, N, \\
\dot{\delta q}_n = \delta p_n, \quad \text{for } n = 1, \ldots, N, \\
\dot{\delta p}_n = - [ e^{2\alpha(q_{n+1}-q_n)} + e^{2\alpha(q_{n}-q_{n-1})}  ]\delta  q_n   + [ e^{2\alpha(q_{n+1}-q_n)} ]\delta  q_{n+1} + [ e^{2\alpha(q_{n}-q_{n-1})}]\delta  q_{n-1},  
\text{ for } n = 1, \ldots, (N), 
\end{cases}
\end{equation}
ABA864 for the HOC chain is easily obtained as first order expansion of the Toda potential.

\subsection{Numerical check of AGP norm saturation in the HOC model}
\label{subapp:HOC_longerT}

\begin{figure}[h]
    \centering    \includegraphics[width=1.02\linewidth]{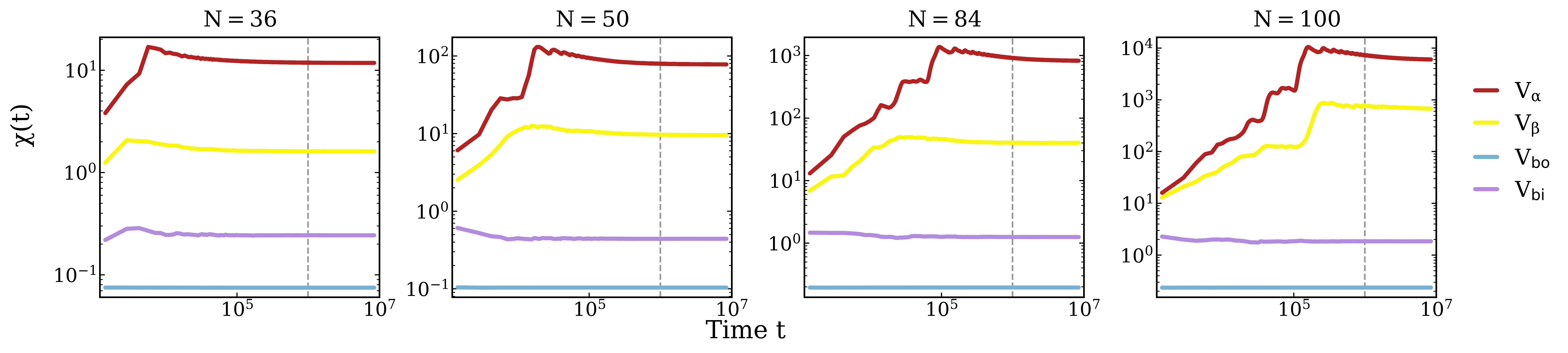}
    \caption{\textit{HOC chain}. Averaged AGP variance $\chi(t)$ for several perturbations and system sizes, evolved to times $T=8.5\times10^6$. 
    The dashed vertical line marks the time $T=1.5\times10^6$ used in the main text to extract the saturation value. 
    The curves remain on the same plateau for more than an order of magnitude longer time.}
\label{fig:ab_longT}
\end{figure}

In the main text, we extract the saturation value of the fidelity susceptibility from simulations performed up to time $T = 1.5 \times 10^6$. 
Figure~\ref{fig:ab_longT}, which is the counterpart of the HOC analysis shown in the left panels of Fig.~\ref{fig:agp-hoc-a-global}, verifies that this simulation window is sufficient. 
We extend the time evolution to $T = 8.5 \times 10^6$, and observe that the plateau reached around $T \sim 10^6$ remains essentially unchanged. 
This demonstrates that the saturation values reported in the main text correspond to the genuine long-time behavior of the system rather than a finite-time transient. 
In particular, despite the presence of multiple numerical time scales --- such as the time beyond which the finite integration step $\delta t$ may effectively act as an integrability-breaking perturbation --- the plateau of the AGP variance $\chi(t)$ remains stable over the extended simulation window.

\subsection{Energy density and Initial parameters}
\label{app:toda_eps}
\begin{figure}[h]
    \centering
\includegraphics[width=1.02\linewidth]{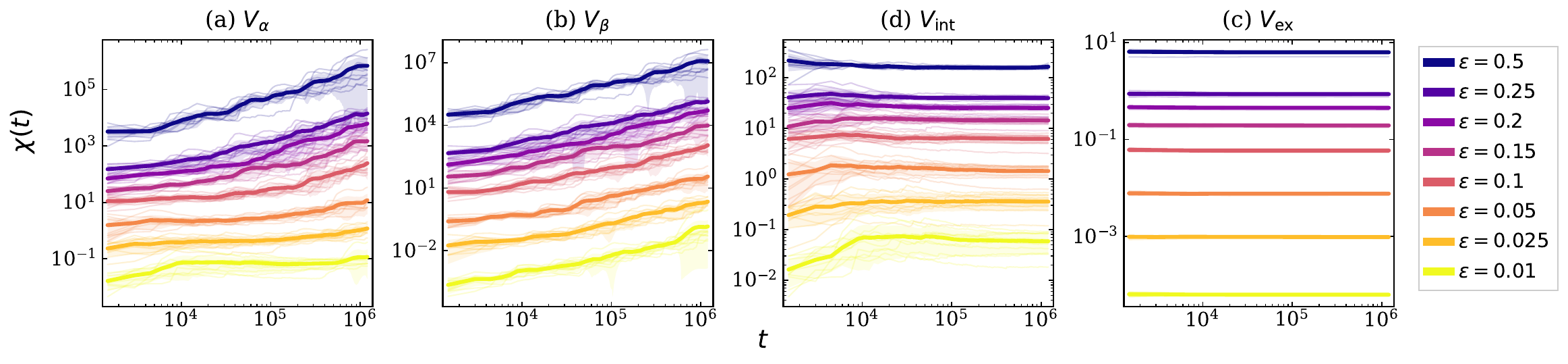}
\caption{Log-log plot of the AGP variance $\chi(t)$ for the Toda chain under $V_{\alpha}$ and $V_{\beta}$ perturbations with $N=100$, and $dt=0.1$. 
For each perturbation strength $\epsilon$, thin translucent lines show all trajectories, while thick lines show their averages. Shows the impact of energy density on the growth of AGP.}
\label{fig:agp_with_eps}
\end{figure}

In numerical studies of FPUT-type models, it is common to discuss \textit{stochasticity thresholds}, in which the apparent degree of chaos depends sensitively on the choice of energy density.
At sufficiently low energy density $\epsilon=E/N$ the $\alpha-$FPUT model lies
very close to the integrable Toda lattice, resulting in dynamics that appear nearly integrable over accessible simulation times. In fact, some earlier works used such arguments to explain the original observations of anomalous thermalization in the $\alpha-$FPUT model (see \cite{Dhar_2020} for review). 
We point this out since this proximity manifests itself in measures of chaos such as maximal Lyapunov exponent and our AGP norm (specifically, in the slow growth of the AGP variance as shown in Fig.~\ref{fig:agp_with_eps}). 
In contrast, while the $\beta$ model also exhibits reduced variance of the AGP at low $\epsilon$, its AGP norm continues to grow, reflecting the presence of genuine chaos.
Note that we are talking about adding $\alpha$ and $\beta$ perturbations on top of the Toda model, and we are discussing the idea that there are some regimes where Toda is particularly close to the $\alpha$-FPUT Hamiltonian. 
Recent work by Karve \textit{et al.} \cite{Karve_2025} has characterized regimes in which the AGP norm grows anomalously slowly as examples of ''weak chaos'', controlled by long Lyapunov times rather than by exact or approximate conservation laws. 
Importantly, this behavior is distinct from the weak integrability-breaking perturbations studied in this work: for WIBs of the form $V={X,H_0}$, the AGP variance saturates at late times no matter what $\epsilon$ we are using, whereas in weakly chaotic regimes it continues to grow if the simulation is extended beyond the Lyapunov time if the perturbation is not a WIB. 
This is why in our work we set $\epsilon=0.1$ for all of our simulations of the Toda and HOC.

\section{Harmonic Oscillator Chain (HOC): Systematic construction of short-range $V_{\text{ex}}$ perturbations }
\label{app:HOC_Vex}
In this appendix, we systematically explore lattice-translation-invariant generators $X_{\text{ex}}$ that are third- and fourth-order polynomials in the phase space variables and that are of range at most 2 (i.e., terms extending to at most nearest neighbors).
Since $H_0$ is bilinear and has only up to the nearest-neighbor terms, the corresponding $V_{\text{ex}}$ are third- and fourth-order polynomials respectively and have a range of at most 3.
We further restrict to perturbations that preserve the total momentum $P_{\text{tot}} = \sum_j p_j = Q_1^{(0)}$.

\subsection{Perturbations that are cubic in the phase space variables}
Table~\ref{tab:SHOCcubicVex} lists such $V_{\text{ex}}$ perturbations that are third-order polynomials in $q$ and $p$.
In most of the entries, the generator $X_{\text{ex}}$ is manifestly momentum-conserving, which guarantees that $V_{\text{ex}}$ is also momentum-conserving.
The last two entries have momentum-non-conserving $X_{\text{ex}}$ that nevertheless produce momentum-conserving $V_{\text{ex}}$; we explain these special occurrences later.
In each case, we write out $V_{\text{ex}} = \sum_j v_{\text{ex}, j}$ in the basis of strings of $q$'s and $p$'s, where $v_{\text{ex}, j}$ contains all such strings that start at $j$ (throughout, $\sum_j$ is a short-hand for $\sum_{j=1}^L$ on a PBC chain with sufficiently large $L$).
While this may obscure the momentum-conserving property of $V_{\text{ex}}$, it allows us to clearly see independent terms present in each case and helps find combinations that can produce simpler $V_{\text{ex}}$.
Furthermore, for easy reference, in each case we also give the time reversal $\Theta_V$ (i.e., parity of the number of $p$'s in $V$) and lattice inversion $I_V$ symmetry numbers; it makes most sense to only consider combinations of $V$'s with the same symmetry numbers.

\onecolumngrid
\begin{table}[h]
\centering
\label{tab:SHOCcubicVex}
\setlength{\tabcolsep}{3pt} 
\begin{tabularx}{\textwidth}{|l|X|c|c|}
\hline
$x_{\text{ex},j}$ 
& $v_{\text{ex},j}$ 
& $\Theta_V$ 
& $I_V$ \\
\hline\hline
$(q_j - q_{j+1})^3$ 
& $3 (p_j q_{j+1}^2 - q_j^2 p_{j+1}) + 6 q_j q_{j+1} (p_{j+1} - p_j)$ 
& $-$ & $-$ \\
$(q_j - q_{j+1})^2 (p_j + p_{j+1})$ 
& $4 q_j p_j^2 - 2 (q_j p_{j+1}^2 + p_j^2 q_{j+1}) - 2 q_j^3 
   + 2 q_j q_{j+1} (q_j + q_{j+1}) + q_j q_{j+2} (q_j + q_{j+2}) 
   - 4 q_j q_{j+1} q_{j+2}$ 
& $+$ & $+$ \\
$(q_j - q_{j+1})^2 (p_j - p_{j+1})$ 
& $4 p_j p_{j+1} (q_{j+1} - q_j) + 2 (q_j p_{j+1}^2 - p_j^2 q_{j+1}) 
   + 8 q_j q_{j+1} (q_j - q_{j+1}) + q_j q_{j+2} (q_{j+2} - q_j)$ 
& $+$ & $-$ \\
$(q_j - q_{j+1}) (p_j^2 + p_{j+1}^2)$ 
& $2 (q_j^2 p_{j+1} - p_j q_{j+1}^2) + 4 q_j q_{j+1} (p_j - p_{j+1}) 
   + p_j p_{j+1} (p_{j+1} - p_j)$ 
& $-$ & $-$ \\
$(q_j - q_{j+1}) p_j p_{j+1}$ 
& $p_j p_{j+1} (p_j - p_{j+1}) - 2 (q_j^2 p_{j+1} - p_j q_{j+1}^2) 
   + 3 q_j q_{j+1} (p_{j+1} - p_j) + q_j q_{j+1} p_{j+2} 
   - p_j q_{j+1} q_{j+2} + q_j q_{j+2} (p_j - p_{j+2})$ 
& $-$ & $-$ \\
$(q_j - q_{j+1}) (p_j^2 - p_{j+1}^2)$ 
& $2 p_j^3 - p_j p_{j+1} (p_j + p_{j+1}) - 8 q_j^2 p_j 
   - 2 (q_j^2 p_{j+1} + p_j q_{j+1}^2) + 8 q_j q_{j+1} (p_j + p_{j+1}) 
   - 4 q_j p_{j+1} q_{j+2}$ 
& $-$ & $+$ \\
$p_j^3$ 
& $-6 q_j p_j^2 + 3 (q_j p_{j+1}^2 + p_j^2 q_{j+1})$ 
& $+$ & $+$ \\
$p_j p_{j+1} (p_j + p_{j+1})$ 
& $2 q_j p_j^2 - 2 (q_j p_{j+1}^2 + p_j^2 q_{j+1}) 
   - 2 p_j p_{j+1} (q_j + q_{j+1}) + q_j p_{j+2}^2 + p_j^2 q_{j+2} 
   + 2 (q_j p_{j+1} p_{j+2} + p_j p_{j+1} q_{j+2})$ 
& $+$ & $+$ \\

$p_j p_{j+1} (p_j - p_{j+1})$ 
& $6 p_j p_{j+1} (q_{j+1} - q_j) 
   + 2 (q_j p_{j+1}^2 - p_j^2 q_{j+1}) 
   - q_j p_{j+2}^2 + p_j^2 q_{j+2} 
   + 2 (q_j p_{j+1} p_{j+2} - p_j p_{j+1} q_{j+2})$ 
& $+$ & $-$ \\
\hline
$q_j^2 p_{j+1} - p_j q_{j+1}^2$ 
& $2 p_j p_{j+1} (q_j - q_{j+1}) + 2 q_j q_{j+1} (q_{j+1} - q_j) 
   + q_j q_{j+2} (q_j - q_{j+2})$ 
& $+$ & $-$ \\ 
\hline
$\begin{aligned}
& (q_j + q_{j+1})(q_j - q_{j+1})^2 \\
& + q_j p_{j+1}^2 + p_j^2 q_{j+1}
\end{aligned}$ 
& $p_j p_{j+1} (p_j + p_{j+1}) + 6 q_j^2 p_j + q_j^2 p_{j+1} 
   + p_j q_{j+1}^2 - 6 q_j q_{j+1} (p_j + p_{j+1}) 
   + 4 q_j p_{j+1} q_{j+2}$ 
& $-$ & $+$ \\
\hline
\end{tabularx}
\caption{Weak integrability breaking perturbations $V_{\text{ex}} = \sum_j v_{\text{ex},j}$ obtained using lattice-translation-invariant extensive local generators $X_{\text{ex}} = \sum_j x_{\text{ex},j}$ that are third-order polynomials in the phase space variables, restricting to generators with range 2 (i.e., only up to nearest-neighbor couplings).
The main block of entries uses generators that conserve $P_{\text{tot}}$, while this is not true for the last two entries; nevertheless, $V_{\text{ex}}$ conserves $P_{\text{tot}}$ in all cases.
For each perturbation, we also indicate its time reversal $\Theta_V$ and lattice inversion $I_V$ symmetry numbers.
}
\label{tab:cubicgenVex4HOC}
\end{table}
We now discuss some interesting combinations.
Thus, we can combine the first two entries with $(\Theta_V, I_V) = (-, -)$ symmetry numbers to obtain the following particularly simple-looking perturbation:
\begin{align}
X_{\text{ex}} = \sum_j \left[ \frac{2}{3}(q_j - q_{j+1})^3 + (q_j - q_{j+1}) (p_j^2 + p_{j+1}^2) \right] ~\implies~
V_{\text{ex}}^{(3),(-,-)} = \sum_j p_j p_{j+1} (p_{j+1} - p_j) = \frac{1}{3} \sum_j (p_j - p_{j+1})^3 ~.
\label{eqapp:HOC_Vex3_m_m}
\end{align}
Here and below, for later referencing, we use superscript labels on the selected $V$, which stand for: ``$(3)$''~=~cubic perturbation; ``$(-,-)$''~=~specific $(\Theta_V, I_V)$.
In fact, this is the simplest perturbation with such symmetry numbers that contains three $p$'s and couples only nearest neighbors.
Thus, the simplest perturbation in this symmetry class is a WIB perturbation of $V_{\text{ex}}$ type.

Next, we can combine the first two entrees with $(\Theta_V, I_V) = (+, +)$ symmetry numbers to obtain the following WIB perturbation:
\begin{equation}
\begin{aligned}
&X_{\text{ex}} = \sum_j (q_j - q_{j+1})^2 (p_j + p_{j+1}) + \frac{2}{3} p_j^3 \\
&V_{\text{ex}}^{(3),(+,+)} = \sum_j \left[-2 q_j^3 + 2 q_j q_{j+1} (q_j + q_{j+1}) + q_j q_{j+2} (q_j + q_{j+2}) - 4 q_j q_{j+1} q_{j+2} \right] \\
& = \frac{1}{3} \sum_j (q_j + q_{j+1} - 2 q_{j+2}) (q_{j+1} + q_{j+2} - 2 q_j) (q_{j+2} + q_j - 2 q_{j+1})  = \sum_j (q_j - q_{j+1})^2 (q_{j-1} - q_j - q_{j+1} + q_{j+2}) ~.
\label{eqapp:HOC_Vex3_p_p}
\end{aligned}
\end{equation}
It is easy to show that this is the only range-3 interaction (i.e., with up to second-nearest-neighbor couplings) with three $q$'s that is invariant under the inversion symmetry and also conserves $P_{\text{tot}}$.
Thus, the simplest cubic in $q$ and momentum-conserving perturbation with such symmetry numbers is given by the above equation, and hence is a $V_{\text{ex}}$-type WIB perturbation.

We now turn to the special second-to-last entry that has $(\Theta_V, I_V) = (+,-)$.
Its generator $X_{\text{ex}}$ does not commute with the momentum $P_{\text{tot}}$, but $V_{\text{ex}}$ does conserve $P_{\text{tot}}$, as becomes manifest by writing $\sum_j q_j q_{j+m} (q_j - q_{j+m}) = \sum_j \frac{1}{3} (q_{j+m} - q_j)^3$ for any $m$:
\begin{equation}
\begin{aligned}
&X_{\text{ex}} = \sum_j \left(q_j^2 p_{j+1} - p_j q_{j+1}^2 \right) ~, \\
&V_{\text{ex}}^{(3),(+,-)} = \sum_j \big[2 p_j p_{j+1} (q_j - q_{j+1}) + \frac{2}{3} (q_j - q_{j+1})^3 + \frac{1}{3}(q_{j+2} - q_j)^3 \big] ~. \label{eqapp:HOC_Vex3_p_m}
\end{aligned}
\end{equation}
Upon further inspection, the underlying reason for $V_{\text{ex}}$ to be momentum-conserving is because $\{ X_{\text{ex}}, P_{\text{tot}} \} = 2 \sum_j (q_j p_{j+1} - p_j q_{j+1}) = 4 Q_3^{(0)}$ happens to be proportional to an IoM.
As we argued in Sec.~\ref{sec:choc}, by Jacobi identity in \ref{eq:jacobi},  $V_{\text{ex}}$ is momentum conserving as long as $\{X_{\text{ex}}, P_\text{tot}\}$ is an IoM or zero.

In the manifold of $V_{\text{ex}}$ perturbations with the symmetry numbers $(\Theta_V, I_V) = (+,-)$, we can reduce the range to strictly nearest-neighbor couplings by considering the following combination:
\begin{equation}
\begin{aligned}
&X_{\text{ex}} \!=\!  \sum_j \left[ (q_j - q_{j+1})^2 (p_j - p_{j+1}) +
q_j^2 p_{j+1} - p_j q_{j+1}^2 \right] ~, \\
&\widetilde{V}_{\text{ex}}^{(3),(+,-)} \!=\! \sum_j \! \big[2 p_j p_{j+1} (q_{j+1} - q_j) + 2 (q_j p_{j+1}^2 - p_j^2 q_{j+1}) + 2 (q_{j+1} - q_j)^3 \big] ~.
\end{aligned}
\end{equation}
However, we cannot eliminate $q p p$-type terms in this manifold of $V_{\text{ex}}$ perturbations; in the main text, we manage to eliminate such terms by combining with a $V_{\text{bi}}$-type WIB perturbation.

Finally, the special last entry that has $(\Theta_V, I_V) = (-,+)$ also has generator $X_{\text{ex}}$ that does not commute with $P_{\text{tot}}$, while $V_{\text{ex}}$ does conserve $P_{\text{tot}}$.
Similarly to the special second-to-last entry considered earlier, the underlying reason is that $\{ X_{\text{ex}}, P_{\text{tot}} \} = 4Q_2^{(0)}$ here also happens to be proportional to an IoM.
There is one more entry in the table with the same symmetry numbers, and the following combination eliminates some $q q p$-type terms:
\begin{equation}
\begin{aligned}
&X_{\text{ex}} = \sum_j
(q_j + q_{j+1}) \left[(q_j - q_{j+1})^2 + \frac{1}{2} (p_j^2 + p_{j+1}^2) \right] ~, \\
&V_{\text{ex}}^{(3),(-,+)} = \sum_j \bigg[p_j^3 + \frac{1}{2} p_j p_{j+1} (p_j + p_{j+1}) + 2 q_j^2 p_j - 2 q_j q_{j+1} (p_j + p_{j+1}) + 2 q_j p_{j+1} q_{j+2} \bigg] ~.
\label{eqapp:HOC_Vex3_m_p}
\end{aligned}
\end{equation}
In the main text, we manage to eliminate all the remaining $q q p$-type terms by combining with a $V_{\text{bi}}$ WIB perturbation.

\subsection{Perturbations that are quartic in the phase space variables}
\begin{table}[h]
\centering
\setlength{\tabcolsep}{2.5pt} 
\begin{tabularx}{1.03\textwidth}{|l|X|c|c|}
\hline
$x_{\text{ex},j}$ & $v_{\text{ex},j}$ & $\Theta_V$ & $I_V$ \\
\hline\hline
$r_j^4$ & $4 p_j \left[ (q_j - q_{j-1})^3 + (q_j - q_{j+1})^3 \right]$& $-$ & $+$ \\
$r_j^3 (p_j + p_{j+1})$ & $3 p_j^2 \left[ -(q_{j-1} - q_j)^2 + (q_j - q_{j+1})^2 \right] + (q_j - q_{j+1})^3 (q_{j-1} -q_j - q_{j+1} + q_{j+2})$ & $+$ & $-$ \\
$r_j^3 (p_j - p_{j+1})$ & $3 p_j^2 \left[ (q_{j-1} - q_j)^2 + (q_j - q_{j+1})^2 \right]  - 6 p_j p_{j+1} (q_j - q_{j+1})^2  + (q_j - q_{j+1})^3 (q_{j-1} - 3 q_j + 3 q_{j+1} - q_{j+2})$ & $+$ & $+$ \\
$r_j^2 (p_j^2 + p_{j+1}^2)$ & $-2\,(q_j-q_{j+1})^{2}\, \left[p_j(2q_j-q_{j-1}-q_{j+1})+p_{j+1}(2q_{j+1}-q_j-q_{j+2})\right] +\,2\,(q_j-q_{j+1})\,(p_j-p_{j+1})\,(p_j^2+p_{j+1}^2)$ & $-$ & $+$ \\
$r_j^2 p_j p_{j+1}$ 
&$-\,2\,(q_j-q_{j+1})\,p_j p_{j+1}(p_j-p_{j+1})\;+\;2\,(q_j-q_{j+1})^{2}\,\left(p_j(2q_j-q_{j-1}-q_{j+1})+p_{j+1}(2q_{j+1}-q_j-q_{j+2})\right )$ & $-$ & $+$ \\
$r_j^2(p_j^2\;-\;p_{j+1}^2)$ 
& $\,2\,(q_j-q_{j+1})\,(p_j-p_{j+1})\,(p_j^2+p_{j+1}^2)\;-\;2\,(q_j-q_{j+1})^{2}\,\big(p_j(2q_j-q_{j-1}-q_{j+1})+p_{j+1}(2q_{j+1}-q_j-q_{j+2})\big)$
& $-$ & $-$ \\
$\,r_j\,(p_j^3 + p_{j+1}^3)$ 
& $p_j p_{j+1} (p_{j+1}^2 - p_j^2) + 3 p_j^2 \left[ (q_{j-1} - q_j)^2 - (q_j - q_{j+1})^2 \right]$ 
& $+$ & $-$ \\
$\,r_j\,(p_j^2p_{j+1}+p_j p_{j+1}^2)$
& $p_j p_{j+1}(p_j^2-p_{j+1}^2)
+ p_j^2 (q_j-q_{j+1})(q_j-2q_{j+1}+q_{j+2})
+ p_{j+1}^2 (q_j-q_{j+1})(q_{j-1}-2q_j+q_{j+1})
+ 2 p_j p_{j+1}(q_j-q_{j+1})(q_{j-1}-q_j-q_{j+1}+q_{j+2})$
& $+$ & $-$ \\
$r_j (p_j^3 - p_{j+1}^3)$ 
& $2 p_j^4 - p_j p_{j+1} (p_j^2 + p_{j+1}^2) 
   - 3 p_j^2 \left[2 (q_{j-1} - q_j)^2 + 2 (q_j - q_{j+1})^2 
   - (q_{j-1} - q_{j+1})^2 \right]$
& $+$ & $+$ \\
$r_j(p_j^2 p_{j+1} - p_j p_{j+1}^2)$ 
& $ p_j p_{j+1} (p_j - p_{j+1})^2  - p_j p_{j+1} \big[ (q_{j-1} - q_j)^2 + (q_{j+1} - q_{j+2})^2 + 6(q_j - q_{j+1})^2 - (q_{j-1} - q_{j+1})^2 - (q_j - q_{j+2})^2 \big]  + \frac{1}{2} p_j^2 \big[(q_{j-2} - q_{j-1})^2 + (q_{j+1} - q_{j+2})^2 + 3 (q_{j-1} - q_j)^2 +  3 (q_j - q_{j+1})^2 - (q_{j-2} - q_j)^2 - (q_j - q_{j+2})^2 \big] $& $+$ & $+$ \\
$p_j^4$ & $4 p_j^3 (q_{j-1} - 2 q_j + q_{j+1})$ & $-$ & $+$ \\
$p_j p_{j+1} (p_j^2 + p_{j+1}^2)$ 
& $p_j^3 (q_{j-2} - 2 q_{j-1} + 2 q_j - 2 q_{j+1} + q_{j+2}) 
   + 3 p_j^2 p_{j+1} (q_{j-1} - 2 q_j + q_{j+1}) 
   + 3 p_j p_{j+1}^2 (q_j - 2 q_{j+1} + q_{j+2})$ 
& $-$ & $+$ \\
$p_j^2 p_{j+1}^2$ 
& $2 p_j p_{j+1}^2 (q_{j-1} - 2 q_j + q_{j+1}) 
   + 2 p_j^2 p_{j+1} (q_j - 2 q_{j+1} + q_{j+2})$ 
& $-$ & $+$ \\
$p_j p_{j+1} (p_j^2 - p_{j+1}^2)$ & $3p_j^2 (p_{j+1} - p_{j-1}) (q_{j-1} - 2 q_j + q_{j+1}) + p_j^3 (-q_{j-2} + 2 q_{j-1} - 2 q_{j+1} + q_{j+2})$
& $-$ & $-$ \\
\hline
\end{tabularx}
\caption{WIBs for HOC obtained using lattice-translation-invariant extensive local generators $X_{\text{ex}} = \sum_j x_{\text{ex},j}$ that are fourth-order polynomials in the phase space variables, restricting to generators with range 2 (i.e., only up to nearest-neighbor terms).
When writing generators, we use a shorthand $r_j = q_j - q_{j+1}$. 
We only use generators that conserve $P_{\text{tot}}$, which produces momentum-conserving $V_{\text{ex}}$. 
For each perturbation, we also indicate its time-reversal $\Theta_V$ and lattice inversion $I_V$ symmetry numbers.
}
\label{tab:SHOCquarticVex}
\end{table}
Table~\ref{tab:SHOCquarticVex} lists momentum-conserving $V_{\text{ex}}$ perturbations that are fourth-order polynomials in $q$ and $p$.
To generate these, we can only use momentum-conserving $X_{\text{ex}}$. (The mechanism that allowed the special momentum-non-conserving cubic $X_{\text{ex}}$ producing momentum-conserving $V_{\text{ex}}$ does not operate for quartic generators, since there are no cubic extensive local IoMs in the HOC.)
All other aspects of the organization of the generated quartic $V_{\text{ex}}$ perturbations are the same as for the cubic perturbations in the previous subsection.

\section{From Toda IoMs to HOC IoM corrections in the $\alpha$-FPUT problem}
\label{app:toda_ioms_to_corr}
Here we show how to use IoMs of the Toda model to construct corrections to IoMs of the HOC perturbed by the $\alpha$-FPUT cubic potential.
To make the connection to the standard form of the $\alpha$-FPUT problem most transparent, we use the single-parameter ($\alpha$) form of the Toda potential, with $V_0(r):= e^{2\alpha r}/(4\alpha^2)$.
Equivalently, in the language of parameters $a$ and $b$ in the main text, we set $b = -2\alpha, ab = 1$. 
As discussed in the main text, Taylor-expanding the Toda potential to third order yields the $\alpha$-FPUT Hamiltonian.
Formally, consider
\begin{align}
H^{\text{Toda}}(\alpha) &= 
\sum_j \left[\frac{p_j^2}{2} + \frac{e^{2\alpha (q_{j+1} - q_j)}}{4\alpha^2} \right]
= \text{const} +
\sum_j \left[\frac{p_j^2}{2} + \frac{1}{2}(q_{j+1} - q_j)^2 + \frac{\alpha}{3} (q_{j+1} - q_j)^3 + O(\alpha^2) \right] \\
&= \text{const} + H_0 + \alpha V_{\text{FPUT}} + O(\alpha^2) ~.
\end{align}
Here and below, the superscript ``Toda'' refers to the Hamiltonian and IoMs of the integrable Toda chain.
We consider these as exact, but Taylor expanded to all orders in powers of $\alpha$, while we only show a few leading terms and express these in terms of the unperturbed HOC Hamiltonian $H_0 \equiv Q_2^{(0)}$, and its IoMs $Q_n^{(0)}$.
We will look for appropriate corrections $Q_n^{(1)}$ to $Q_n^{(0)}$ under the FPUT perturbation $V_{\text{FPUT}}$, as will become more clear below.
In this section,  $Q_n^{(0)}$ and $Q_n^{(1)}$ refer solely to the HOC system under the $\alpha$-FPUT perturbation in the setup of Sec.~\ref{sec:HOC_VFPUT}.
Since we only consider the $\alpha$-FPUT perturbation, we will drop ``$\alpha$'' from the name for the rest of the appendix, and note that $V_{\text{FPUT}}$ refers to Eq.~(\ref{eq:valpha}).
We have
\begin{align}
& 0 = \{H^{\text{Toda}}(\alpha), H^{\text{Toda}}(\alpha)\} = \alpha \left(\{H_0, V_{\text{FPUT}}\} + \{V_{\text{FPUT}}, H_0\}\right) + O(\alpha^2) ~, \\ &\implies~~~ \{Q_2^{(1)}, H_0\} = \{V_{\text{FPUT}}, Q_2^{(0)}\} ~, \qquad Q_2^{(1)} = V_{\text{FPUT}} ~.
\label{eq:Q21def}
\end{align}
Note that we did not need to use commutation of the Toda Hamiltonian with itself, and the final equation is trivial.
However, this manipulation showcases the general idea using exact commutations of IoMs in the Toda model to derive relations useful for the perturbed HOC study.

We already showed in the main text that we can take Toda's $Q_3^{\text{Toda}}$, Taylor expand it to third power in $\alpha r$ and get the correction to the HOC $Q_3^{(0)}$ in the HOC + $V_{\text{FPUT}}$ problem. 
Here, we provide the details we omitted and the exact recipe for computing corrections to higher IoMs.
We consider
\begin{align}
    Q_3^{\text{Toda}}(\alpha) &= \sum_j \left[\frac{p_j^3}{3} + (p_{j+1} + p_j) \frac{e^{2\alpha (q_{j+1} - q_j)}}{4\alpha^2} \right] \\
    &= \sum_j \left[\frac{1}{4\alpha^2}(p_{j+1} + p_j) + \frac{1}{2\alpha}(p_{j+1} + p_j)(q_{j+1} - q_j) + \frac{1}{2}(p_{j+1} + p_j)(q_{j+1} - q_j)^2 + \frac{p_j^3}{3} + O(\alpha) \right] \\
    &  = \frac{1}{2\alpha^2} Q_1^{(0)} + \frac{1}{\alpha} G_3^{(0)} + G_3^{(1)} + O(\alpha) ~, \quad G_3^{(0)} := -Q_3^{(0)}~, G_3^{(1)} := -Q_3^{(1)} ~,
\end{align}
where $Q_1^{(0)} = P_{\text{tot}}$ is the total momentum given by Eq.~(\ref{eq:HOC_Q1}), $Q_3^{(0)}$ is the HOC IoM given by Eq.~(\ref{eq:HOC_Q3}), and $Q_3^{(1)}$ is given by Eq.~(\ref{eq:Q31aFPUT}).
We have
\begin{equation}
0 = \{Q_3^{\text{Toda}}(\alpha), H^{\text{Toda}}(\alpha)\} = -\{Q_3^{(0)}, V_{\text{FPUT}}\} - \{Q_3^{(1)}, H_0\} + O(\alpha) ~, ~~~\implies~~~ \{Q_3^{(1)}, H_0\} = \{V_{\text{FPUT}}, Q_3^{(0)}\} ~,
\end{equation}
where we have used that $Q_1^{(0)}$ is conserved by $H^{\text{Toda}}(\alpha)$.
Hence, we have rederived the quasi-IoM condition Eq.~(\ref{eq:HOC_VPFUT_Q3quasiIoM}) in the main text from this perspective, and we have recovered exactly the correction $Q_3^{(1)}$ we found earlier in Eq.~(\ref{eq:Q31aFPUT}).

\begin{equation}
\boxed{Q_3^{(1)} = -\sum_j \left[ \frac{1}{3} p_j^3 + \frac{1}{2}(p_j + p_{j+1})(q_j - q_{j+1})^2  \right]} ~,
\end{equation}

Let us do the same for the HOC IoMs $Q_4^{(0)}$ and  $Q_5^{(0)}$.
We start from the Toda's IoM $Q_4^{\text{Toda}}(\alpha)$ and Taylor expand in formal powers of $\alpha$ (where we can stop exhibiting at cubic terms in the phase space variables, since this is the form expected for the IoM corrections):
\begin{align}
Q_4^{\text{Toda}}(\alpha) &= \sum_j \left[ \frac{p_j^4}{4} + (p_{j+1}^2 + p_j p_{j+1} + p_j^2) \frac{1}{4\alpha^2} e^{2\alpha (q_{j+1} - q_{j})} + \frac{1}{32\alpha^4} e^{4\alpha (q_{j+1} - q_{j})} + \frac{1}{16\alpha^4}e^{2\alpha (q_{j+2} - q_{j})} \right] \\
&= \text{const} + \frac{1}{\alpha^2} G_4^{(0)} + \frac{1}{\alpha} G_4^{(1)} + O(\alpha^0) ~,
\label{eq:q41}
\end{align}
where
\begin{align}
G_4^{(0)} &:= \frac{1}{4} \sum_j \left[p_{j+1}^2 + p_j p_{j+1} + p_j^2 + (q_{j+1} - q_j)^2 + \frac{1}{2}(q_{j+2} - q_j)^2 \right]= Q_2^{(0)} + \frac{1}{2}Q_4^{(0)} ~, \\
G_4^{(1)} &:= \frac{1}{2} \sum_j \left[(p_{j+1}^2 + p_j p_{j+1} + p_j^2)(q_{j+1} - q_j) + \frac{2}{3} (q_{j+1} - q_j)^3 + \frac{1}{6}(q_{j+2} - q_j)^3 \right] =: Q_2^{(1)} + \frac{1}{2}Q_4^{(1)} ~.
\end{align}
Here $Q_2^{(0)}$ and $Q_4^{(0)}$ are the HOC IoMs given by Eqs.~(\ref{eq:HOC_Q2}) and (\ref{eq:HOC_Q4}) respectively.
The correction $Q_2^{(1)}$ is given by Eq.~(\ref{eq:Q21def}), and the remaining contribution to the $O(\frac{1}{\alpha})$ term we call $Q_4^{(1)}$.

Now, to check if our correction to $Q_4^{(0)}$ is correct, we can check if the following relation is satisfied
\begin{align}
&\{G_4^{(0)} + \alpha G_4^{(1)}, H_0 + \alpha  V_{\text{FPUT}} \} = O(\alpha^2), \\
& \implies \quad
\{G_4^{(1)}, H_0\} \;=\; \{ V_{\text{FPUT}}, G_4^{(0)}\} = \{ V_{\text{FPUT}}, Q_2^{(0)} + \frac{1}{2}Q_4^{(0)}\} \\
& \text{since~} Q_2^{(0)} = H_0 \quad\implies\quad
\{G_4^{(1)} -  V_{\text{FPUT}}, H_0\}   = \{ V_{\text{FPUT}}, \frac{1}{2}Q_4^{(0)}\} 
=:\{\frac{1}{2}Q_4^{(1)}, H_0\} ~.
\end{align}
\begin{align}
\boxed{Q_4^{(1)} = 2[G_4^{(1)} - V_{\text{FPUT}}] = \sum_j \left[(p_j^2 + p_j p_{j+1} + p_{j+1}^2) (q_{j+1} - q_{j}) + \frac{1}{6}(q_{j+2} - q_{j})^3 \right] } ~.
\label{eq:Q41app}
\end{align}

We can perform a similar analysis to obtain the correction $Q_5^{(1)}$.
We first write $Q_5^{\text{Toda}}$ from Ref.~\cite{Henon1974}, and consider its Taylor series in $\alpha$:
\begin{align}
Q_5^{\text{Toda}}(\alpha) &= \sum_j \bigg[ \frac{p_j^5}{5} + (p_{j+1}^3 + p_j^2 p_{j+1} + p_j p_{j+1}^2 + p_j^3) V_0(q_{j+1} - q_j) + (p_{j+1} + p_j) V_0(q_{j+1} - q_j)^2 \notag \\
& \qquad\quad~ + (p_j + 2p_{j+1} + p_{j+2}) V_0(q_{j+2} - q_{j+1}) V_0(q_{j+1} - q_{j}) \bigg] = \notag  \\
& \quad \quad = \frac{3}{8 \alpha^4} Q_1^{(0)} + \frac{1}{\alpha^3} G_5^{(0)} + \frac{1}{\alpha^2}G_5^{(1)} ~,  
\end{align}
where
\begin{align}
G_5^{(0)} &:= \sum_j\frac{1}{4} \left[ (p_{j+1} + p_j) (q_{j+1} - q_j) + \frac{1}{ 2}(p_j + 2p_{j+1} + p_{j+2})(q_{j+2} - q_j) \right] = -Q_3^{(0)} -\frac{1}{4}Q_5^{(0)} ~, \\
G_5^{(1)} &:= \sum_j \bigg[ \frac{1}{4}(p_{j+1}^3 + p_j^2 p_{j+1} + p_j p_{j+1}^2 + p_j^3) + \frac{1}{2}(p_{j+1} + p_j) (q_{j+1} - q_j)^2 +\frac{1}{8}(p_j + 2p_{j+1} + p_{j+2})(q_{j+2} - q_j)^2 \bigg ]  \notag \\
& =:- Q_3^{(1)} - \frac{1}{4}Q_5^{(1)} ~.
\end{align}
Like earlier, we can check if the correction we found is correct by checking that the following relation holds.
\begin{align}
&\{G_5^{(0)} + \lambda G_5^{(1)}, H_0+ \alpha V_{\text{FPUT}} \} = O(\alpha^2) \\
& \implies \quad \{G_5^{(1)}, H_0\} \;=\; \{V_{\text{FPUT}}, G_5^{(0)}\}  = \{V_{\text{FPUT}}, \frac{1}{4}Q_5^{(0)} + Q_3^{(0)}\} \\
&\implies \quad \{G_5^{(1)} - Q_3^{(1)}, H_0\} \;=\; \{ \frac{1}{4}Q_5^{(1)}, H_0\} = \{V_{\text{FPUT}}, \frac{1}{4}Q_5^{(0)}\}
\end{align}
\begin{align}
    \boxed{Q_5^{(1)} = 4 [G_5^{(1)} - Q_3^{(1)}]= -\sum_j \left[ \frac{1}{3} (p_j + p_{j+1})^3 + \frac{1}{2}(p_j + 2p_{j+1}+p_{j+2})(q_j - q_{j+2})^2 \right]}
\end{align}

\subsection{Higher-order WIB perturbations of HOC and the corresponding corrections}
\label{app:toda_ioms_to_corr_high}
The idea we use in this work is that one can write a long-range Hamiltonian that depends on some parameter $\lambda$ as a power series 
\begin{align}
    \mathcal{H}(\lambda) = H^{(0)} + \lambda  H^{(1)} + \lambda^2 H^{(2)} + ...
\end{align}
Usually, we have constructed only perturbations that are WIB to the leading $\lambda$, but we can also construct them for one order higher.
Consider the Toda Hamiltonian Taylor-expanded and truncated at power $\alpha^2$:
\begin{align}
    \mathcal{H}(\lambda) := H^{(0)} + \lambda  H^{(1)} + \lambda^2  H^{(2)}  = \sum_j \left[\frac{1}{2} p_j^2 + \frac{1}{2}(q_j - q_{j+1})^2 + \lambda \frac{1}{3}(q_j - q_{j+1})^3 + \lambda^2 \frac{1}{6}(q_j - q_{j+1})^4 \right] ~.
\end{align}
We can view $\lambda H^{(1)} + \lambda^2 H^{(2)}$ as a higher-order WIB to the HOC model, in the sense that each original IoM can be corrected to satisfy near-conservation to order $O(\lambda^3)$:
There exist $Q_n^{(1)}$ and $Q_n^{(2)}$ such that
\begin{equation}
\{Q_n^{(0)} + \lambda Q_n^{(1)} + \lambda^2 Q_n^{(2)}, H(\lambda)\} = O(\lambda^3) ~.
\end{equation}
We refer to Ref.~\cite{Surace2023} for a general definition and construction of such higher WIB from truncated long-range deformations and an example for the quantum Heisenberg chain.

By the above condition, the following identities should hold and can be used to find corrections.
\begin{align}
    &\text{0th order} & [ Q_3^{(0)}, H^{(0)}] = 0   \\
    & \text{1st order} & [ Q_3^{(0)}, H^{(1)}] + [Q_3^{(1)}, H^{(0)} ] = 0 \\
    & \text{2nd order} & [ Q_3^{(0)}, H^{(2)}] + [Q_3^{(1)}, H^{(1)}] + [Q_3^{(2)}, H^{(0)} ] = 0  
\end{align}
where $Q_3^{(0)}$, and $Q_3^{(1)}$ we already defined and we want to find $Q_3^{(2)}$ that satisfies the relation above.

In the present case, generalizing our discussion of finding corrections $Q_n^{(1)}$ starting from the Toda chain IoMs, it is clear that we can continue to produce and demonstrate the higher-order WIBs.
Taking $Q_3^{(0)}$ as an example, we generalize
\begin{align}
& Q_3^{(Toda)}(\alpha) = \frac{1}{2\alpha^2} Q_1^{(0)} + \frac{1}{\alpha} G_3^{(0)} + G_3^{(1)} + \alpha G_3^{(2)} + O(\alpha^2) ~, \\
& \implies \{G_3^{(0)} + \lambda G_3^{(1)} + \lambda^2 G_3^{(2)}, H_0 + \lambda H^{(1)} + \lambda^2 H^{(2)} \} = O(\lambda^3) ~,
\end{align}
with $G_3^{(2)} = \sum_j \frac{2}{3}(q_{j+1} - q_j)^3 (p_j + p_{j+1})$.
Since we have earlier identified $Q_3^{(0)} = -G_3^{(0)}, Q_3^{(1)} = -G_3^{(1)}$, we have $Q_3^{(2)} = -G_3^{(2)}$. ...
From direct expansion of $Q_3^{(0), Toda}$ I get
\begin{align}
    Q_3^{(2)} = \alpha\frac{1}{3}(p_{j+1} + p_{j})(q_{j+1} - q_{j})^3
\end{align}
so that
\begin{align}
    [H^{(2)}, Q_3^{(0)}] + [H^{(1)},Q_3^{(1)} ] + [H^{(0)},Q_3^{(2)} ] = 0 
\end{align}
Note that a similar analysis can be used to argue why $\beta-$FPUT is not a WIB wrt HOC. 

\section{Generator for cubic potential perturbations of the HOC}
\label{app:AGPkspace}
We here show how to derive the expression of the generator in momentum space for a generic translational-invariant cubic perturbation of the HOC.
We start by writing our Hamiltonian in the momentum space:
\begin{equation}
    H_0 = \sum_k \left( \frac{1}{2} \tilde{p}_k \tilde{p}_{-k} + \frac{\omega_k^2}{2} \tilde{q}_k \tilde{q}_{-k} \right) ~, \qquad \omega_k^2 = 2[1 - \cos(k)] = \omega_{-k}^2 ~,
\end{equation}
where the new variables are defined by
\begin{equation}
    q_j(t) = \frac{1}{\sqrt{L}}\sum_k\tilde{q}_k(t)e^{ikj}, \quad \quad p_j(t) = \frac{1}{\sqrt{L}}\sum_k\tilde{p}_k(t)e^{ikj},
\end{equation}
with $k=\frac{2\pi}{L}n, \quad n=0,1,\dots L-1$. They obey the basic Poisson bracket $\{\tilde{q}_k, \tilde{p}_{k'}\} = \delta_{k,-k'}$.
We hence have $\{\tilde{q}_k, H_0\} = \tilde{p}_k$ and $\{\tilde{p}_k, H_0\} = -\omega_k^2 \tilde{q}_k$.

Before diving into calculations, let us set up some conventions for symmetric cubic terms and their expressions in terms of Fourier modes. 
Assume the perturbation $V$ is cubic, of the form
\begin{equation}
V=\sum_{j_1,j_2,j_3} V(j_1, j_2, j_3) q_{j_1} q_{j_2} q_{j_3} ~.
\end{equation}
W.l.o.g., we can assume that $V(j_1,j_2,j_3)$ is invariant under all permutations [it is enough to require $V(j_1, j_2, j_3) = V(j_2, j_1, j_3) = V(j_1, j_3, j_2)$  for any $j_1, j_2, j_3$].
In a translationally invariant PBC chain, we further expect that $V(j_1,j_2,j_3)$ depends only on relative distances. 
Thus, we assume that we can write $V(j_1, j_2, j_3) = \Upsilon(j_1 - j_3, j_2 - j_3)$, with $\Upsilon(r_1 + mL, r_2 + nL) = \Upsilon(r_1, r_2)$ for any integers $m, n$.
Furthermore, we assume that $\Upsilon$ satisfies $\Upsilon(r_1, r_2) = \Upsilon(r_2, r_1) = \Upsilon(r_1 - r_2, -r_2)$ for any $r_1, r_2$; this guarantees that $V$ is invariant under all permutations. 
Going to Fourier modes, we can write $V$ as 
\begin{equation}
\label{eq:Vk_app}
V = \sum_{j_1,j_2,j_3} \Upsilon(j_1-j_3, j_2-j_3) q_{j_1} q_{j_2} q_{j_3} 
= \sum_{k,k'} v(k,k') \tilde q_k \tilde q_{k'} \tilde q_{-k-k'} ~,
\quad
\Upsilon(r_1, r_2) = \frac{1}{L^{3/2}} \sum_{k,k'} v(k,k') e^{-i k r_1 - i k' r_2} ~,
\end{equation}
where we chose the specific Fourier transform convention and normalization to simplify writing of $V$ and manipulations below.
The invariance of $V$ under permutations translates to the following properties for $v$:
\begin{align}
v(k,k') = v(k',k) = v(k,-k-k') = v(-k-k',k) = v(-k-k',k') = v(k',-k-k') ~.
\label{eq:vkkp_symm}
\end{align}
We now look for an $X$ that satisfies $V = \{X,H_0\}$.
$X$ has to be cubic, translationally invariant, and with an odd number of $\tilde{p}_k$'s (i.e., time-reversal odd). 
The most general ansatz is of the form
\begin{equation}
\label{eq:Xkspace}
X = \sum_{k,k'} \left[g(k,k') \tilde{p}_k \tilde{p}_{k'} \tilde{ p}_{-k-k'} + f(k,k') \tilde{q}_k \tilde q_{k'} \tilde{p}_{-k-k'} \right].
\end{equation}
Using the same argument used above for $v$, we can assume, w.l.o.g., that $g(k,k') = g(k',k) = g(k,-k-k') = \dots$.
On the other hand, in the $\sum_{j_1,j_2,j_3} F(j_1, j_2, j_3) q_{j_1} q_{j_2} p_{j_3}$ part of $X$ we only require symmetry under $j_1 \leftrightarrow j_2$; hence,
we can assume that $f$ satisfies $f(k, k') = f(k',k)$.
Our goal is to find $g$ and $f$.
We have
\begin{align}
    \{X, H_0\} &= \sum_{k,k'} \left[ -3\omega_k^2 g(k,k') \tilde{q}_k \tilde{p}_{k'} \tilde{p}_{-k-k'} 
    - \omega_{-k-k'}^2 f(k,k') \tilde{q}_k \tilde{q}_{k'} \tilde{q}_{-k-k'} + 2f(k,k') \tilde{q}_k \tilde{p}_{k'} \tilde{p}_{-k-k'} \right] = \notag \\
    &= \sum_{k,k'}\left[\Big(- 3\omega_{k}^2 g(k,k') + f(k,k') + f(k,-k-k') \Big) \tilde{q}_k \tilde{p}_{k'} \tilde{p}_{-k-k'} \right. \notag \\
    &\qquad\quad \left.-\frac{1}{3}\Big(\omega_{-k-k'}^2 f(k,k')+\omega_{k}^2 f(-k-k',k')+\omega_{k'}^2 f(k,-k-k')\Big) \tilde{q}_k \tilde{q}_{k'} \tilde{q}_{-k-k'} \right] ~.
    \label{eq:XH0}
\end{align}
In the first line, we have used the fact that $ppp$ terms in $X$ give $qpp$ terms in $\{X, H_0\}$, while $qqp$ terms give $qpp$ and $qqq$ terms, and then we used the symmetry properties of the $g$ and $f$ functions to group and relabel terms. 
In the second line, we collected $qpp$ terms and organized the $f$ contributions so that the exhibited combination $a(k,k')= f(k,k') + f(k,-k-k')$ has the property $a(k,-k-k') = a(k,k')$ (the $g$ contribution already has this property).
In the last line, we organized $qqq$ terms so that the amplitudes exhibit the same property as $v(k,k')$ in Eq.~(\ref{eq:vkkp_symm}).
 
As we explain below, for a given triple $k,k',-k-k'$, the condition that $\{X,H_0\}$ contains no $qpp$ terms [from comparing Eqs.~(\ref{eq:Vk_app}) and (\ref{eq:XH0})] gives three linear constraints
in the first three lines below, while matching the $qqq$ vertex gives the final constraint:
\begin{equation}
\begin{cases}
    -3\omega_{k}^2 g + f(k, k') + f(k, -k-k') = 0 ~, \\
    -3\omega_{k'}^2 g + f(k, k') + f(-k-k', k') =0 ~, \\
    -3\omega_{-k-k'}^2 g + f(k,-k-k') + f(-k-k', k') = 0 ~, \\
    -\frac{1}{3} \Big(\omega_{-k-k'}^2 f(k,k') + \omega_{k}^2 f(-k-k',k') + \omega_{k'}^2 f(k,-k-k') \Big) = v ~.
    \label{eq:fgsyst}
\end{cases}
\end{equation}
To streamline notation, we dropped the $ k$ and $ k'$ dependence of $g$ and $v$.
The first equation follows directly from the expression for $\{X, H_0\}$ and needs to hold for every $k$, $k'$, i.e., every term in the sum (note that because of the chosen writing, the term with $k' \to -k-k'$ gives an identical equation).
However, the same $f$'s enter also in some other equations, i.e., in some other terms in the sum.
We find these as the second and third equations, which we can formally obtain from the first by substitutions $k \to k'$, $k' \to k$ or $-k-k'$ and $k \to -k-k'$, $k' \to k'$ or $k$.
We have now listed all equations where $k,k',-k-k'$ enter.
Note that in general there are three independent $f$'s: $f(k,k')=f(k',k)$, $f(k,-k-k') = f(-k-k',k)$, and $f(-k-k',k') = f(k',-k-k')$, and we have already used these symmetry relations to show only three unique ones.
On the other hand, there is only one $g$: $g(k,k') = g(k',k) = g(k,-k-k') = g(-k-k',k) = g(-k-k',k') = g(k',-k-k')$, which is similar to the symmetries of $v$; to emphasize this, we did not show arguments for $g$ and $v$.
Finally, the last equation containing $f$'s and $v$ is the only independent such equation for the triple $k,k',-k-k'$, since it has already been fully symmetrized.

Using the first three equations, we can uniquely determine the three $f$'s in terms of the single $g$:
\begin{align}
& f(k,k') = \frac{3g}{2} (\omega_k^2 + \omega_{k'}^2 - \omega_{-k-k'}^2) ~, \\
& f(k,-k-k') = \frac{3g}{2} (\omega_k^2 + \omega_{-k-k'}^2 - \omega_{k'}^2) ~, \\
& f(-k-k',k') = \frac{3g}{2} (\omega_{-k-k'}^2 + \omega_{k'}^2 - \omega_{k}^2) ~.
\end{align}
Substituting these into the last equation (containing $f$'s and $v$), we obtain an equation for $g$:
\begin{equation}
-\frac{1}{2} \left[\omega_{-k-k'}^2 (\omega_k^2 +\omega_{k'}^2 - \omega^2_{-k-k'}) +\omega_k^2 (\omega_{-k-k'}^2 + \omega_{k'}^2 -\omega_k^2) + \omega_{k'}^2 (\omega_k^2 +\omega_{-k-k'}^2 -\omega_{k'}^2) \right] g = v ~.
\end{equation}
We can simplify the expression in the square brackets on the l.h.s.\ by using the following identity
\begin{equation}
    2\omega_k^2 \omega_{k'}^2 + 2\omega_k^2 \omega_{-k-k'}^2 + 2\omega_{k'}^2 \omega_{-k-k'}^2 -\omega_k^4 - \omega_{k'}^4 - \omega_{-k-k'}^4
    = 64 \sin^2\left(\frac{k}{2}\right) \sin^2\left(\frac{k'}{2}\right) \sin^2\left(\frac{-k-k'}{2}\right) ~.
\end{equation}
and we get the final result for $g$ and $f$
\begin{equation}
\label{eq:gk}
    \boxed{g(k,k')=-\frac{v}{32 \sin^2\left(\frac{k}{2}\right)\sin^2\left(\frac{k'}{2}\right)\sin^2\left(\frac{k+k'}{2}\right)}},
\end{equation}
\begin{equation}
\label{eq:fk}
\boxed{f(k,k') = \frac{3}{2} (\omega^2_k + \omega^2_{k'} - \omega^2_{-k-k'}) g(k,k').}
\end{equation}
For later convenience, we also show the expression for the frequency combination in the last expression:
\begin{equation}
\omega^2_k + \omega^2_{k'} - \omega^2_{-k-k'} = - 8 \sin\left(\frac{k}{2}\right) \sin\left(\frac{k'}{2}\right) \cos\left(\frac{-k-k'}{2}\right) ~.
\end{equation}

It is now useful to separately discuss two cases: (i) when the momenta $k$, $k'$, and $-k-k'$ are all non-zero; and (ii) when at least one of them is zero. 
Let us begin with the latter case, and take $k'=0$ for concreteness.
In this situation $\omega_{k'} = 0$, and using the second and last lines of Eq.~(\ref{eq:fgsyst}) we obtain $v(k,0) = 0$.
This represents a constraint on the perturbation $V$ for the equation $V = \{X, H_0\}$ to admit a solution.
If this condition is satisfied, the system can be solved, e.g., by setting $g(k,0) = f(k,0) = 0$.
We remark that if $v(k,0) = 0$, the equation for $g$ can in fact be satisfied by any value of $g(k,0)$.
We then obtain $f(k,0) = f(-k,0) = 0$, while $f(k,-k) = 3\omega_k^2 g(k,0)$.
We can understand this solution and the arbitrariness of $g(k,0)$ as follows.
Assuming $k \neq 0$ and collecting all the corresponding contributions to $X$, we obtain $6 g(k,0) \tilde{p}_0 (\tilde{p}_k \tilde{p}_{-k} + \omega_k^2 \tilde{q}_k \tilde{q}_{-k})$, which we recognize is a product of familiar IoMs $\tilde{p}_0$ (proportional to the total momentum) and $\tilde{p}_k \tilde{p}_{-k} + \omega_k^2 \tilde{q}_k \tilde{q}_{-k}$ (proportional to an eigenmode energy), and such contributions to $X$ produce zero contribution to $\{X, H_0\}$.
[When $k = 0$, we simply obtain a contribution to $X$ of $g(0,0) \tilde{p}_0^3$, which is again a product of IoMs.]
Thus, from this analysis, we have also recovered the known freedom in $X$, namely, admitting the addition of an arbitrary conserved observable.
The simplest choice is to set $g(k,0) = 0$, and we will do this in what follows.

Let us now consider a few important examples of cubic potentials/perturbations. 

\subsubsection{The $\alpha$-FPUT potential}
The natural starting point is the $\alpha$-FPUT potential given by 
\begin{equation}
    V_{\text{FPUT}}=    \sum_{k,k'}\underbrace{\frac{(2i)^3}{3\sqrt{L}}\sin\left(\frac{k}{2}\right)\sin\left(\frac{k'}{2}\right)\sin\left(\frac{-k-k'}{2}\right)}
    _{v_{\text{FPUT}}}\tilde q_k \tilde q_{k'} \tilde q_{-k-k'}.
\end{equation}
Note that $v_{\text{FPUT}}(k,0) = 0$, so this perturbation admits a well-defined generator, as discussed previously.
This means that we can focus on the case when $k,k',-k-k'\neq 0$.
Using Eqs.~(\ref{eq:gk}) and (\ref{eq:fk}) we find
\begin{equation}
\label{eq:gfFPUT}
    g_{\text{FPUT}}(k,k') = \frac{i}{12\sqrt{L} \sin\left(\frac{k}{2}\right)\sin\left(\frac{k'}{2}\right)\sin\left(\frac{-k-k'}{2}\right)} ~, \qquad\qquad
    f_{\text{FPUT}}(k,k') 
   = \frac{-i \cos\left(\frac{-k-k'}{2}\right)}{\sqrt{L} \sin\left(\frac{-k-k'}{2}\right)}~.
\end{equation}
We note that both $g_{\text{FPUT}}$ and $f_{\text{FPUT}}$ have poles of order 1, and we will use this fact to argue later that all IoMs in the case of the $\alpha$-FPUT can be corrected. 

\subsubsection{Cubic, momentum-conserving
potentials of arbitrary range}
As another example of a well-defined generator, we can consider the family of cubic perturbations with an increasing range given by $m$ where $m \geq 1$ ($m=1$ just recovers the FPUT result):
\begin{equation}
    V_{m}=\frac{1}{3} \sum_j (q_{j+m} - q_j)^3 = \sum_{k,k'} \underbrace{\frac{(2i)^3}{3\sqrt{L}}\sin\left(m\frac{k}{2}\right)\sin\left(m\frac{k'}{2}\right)\sin\left(m\frac{-k-k'}{2}\right)}_{v_{m}} \tilde q_k \tilde q_{k'} \tilde q_{-k-k'} ~.
\end{equation}
This perturbation also satisfies the condition $v_m(k,0) = 0$, and admits a well-defined generator.
The corresponding generator coefficients follow immediately from
Eqs.~(\ref{eq:gk}) and~(\ref{eq:fk}):
\begin{equation}
g_m(k,k') = -\frac{v_m(k,k')}{32\sin^2\left(\frac{k}{2}\right)\sin^2\left(\frac{k'}{2}\right)\sin^2\left(\frac{k+k'}{2}\right)} ~, \quad \quad f_m(k,k') = \frac{3}{2}\big(\omega_k^2+\omega_{k'}^2-\omega_{-k-k'}^2\big)\, g_m(k,k') ~.
\end{equation}
One can observe the same type of pole structure as in $v_{\text{FPUT}}$ by noting that the individual terms in $\sin\left(m\frac{k}{2}\right) \sin\left(m\frac{k'}{2}\right) \sin\left(m\frac{-k-k'}{2}\right)$ for $m>1$ can be written as 
\begin{equation}
    \sin\left(m\frac{k}{2}\right) = \sin\left(\frac{k}{2}\right)U_{m-1}\left(\cos\left(\frac{k}{2}\right)\right) ~,
\end{equation}
where $U_{m}(x)$ is the Chebyshev polynomial of the second kind given by 
\begin{equation}
    U_m(\cos \theta) \sin \theta = \sin((m+1)\theta), \quad m = 0, 1, 2, 3, \dots
\end{equation}
with the following recurrence relation 
\begin{align}
    &U_0(x) = 1 ~,\\
    &U_1(x) = 2x ~,\\
    &U_{n+1}(x) = 2xU_n(x) - U_{n-1}.
\end{align}
For example, for $m=2$,
one can see right away that in $g_2(k,k')$ the second-order poles are reduced to first order by $v$ in the same way as for the $v_{\text{FPUT}}$ case, since 
\begin{align}
    v_2(k,k') &= \frac{(2i)^3}{3\sqrt{L}}\sin\left(2\frac{k}{2}\right)\sin\left(2\frac{k'}{2}\right)\sin\left(2\frac{-k-k'}{2}\right) = \notag \\
    & = \frac{(2i)^3}{3\sqrt{L}}\sin\left(\frac{k}{2}\right)\sin\left(\frac{k'}{2}\right)\sin\left(\frac{-k-k'}{2}\right)U_{1}\left(\cos\left(\frac{k}{2}\right)\right)U_{1}\left(\cos\left(\frac{k'}{2}\right)\right)U_{1}\left(\cos\left(\frac{-k-k'}{2}\right)\right) = \notag \\
    & = \frac{(2i)^3}{3\sqrt{L}}\underbrace{\sin\left(\frac{k}{2}\right)\sin\left(\frac{k'}{2}\right)\sin\left(\frac{-k-k'}{2}\right)}_{\text{reduces 2nd order poles in $g_2$ and $f_2$}} 2 \cos\left(\frac{k}{2}\right) 2\cos\left(\frac{k'}{2}\right) 2\cos\left(\frac{-k-k'}{2}\right).
\end{align}
A similar argument can be applied to generic cubic perturbations of the form
\begin{equation}
    V_{\mathbf{m},\mathbf{n}}= \frac{1}{3} \sum_j (q_{j+m_1}-q_{j+n_1})(q_{j+m_2}-q_{j+n_2})(q_{j+m_3}-q_{j}) ~,
\end{equation}
where $\mathbf{m} = (m_1, m_2, m_3)$ and $\mathbf{n} = (n_1, n_2)$ contain arbitrary integers.
In all these cases, the second-order poles are reduced to first-order poles in the final result for $g(k,k')$.

\subsubsection{Cubic, momentum-conserving perturbations from bi-local generator}
Perturbations of the form above can be combined to obtain other WIBs. 
However, there are special combinations that correspond to WIBs generated from the bilocal generator presented in the main text (see \ref{subsec:bilocal}).
Take the following example
\begin{equation}
    V_{\text{bi}} = \frac{1}{3} \sum_j (q_{j+1} - q_j)^3 - \frac{1}{24}\sum_j(q_{j+2} - q_j)^3 ~,
\end{equation}
which was derived from our bilocal generator defined in Eq.~(\ref{eq:bilocal}), by setting $\beta=1, \gamma=2$, see Eq.~(\ref{eq:tVbi12}). 
\begin{align}
    v_{\text{bi}} &= \frac{(2i)^3}{3\sqrt{L}}\sin\left(\frac{k}{2}\right)\sin\left(\frac{k'}{2}\right)\sin\left(\frac{-k-k'}{2}\right)-\frac{(2i)^3}{24\sqrt{L}}\sin\left(k\right)\sin\left(k'\right)\sin\left(-k-k'\right)\\
    &= v_{\text{FPUT}}\times\left[1 - \cos\left(\frac{k}{2}\right)\cos\left(\frac{k'}{2}\right)\cos\left(\frac{-k-k'}{2}\right)\right] ~.
    \label{eq:vbi_from_vfput}
\end{align}
Finally, we note that for a generic $V$ that does not preserve momentum, functions $f$ and $g$ have poles of order 2 in $k=0$, $k'=0$, $k+k'=0$, while for all momentum-conserving potentials, the poles are of order 1. 

\subsection{Corrections to IoMs of the HOC}
\label{subapp:HOC_AGP_kspace_IoMcorrs}
Because of the poles, we can already anticipate that the AGP of the $\alpha$-FPUT chain is not a local operator in real space.
An explicit (``trilocal'') real-space expression will be discussed in Sec.~\ref{app:AGPFPUT}.
Similar considerations hold for the generic cubic momentum-conserving perturbations discussed above.
The non-locality of the generator is not an issue for defining WIBs: generators of the boosted or bilocal type are non-local, but produce local corrections to the IoMs.
Crucially, we need to check if the corrections to the conserved quantities for the case of cubic perturbations are also singular in momentum space, like the AGP, or whether they have no singularities and therefore produce local corrections to the IoMs.
In the latter case, these perturbations are WIBs.

\subsubsection{Corrections to odd IoMs}
Let us start with odd IoMs.
These are defined in Eq.~(\ref{eq:HOC_oddIoMs}) and translate to Fourier modes as 
\begin{equation}
Q_{2n+1}^{(0)} = \frac{1}{2} \sum_j (q_j p_{j+n} - p_j q_{j+n}) = \frac{1}{2} \sum_k \tilde q_k \tilde p_{-k}(e^{-ikn} - e^{ikn}) = -i \sum_k \sin(nk) \tilde q_k \tilde p_{-k} ~.
\end{equation}
We now compute the correction, which is easy to do using $\{\tilde{q}_k, Q_{2n+1}^{(0)}\} = -i\sin(nk) \tilde{q}_k$, $\{\tilde{p}_k, Q_{2n+1}^{(0)}\} = -i\sin(nk) \tilde{p}_k$:
\begin{align}
Q_{2n+1}^{(1)} = \{X, Q_{2n+1}^{(0)}\}&= 
-i\sum_{k,k'}\big(\sin(nk)+\sin(nk')+\sin(-nk-nk')\big)[g(k,k')\tilde p_k \tilde p_{k'}\tilde p_{-k-k'}+f(k,k') \tilde q_k \tilde q_{k'} \tilde p_{-k-k'}] \notag \\
&= 4i \sum_{k,k'}\sin\left(\frac{nk}{2}\right)\sin\left(\frac{nk'}{2}\right)\sin\left(\frac{n(-k-k')}{2}\right)[g(k,k')\tilde p_k \tilde p_{k'}\tilde p_{-k-k'} + f(k,k') \tilde q_k \tilde q_{k'} \tilde p_{-k-k'}] ~.
\end{align}
From this expression, we see that if $g$ and $f$ have poles of order 1 (as is the case for all momentum-conserving perturbations such as$V_{\text{FPUT}}$, $V_m$, and $V_{\text{bi}}$), then the correction to $Q_{2n+1}^{(0)}$ is not singular in $k$-space, and hence it is (extensive) local in real space.

As an example, for the $\alpha$-FPUT potential, we obtain for $n=1$:
\begin{equation}
Q_3^{(1)} = -\frac{1}{3\sqrt{L}} \sum_{k,k'} \tilde{p}_k \tilde{p}_{k'} \tilde{p}_{-k-k'} + \frac{4}{\sqrt{L}} \sum_{k,k'} \sin\left(\frac{k}{2}\right) \sin\left(\frac{k'}{2}\right) \cos\left(\frac{-k-k'}{2}\right) \tilde{q}_k \tilde{q}_{k'} \tilde{p}_{-k-k'} ~.
\end{equation}

This reproduces exactly our correction to $Q_3^{(0)}$ in the FPUT case we discussed in the main text in Eq.~(\ref{eq:Q31aFPUT}).
Note that there is a small subtlety of whether we are allowed to include $k$-points with $k=0$, or $k'=0$, or $-k-k'=0$ in the sum in the last expression, since earlier we defined $g$ and $f$ to be zero at such wavevectors for $X$ to be well-defined on finite chains.
We will fully resolve this subtlety later.

\subsubsection{Corrections to even IoMs}
Now, let us consider even IoMs. These are defined in Eq.~(\ref{eq:HOC_evenIoMs}) and are expressed in $k$-space as 
\begin{equation}
Q_{2n+2}^{(0)} = \frac{1}{2} \sum_j (p_j p_{j+n} + 2q_jq_{j+n} - q_{j}q_{j+n-1} - q_jq_{j+n+1})
= \frac{1}{2} \sum_k \cos(nk) [\tilde p_k \tilde p_{-k} +\omega_k^2\tilde q_k \tilde q_{-k}] ~.
\end{equation}
We again compute the correction, which is easy to do using $\{\tilde{q}_k, Q_{2n+2}^{(0)}\} = \cos(nk) \tilde{p}_k$, $\{\tilde{p}_k, Q_{2n+2}^{(0)}\} = -\cos(nk) \omega_k^2 \tilde{q}_k$:
\begin{align}
\{X, Q_{2n+2}^{(0)}\}&= \sum_{k,k'} [-3g(k,k')\cos(nk)\omega_k^2 \tilde q_k \tilde p_{k'}\tilde p_{-k-k'} + 2f(k,k')\cos(nk') \tilde q_k \tilde p_{k'} \tilde p_{-k-k'} \notag \\
&\qquad\quad - f(k,k')\cos(n(-k-k'))\omega_{-k-k'}^2 \tilde q_k \tilde q_{k'} \tilde q_{-k-k'} ] \notag \\
&= \sum_{k,k'} \Big\{ \big[-3g(k,k')\cos(nk)\omega^2_k + f(k, k')\cos(nk') + f(k, -k-k')\cos(n(-k-k'))\big] \tilde q_k \tilde p_{k'} \tilde p_{-k-k'} \notag \\
&- \frac{1}{3} \big[f(k,k')\cos(n(-k-k'))\omega_{-k-k'}^2 + f(k, -k-k')\cos(nk')\omega_{k'}^2 + f(-k-k',k')\cos(nk)\omega_k^2 \big] \tilde q_k \tilde q_{k'} \tilde q_{-k-k'}\Big\}, \notag
\end{align}
where we used symmetries of $g$ and $f$ to organize/regroup terms for convenience.
Let us first consider coefficient of the $\tilde q_k \tilde q_{k'}\tilde q_{-k-k'}$ term:
\begin{equation}
f(k,k') \omega_{-k-k'}^2 = \frac{3}{2} g(k,k') (\omega^2_k + \omega^2_{k'} - \omega^2_{-k-k'}) \omega_{-k-k'}^2 = -24 g(k,k') \sin\left(\frac{k}{2}\right) \sin\left(\frac{k'}{2}\right) \sin\left(\frac{-k-k'}{2}\right) \sin\left(-k-k'\right) ~, \notag
\end{equation}
and all the poles in $g$ are manifestly canceled (and similarly for all contributions to the coefficient of the $qqq$ term).
For later use, the full coefficient of $\tilde{q}_k \tilde{q}_{k'} \tilde{q}_{-k-k'}$ reads
\begin{equation}
8 g(k,k') \sin\left(\frac{k}{2}\right) \sin\left(\frac{k'}{2}\right) \sin\left(\frac{-k-k'}{2}\right) \big[\sin(-k-k') \cos(n(-k-k')) + \sin(k') \cos(nk') + \sin(k) \cos(nk) \big] ~.
\label{eq:HOC_even_corr_qqq}
\end{equation}

Next let us consider the coefficient of $\tilde q_k \tilde p_{k'}\tilde p_{-k-k'}$:
\begin{align}
&-3g(k,k') \cos(nk)\omega^2_k + f(k,k')\cos(nk') + f(k,-k-k')\cos(n(-k-k')) = \label{eq:HOC_even_corr_qpp}\\
& = \frac{3g}{2}\bigg[\bigg(\cos(nk') - \cos(nk)\bigg)(\omega_k^2+\omega_{k'}^2-\omega_{-k-k'}^2) + \bigg (\cos(n(-k-k')) - \cos(nk)\bigg)(\omega_k^2+\omega_{-k-k'}^2 - \omega_{k'}^2) \bigg] = \notag \\
& = 24 g \sin\left(\frac{k}{2}\right)\sin\left(\frac{k'}{2}\right)\sin\left(\frac{-k-k'}{2}\right) \cdot\bigg\{\sin\left(\frac{n(k-k')}{2}\right) \cos\left(\frac{-k-k'}{2}\right)U_{n-1}\bigg[\cos\left(\frac{-k-k'}{2}\right)\bigg] + \notag \\
& \qquad \qquad \qquad  \qquad \qquad \qquad \qquad \qquad \qquad  +\sin\left(\frac{n(2k+k')}{2}\right) \cos\left(\frac{k'}{2}\right)U_{n-1}\bigg[\cos\left(\frac{k'}{2}\right)\bigg]  \bigg\}. \notag
\end{align}
We see that the poles in $g$ are manifestly canceled.

As an example, let $n=1$, so we can find the correction to $Q_4^{(0)}$.
We obtain for the $\alpha$-FPUT perturbation
\begin{equation}
Q_4^{(1)} = \frac{i}{\sqrt{L}} \sum_{k,k'} \left\{ [2\sin(k) - \sin(k') - \sin(-k-k')] \tilde{q}_k \tilde{p}_{k'} \tilde{p}_{-k-k'}
+ \frac{1}{3} [\sin(2k) + \sin(2k') + \sin(2(-k-k'))] \tilde{q}_k \tilde{q}_{k'} \tilde{q}_{-k-k'} \right\} ~.
\end{equation}
After simple manipulations, this matches exactly $Q_4^{(1)}$ calculated earlier, Eq.~(\ref{eq:Q41aFPUT}), from the Toda connection (see also App.~\ref{app:toda_ioms_to_corr}).

\subsubsection{Correction to IoMs of the HOC: ``gauge freedom''}
In summary, our discussion of the corrections of odd and even IoMs showed that it is sufficient for $v(k,k')$ to have simple zeros in $k, k', -k-k'$, in order to get local IoM corrections, even though the generator may be non-local.
All the momentum-conserving cubic perturbations have zeros in $k, k', -k-k'$, and therefore they are WIBs.

A subtle aspect that we have not explicitly addressed so far is the following.
We set $g(k,0) = g(0,k') = g(k,-k) = 0$, to deal with well-defined expressions for $X$ in all our calculations.
However, the residue of $g$ at these poles may be finite, and setting those values to zero would lead to discontinuities in the momentum-space expressions of the IoM corrections.
At first sight, such discontinuities might suggest the emergence of non-local contributions to the corrected IoMs.
The resolution of this apparent issue lies in exploiting the freedom to add arbitrary conserved quantities to the IoM corrections.

Consider, e.g., a correction to the odd IoM with $n=1$:
\begin{align}
Q_{3}^{(1)}&= 4i \sum_{k,k'} \sin\left(\frac{k}{2}\right) \sin\left(\frac{k'}{2}\right) \sin\left(\frac{-k-k'}{2}\right) [g(k,k')\tilde p_k \tilde p_{k'} \tilde p_{-k-k'} + f(k,k') \tilde q_k \tilde q_{k'} \tilde p_{-k-k'}] \\
&=   -\frac{i}{8} \sum_{\substack{k\neq0,k'\neq0 ,\\-k-k'\neq 0}} \frac{v(k,k')}{\sin\left(\frac{k}{2}\right) \sin\left(\frac{k'}{2}\right) \sin\left(\frac{-k-k'}{2}\right)} \left[\tilde p_k \tilde p_{k'} \tilde p_{-k-k'} + \frac{3}{2}(\omega_k^2 + \omega_{k'}^2 - \omega_{-k-k'}^2) \tilde q_k \tilde q_{k'} \tilde p_{-k-k'} \right] ~.
\end{align}
In all the cases we consider, $v(k,k')$ has zeros in $k=0, k'=0, -k-k'=0$, and we can define an analytic function $R(k,k')$
\begin{equation}
    R(k,k') = -\frac{i}{8} \frac{v(k,k')}{\sin\left(\frac{k}{2}\right) \sin\left(\frac{k'}{2}\right) \sin\left(\frac{-k-k'}{2}\right)} ~,
\end{equation}
with the property $R(k,k')=R(k',k)=R(k,-k-k')=R(-k-k',k)=R(k',-k-k')=R(-k-k',k')$.
Although the denominator vanishes when any of the momenta is zero, the numerator removes these singularities, and $R(k,k')$ can be defined there by analytic extension.
We want to prove that we can define a new IoM correction $\tilde Q_3^{(1)}$ that differs from $Q_3^{(1)}$ only by conserved quantities and has the form (note the absence of restrictions in the sum):
\begin{equation}
    \tilde{Q}_3^{(1)} := \sum_{k,k'}R(k,k')\left[\tilde p_k \tilde p_{k'} \tilde p_{-k-k'} + \frac{3}{2}(\omega_k^2 + \omega_{k'}^2 - \omega_{-k-k'}^2) \tilde q_k \tilde q_{k'} \tilde p_{-k-k'} \right] = Q_3^{(1)} + \text{conserved quantities} ~.\label{eq:gauge}
\end{equation}
If this is true, then $\tilde Q_{3}^{(1)}$ is also an IoM correction in the sense of satisfying Eq.~(\ref{eq:corrections}) and is explicitly local in real space.
It is such $\tilde{Q}^{(1)}_m$ expressions with no restrictions in the sum that we meant implicitly as the final extensive local corrections in the previous two subsections (and we may loosely call this ambiguity in the definition of the corrections as ``gauge freedom.''

To prove the above, we compute $\tilde Q_{3}^{(1)} - Q_{3}^{(1)}$ and show that it is a conserved quantity. There are four cases we need to consider i) $k=0, k'\neq 0, -k-k'\neq 0$; ii) $k\neq 0, k'= 0, -k-k'\neq 0$; iii) $k\neq 0, k'\neq 0, -k-k'=0$; and iv) $k=k'=-k-k'=0$:
\begin{align}
   \tilde Q_{3}^{(1)} - Q_{3}^{(1)} &= \sum_{k'\neq0} R(0,k')\tilde p_0 \tilde p_{k'} \tilde p_{-k'} + \sum_{k\neq 0} R(k,0) \tilde p_{k} \tilde p_0 \tilde p_{-k} + \sum_{k\neq 0} R(k,-k)[\tilde p_{k} \tilde p_{-k} \tilde p_0 + 3\omega_k^2 \tilde q_k \tilde q_{-k} \tilde p_0] + R(0,0) \tilde p_0^3 \notag\\
   &=\sum_{k\neq 0} 3R(k,0)[\tilde p_0 (\tilde p_{k} \tilde p_{-k} + \omega_k^2 \tilde q_k \tilde q_{-k})] + R(0,0) \tilde p_0^3 ~.
\end{align}
Since $\tilde p_0$ and $\tilde p_{k} \tilde p_{-k} + \omega_k^2 \tilde q_k \tilde q_{-k}$ are all conserved quantities, we thus proved Eq.~(\ref{eq:gauge}).
This proof can be straightforwardly extended to all odd integrals of motion.

It can also be generalized to the even integrals of motion.
Indeed, consider terms $\tilde{q}_k \tilde{q}_{k'} \tilde{q}_{-k-k'}$ in the correction to $Q_{2n+2}^{(0)}$, with coefficients given by Eq.~(\ref{eq:HOC_even_corr_qqq}).
The square bracket in this expression vanishes when $k=0$ or $k'=0$ or $-k-k'=0$, so the residue is zero in this case.
On the other hand, for terms $\tilde{q}_k \tilde{p}_{k'} \tilde{p}_{-k-k'}$ with coefficients given by Eq.~(\ref{eq:HOC_even_corr_qpp}), the residue vanishes when $k=0$ but is non-vanishing when $k'=0$ or $-k-k'=0$.
However, in the latter cases we see that they would give contributions that are products of conserved quantities $\tilde{p}_0$ and $\tilde{q}_k \tilde{p}_{-k} - \tilde{q}_k \tilde{p}_{-k}$, completing the resolution of special wavevectors for corrections to even IoMs as well.

\section{Real-space generator for $\alpha$-FPUT}
\label{app:AGPFPUT}
\subsection{Finite sizes}
Using the momentum space expression for $ X_{\text{FPUT}}$ in Eqs.~(\ref{eq:Xkspace}) and (\ref{eq:gfFPUT}), the real space solution for finite sizes can be found via a Fourier transform,
\begin{equation}
    X_{\text{FPUT}} = \sum_{j_1,j_2,j_3} \Gamma(j_1-j_3, j_2 - j_3) p_{j_1} p_{j_2} p_{j_3} + \Phi(j_1-j_3, j_2-j_3) q_{j_1} q_{j_2} p_{j_3},
\end{equation}
where we have defined 
\begin{align}
    &\Gamma(r_1, r_2)=\frac{i}{12L^2}\sum_{k,k'}\frac{e^{-ikr_1} e^{-ik'r_2}}{\sin\left(\frac{k}{2}\right)\sin\left(\frac{k'}{2}\right)\sin\left(\frac{-k-k'}{2}\right)} ~,\\
    &\Phi(r_1, r_2)=\frac{i}{4L^2}\sum_{k,k'}\frac{e^{-ikr_1} e^{-ik'r_2} \left[1-\cos(k)-\cos(k')+\cos(-k-k') \right]}{\sin\left(\frac{k}{2}\right)\sin\left(\frac{k'}{2}\right)\sin\left(\frac{-k-k'}{2}\right)} ~.
\end{align}
In the sums over momenta, we exclude the terms with $k=0$, $k'=0$, and $-k-k'=0$ (as discussed in App.~\ref{app:AGPkspace}, those contributions can be set to zero).
Note that these satisfy $\Gamma(r_1, r_2) = \Gamma(r_2, r_1) = \Gamma(r_1 - r_2, -r_2)$ [which guarantees that $\Gamma(j_1-j_3, j_2-j_3)$ 
is symmetric under all permutations of $j_1,j_2,j_3$] and $\Phi(r_2, r_1) = \Phi(r_1, r_2)$ [which gives $\Phi(j_1-j_3, j_2-j_3)$ 
symmetric under $j_1 \leftrightarrow j_2$]. 

\begin{figure}
    \centering
    \includegraphics[width=0.8\linewidth]{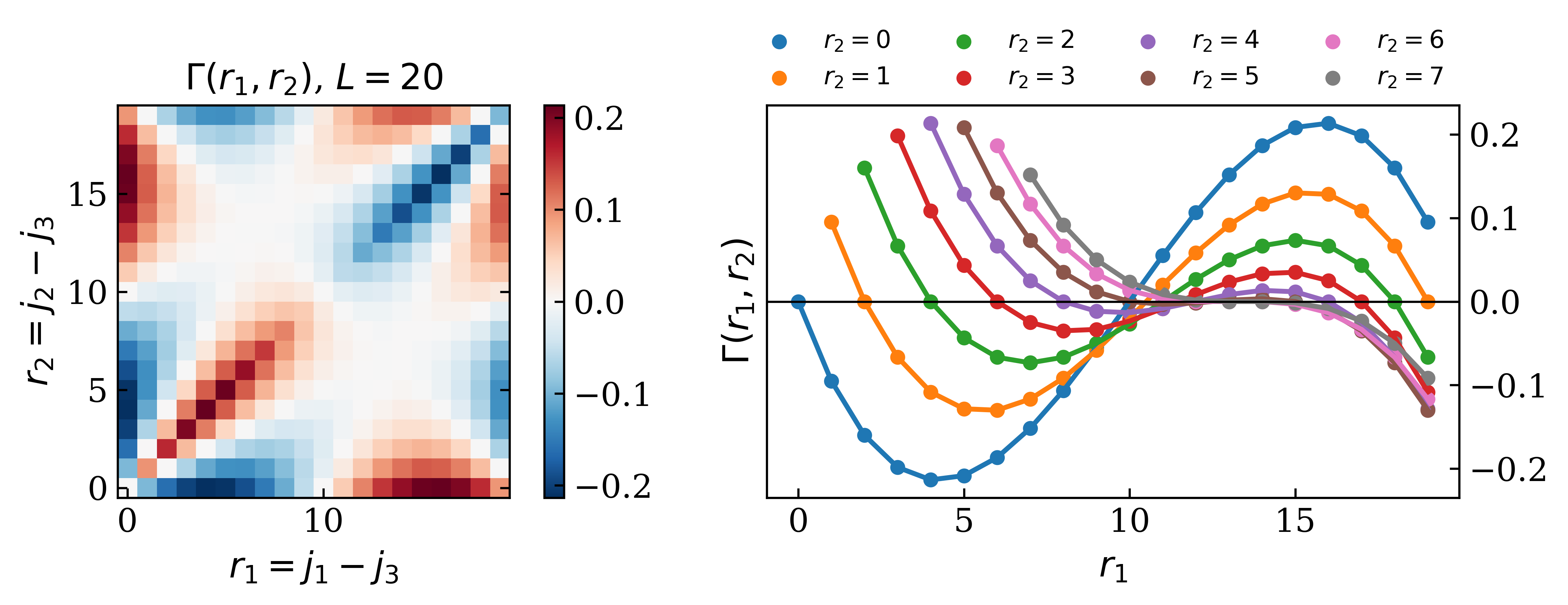}
    \caption{Left: 2D plot of the function $\Gamma(r_1, r_2)$ for $L=20$. Right: sections of $\Gamma(r_1, r_2)$ vs $r_1$ for fixed $r_2\le r_1$. The dots are obtained numerically while the solid lines are the analytical expression in Eq.~(\ref{eq:Gfinite}).
    }
\label{fig:Gfinite}
\end{figure}

The function $\Gamma$ is plotted in Fig.~\ref{fig:Gfinite} for $L=20$. Let us focus on the region $0\le r_2 \le r_1 < L$ [the other region can be obtained from the symmetry $\Gamma(r_1, r_2)=\Gamma(r_2, r_1)$].
We find that $\Gamma(r_1, r_2)$ is a cubic function in that region, with the expression
\begin{equation}
\label{eq:Gfinite}
    \Gamma(r_1, r_2)=\frac{1}{9}(2r_2-r_1)-\frac{r_1}{3L}(2r_2-r_1)-\frac{1}{9L^2}(2r_1-r_2)(2r_2-r_1)(-r_1-r_2) ~.
\end{equation}
In addition to $\Gamma(r_1, r_2) = \Gamma(r_2, r_1)$, the kernel $\Gamma$ satisfies the nontrivial discrete symmetry $\Gamma(r_1 - r_2, -r_2) = \Gamma(r_1, r_2)$ where all arguments are understood modulo $L$. In particular, for $0 \leq r_2 \leq r_1 < L$, this implies $\Gamma(r_1 - r_2, -r_2) \equiv \Gamma(r_1 - r_2, L - r_2) = \Gamma(L - r_2, r_1 - r_2)$ which maps the representative region onto itself. 
This symmetry reflects the full permutation invariance of the cubic generator under exchanges of the site indices $j_1, j_2, j_3$, and can be verified explicitly.

\begin{figure}
    \centering
    \includegraphics[width=0.7\linewidth]{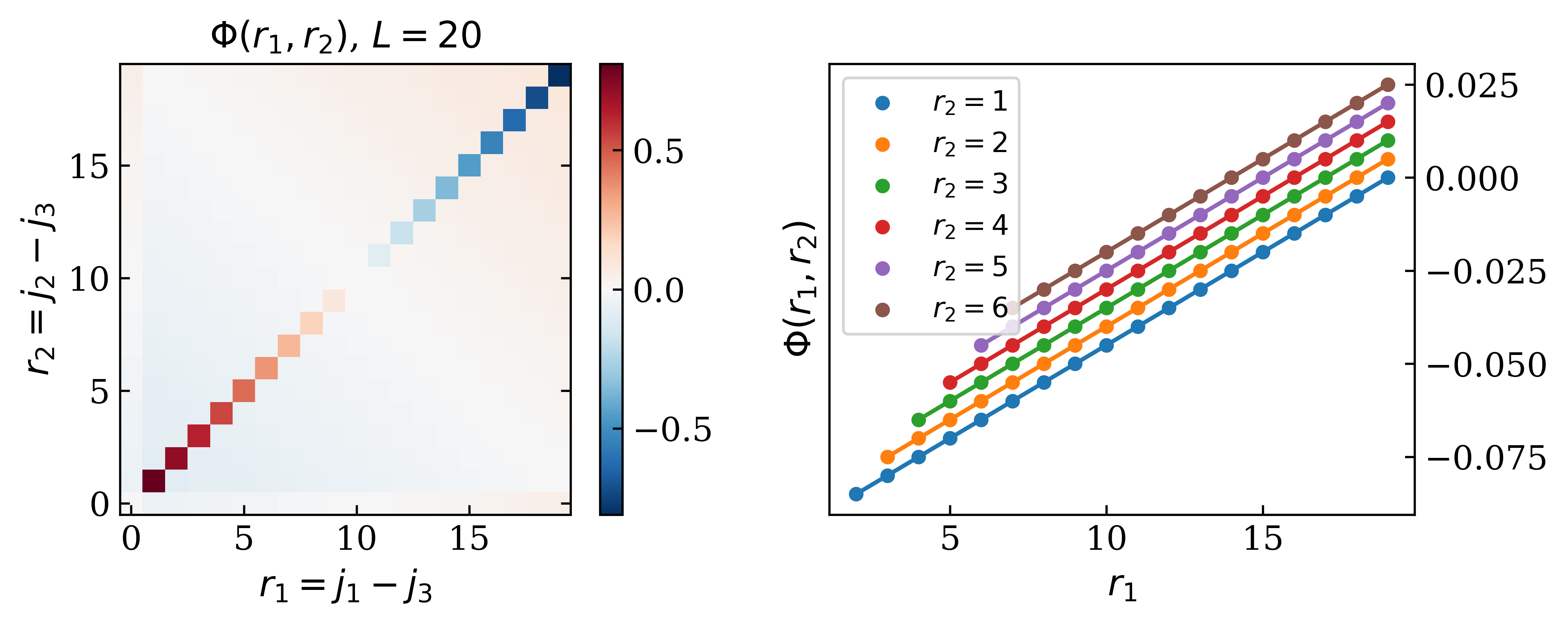}
    \includegraphics[width=0.29\linewidth]{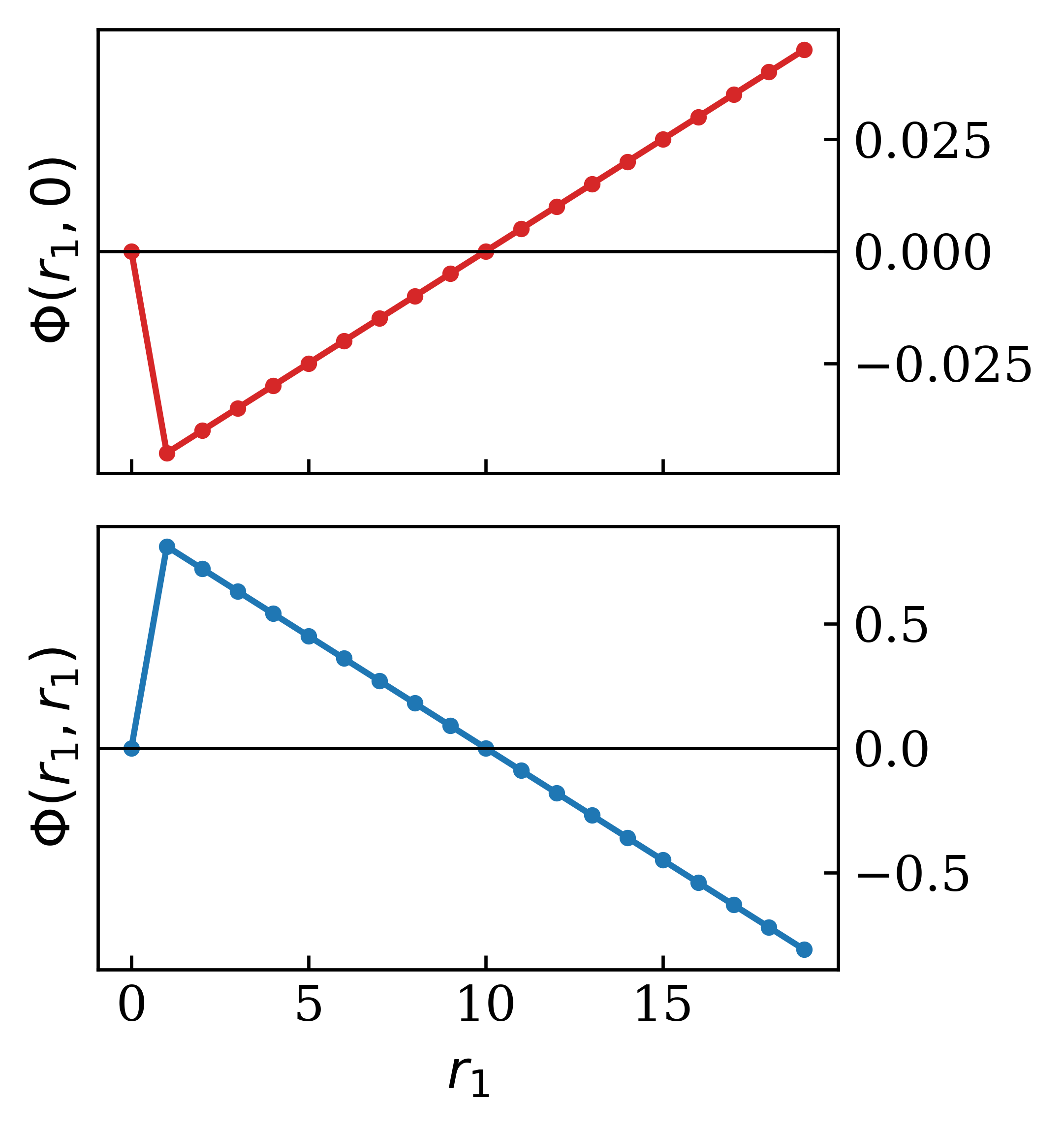}
    \caption{Left: 2D plot of the function $\Phi(r_1, r_2)$ for $L=20$. Center: sections of $\Gamma(r_1, r_2)$ vs $r_1$ for fixed $r_2< r_1$. The dots are obtained numerically while the solid lines are the analytical expressions in Eq.~(\ref{eq:Ffinite}). Right top: section of $\Phi$ vs $r_1$ for $r_2=0$. Right bottom: section of $\Phi$ vs $r_1$ for $r_2=r_1$ (diagonal).
    }
    \label{fig:Ffinite}
\end{figure}

The function $\Phi$ is plotted in Fig.~\ref{fig:Ffinite} for $L=20$. Similarly to the case of $\Gamma$, we can use the symmetry $\Phi(r_1, r_2)=\Phi(r_2, r_1)$ and focus on $0\le r_2 \le r_1 < L$. We find that $\Phi$ has the following expression
\begin{equation}
\Phi(r_1, r_2) = \left(1 - \frac{2r_1}{L} \right) \delta_{r_1 = r_2 \neq 0} + \frac{1}{L}(\delta_{r_1 = 0} + \delta_{r_2 = 0} - 2) + \frac{2}{L^2} (r_1 + r_2) ~.\label{eq:Ffinite}
\end{equation}
This expression is valid for any $r_1, r_2 \in [0, \dots, L-1]$. 

\subsection{Infinite size}
From the analytical expressions of $\Gamma$ and $\Phi$ in Eqs.~(\ref{eq:Gfinite}) and (\ref{eq:Ffinite}), we can now derive infinite size limits.
We get:
\begin{equation}
    \Gamma(r_1, r_2) \rightarrow \frac{1}{9}(2r_2-r_1) \qquad \text{  for } 0 \le r_2 \le r_1 ~,
\end{equation}
and
\begin{equation}
    \Phi(r_1, r_2) \rightarrow \delta_{r_1,r_2} \mathrm{sgn}(r_1) ~.
\end{equation}
Note that in these infinite-system expressions, we want to allow the relative coordinates $r_1 = j_1 - j_3$, $r_2 = j_2 - j_3$ to be of either sign.
This requires slight care for the $\Phi$ function, since the exhibited finite PBC chain solutions are given for the domain $r_1, r_2 \in [0, \dots, L-1]$:
To access finite negative $r_1$ (and $r_2 = r_1$) in the $L \to \infty$ limit, we have used instead $L + r_1$ in the finite chain expression.

\begin{figure}[h]
    \centering
    \includegraphics[width=0.75\linewidth]{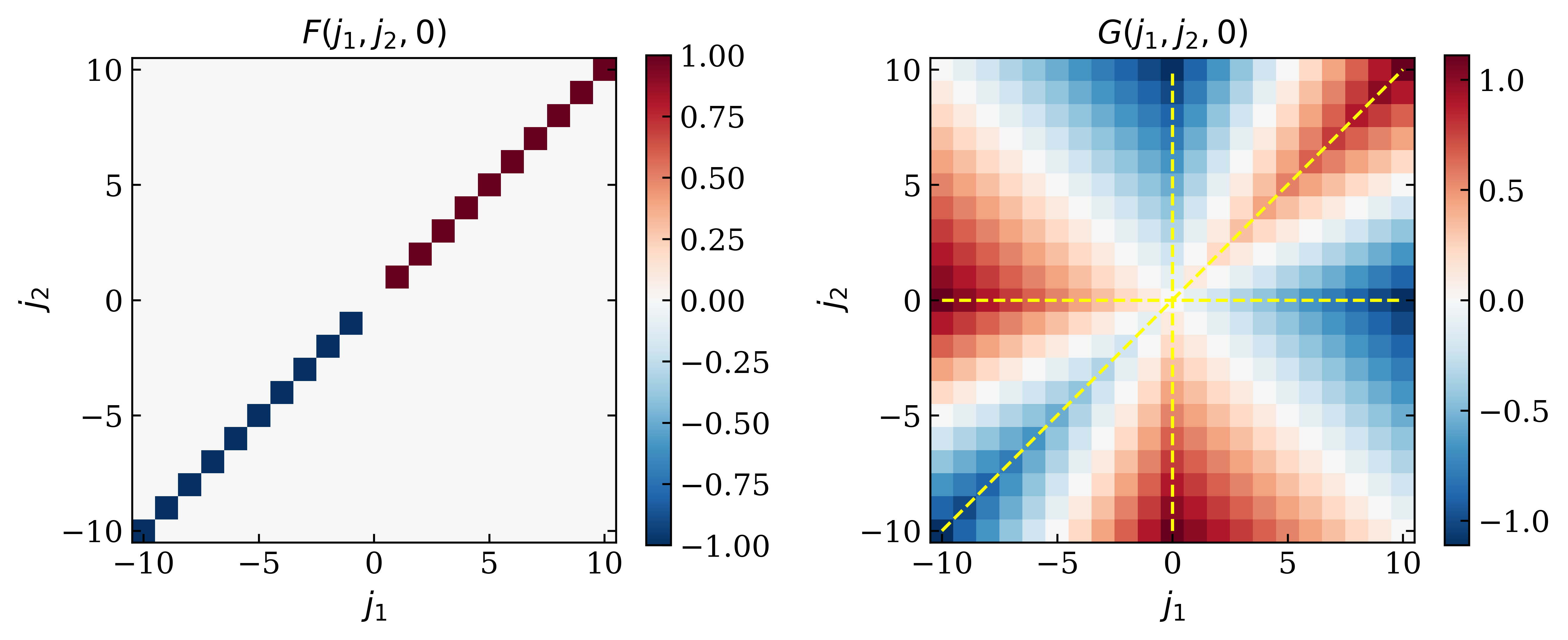}
    \caption{Plots of the functions $F$ in Eq.~(\ref{eq:Finf}) and $G$ in Eq.~(\ref{eq:Ginf2}). Using translational invariance, we set $j_3=0$ and plot $F(j_1, j_2, 0)$, $G(j_1, j_2, 0)$. The dashed yellow lines separate regions where $G$ is linear.}
    \label{fig:Gamma_all}
\end{figure}
We here show an alternative derivation, using an ansatz for the expression of the trilocal operator and checking explicitly that it reproduces $V_{\text{FPUT}}$.
We consider the following ansatz for $X_{\text{FPUT}}$:
\begin{equation}
    X_{\text{FPUT}}= \sum_{j_1, j_2, j_3} G(j_1, j_2, j_3) p_{j_1}p_{j_2} p_{j_3} + \sum_{j_1, j_2, j_3} F(j_1, j_2, j_3) q_{j_1}q_{j_2} p_{j_3} ~.
\end{equation}
This represents the most general time-reversal-odd cubic generator.
We can assume, without loss of generality, that the function $G(j_1, j_2, j_3)$ is invariant under any permutation of its arguments, and that $F(j_1, j_2, j_3) = F(j_2, j_1, j_3)$.
We obtain
\begin{equation}
\begin{aligned}
\{X_{\text{FPUT}}, H_0\} = & \sum_{j_1, j_2, j_3} 3G(j_1, j_2, j_3) (q_{j_1-1} + q_{j_1+1} - 2q_{j_1})p_{j_2} p_{j_3}  \\
& + \sum_{j_1, j_2, j_3} \big[ F(j_1, j_2, j_3)   q_{j_1} q_{j_2} (q_{j_3-1} + q_{j_3+1} - 2q_{j_3}) + 2F(j_1, j_2, j_3) q_{j_1}p_{j_2} p_{j_3} \big] ~.
\end{aligned}
\end{equation}
Here, we have used the corresponding symmetries of $G$ and $F$ under exchanges to obtain a more compact expression.

Let us first consider the terms containing three $q$'s.
We can reproduce the $\alpha$-FPUT cubic potential if we take 
\begin{equation}
\label{eq:Finf}
F(j_1, j_2, j_3) = \delta_{j_1, j_2} \mathrm{sgn}(j_1 - j_3) ~,
\end{equation}
where the sign function $\mathrm{sgn}(r)$ is defined with convention $\mathrm{sgn}(r=0) := 0$.
This $F$ is manifestly invariant under $j_1 \leftrightarrow j_2$.
Then, we indeed have
\begin{equation}
\sum_{j_1, j_3} \mathrm{sgn}(j_1 -j_3)\, q_{j_1}^2 (q_{j_3-1} + q_{j_3+1} - 2q_{j_3}) = \sum_{j_1} q_{j_1}^2 (q_{j_1+1} - q_{j_1-1}) = \sum_j \frac{1}{3} (q_{j+1} - q_{j})^3 ~,
\end{equation}
where we have performed the usual infinite-system manipulations dropping the ``boundary terms'' at $\pm \infty$.

Let us now consider the terms containing both $p$ and $q$, which should cancel. We can write this condition as
\begin{equation}
    \sum_{j_1, j_2, j_3} \big[3G(j_1+1, j_2, j_3) + 3G(j_1-1, j_2, j_3) - 6G(j_1, j_2, j_3) + F(j_1, j_2, j_3) + F(j_1, j_3, j_2) \big] \, q_{j_1} p_{j_2} p_{j_3} = 0 ~.
\label{eq:GFcondition}
\end{equation}
Note that we have further organized the {contributing $F$ terms so that the expression in the square brackets is symmetric under exchanging $j_2$ and $j_3$ (the $G$ terms were already symmetric this way); then satisfying the above condition is equivalent to the expression in the square brackets being $0$ for each $j_1, j_2, j_3$.
We now show an ansatz for $G$ that satisfies these conditions.
We consider the following $G$:
\begin{equation}
\label{eq:Ginf}
    G(j_1, j_2, j_3) =\frac{1}{9} (2j_2 - j_1 - j_3), 
    \qquad \text{for~~} j_1 \le j_2 \le j_3 \text{~~or~~} j_3 \le j_2 \le j_1 ~.
\end{equation}
In the other cases, $G(j_1, j_2, j_3)$ is defined using the invariance under permutations: $G(j_1, j_2, j_3) = G(j_2, j_1, j_3) = G(j_1, j_3, j_2) = \dots $.
Alternatively, we can use the following compact expression
\begin{equation}
\label{eq:Ginf2}
G(j_1, j_2, j_3) = \frac{2}{9}(j_1 + j_2 + j_3) - \frac{1}{3}\max(j_1, j_2, j_3) - \frac{1}{3}\min(j_1, j_2, j_3) ~,
\end{equation}
and check that the exhibited $G$ and $F$ indeed give zero for the expression in the square brackets in Eq.~(\ref{eq:GFcondition}).
Note that one only needs to check this for the $\max + \min$ part of $G$, since the $j_1 + j_2 + j_3$ part gives trivially zero in the ``second derivative'' of $G$ w.r.t.\ $j_1$.
The $j_1 + j_2 + j_3$ part is included so that $G(j_1, j_2, j_3)$ is invariant under translating $j_1, j_2, j_3$ by the same amount (equivalently, depends only on two relative coordinates rather than three absolute coordinates).

In conclusion, we have proved that the following  AGP $X_{\text{FPUT}}$ [equivalent to Eq.~(\ref{eq:XFPUTr})] generates $V_{\text{FPUT}}$:
\begin{equation}
    X_{\text{FPUT}} = \sum_{j_1, j_2, j_3} \frac{1}{3} \left[\frac{2}{3}(j_1 + j_2 + j_3) - \max(j_1, j_2, j_3) - \min(j_1, j_2, j_3) \right] p_{j_1} p_{j_2} p_{j_3} + \sum_{j_1, j_2}  \mathrm{sgn}(j_1-j_3) p_{j_3} q_{j_1}^2 ~.
\end{equation}

\section{Finite size expression for a subset of $X_{\text{bi}}^{[\beta,\gamma]}$ generators in the HOC model}
We now demonstrate that the generator for the perturbation $ V_\mathrm{bi}^{[1,2]}$ can be derived exactly for a finite system. Since $ V_\mathrm{bi}^{[1,2]}$ is a cubic perturbation, one could follow the momentum-space method used for the $\alpha$-FPUT and then Fourier-transform back to real space. Here we opt for a more direct approach using a real-space ansatz previously applied in the quantum regime \cite{Vanovac2024, Pozsgay_notes2023}.

We consider a general ansatz of the form
\begin{equation}
\widetilde{X}^{\beta\gamma} =2 \sum_{j=1}^L \sum_{k=1}^L c_{j,k} \, q_{\beta,j}^{(0)} q_{\gamma,k}^{(0)} ~,
\end{equation}
where $q_{\beta,j}^{(0)}$ is the local density of the charge $Q_\beta^{(0)}$, and similarly for $q_{\gamma,j}^{(0)}$.
Using the definition of the currents, Eq.~(\ref{eq:Jdef}), we calculate the Poisson bracket with the charge $Q_\alpha^{(0)}$,
\begin{align}
\{\widetilde{X}^{\beta\gamma}, Q_\alpha^{(0)}\} &=
2 \sum_{j=1}^L \sum_{k=1}^L c_{j,k} \left[ (J_{\beta\alpha,j} - J_{\beta\alpha,j+1}) q_{\gamma,k}^{(0)} + q_{\beta,j}^{(0)} (J_{\gamma\alpha,k} - J_{\gamma\alpha,k+1})
\right] = \\
&= 2 \sum_{j=1}^L \left[ -q_{\gamma,j}^{(0)} \left(\sum_{k=1}^L J_{\beta\alpha,k}^{(0)} (c_{k-1,j} - c_{k,j})\right) + q_{\beta,j}^{(0)} \left(\sum_{k=1}^L J_{\gamma\alpha,k}^{(0)} (c_{j,k} - c_{j,k-1})\right)  \right],
\end{align}
where we have obtained the second line after some resummations and relabelings.
Inspired by the treatment in the quantum case~\cite{Vanovac2024}, adopting the convention in this paper, we take an ansatz where the coefficients $c_{j,k}$ are defined by the $L$-periodic function
\begin{equation}
\label{eq:cjkPBC}
c_{j,k} = t(k-j) = t(k-j \pm L) ~,\qquad \qquad 
t(n) = 
\begin{cases}
0, & \text{if $n=0$} \\
- \frac{1}{2} + \frac{n}{L}, & \text{if $n=1,2,\dots, L-1$} ~,
\end{cases}
\end{equation}
such that 
\begin{equation}
t(n) - t(n-1) = \frac{1}{L} -\frac{1}{2} \delta_{n,0} - \frac{1}{2} \delta_{n,1} ~.
\end{equation}
We obtain
\begin{align}
\{\widetilde{X}^{\beta\gamma}, Q_\alpha^{(0)}\} = \sum_{j=1}^L \left[ q_{\gamma,j}^{(0)} \left( J_{\beta\alpha,j}^{(0)} +J_{\beta\alpha,j+1}^{(0)}\right) -  q_{\beta,j}^{(0)} \left( J_{\gamma\alpha,j}^{(0)} +J_{\gamma\alpha,j+1}^{(0)}\right) \right] + \frac{2}{L} \left( Q_\beta^{(0)} J_{\gamma,\alpha;\text{tot}}^{(0)}  -  Q_\gamma^{(0)} J_{\beta,\alpha;\text{tot}}^{(0)}  \right).
\end{align}
By fixing $\alpha = 2$ (since $H_0 = Q_2^{(0)}$) and identifying the perturbation  $V_{\text{bi}}^{\beta\gamma}$  from Eq.~(\ref{eq:bilocal}), we obtain
\begin{align}
\{\widetilde{X}^{\beta\gamma}, H_0\} = V_{\text{bi}}^{\beta\gamma} + \frac{2}{L} \left( Q_\beta^{(0)} J_{\gamma,2;\text{tot}}^{(0)}  -  Q_\gamma^{(0)} J_{\beta,2;\text{tot}}^{(0)}  \right).
\end{align}
The above result is completely general, but the extra term is in general, not conserved and also is not local, so it is not as useful. We now specialize to the HOC model and consider different cases.

\subsection{$\beta=1$ and $\gamma$ even}
While the above result is general, we now focus on the HOC case where $\beta = 1$ and $\gamma = 2$.
As established in the main text in our discuss of Eqs.~(\ref{eq:HOC_J1j})-(\ref{eq:HOC_J4j}), $J_{1,2;\text{tot}}^{(0)} = 0$, while  $J_{2,2;\text{tot}}^{(0)} = Q_3^{(0)}$ is a conserved quantity.
Hence,
\begin{equation}
\{\widetilde{X}^{1,2}, H_0\} = V_{\text{bi}}^{[1,2]} - \frac{4}{L}Q_2^{(0)} Q_3^{(0)}.
\end{equation}
Consequently, $\widetilde{X}^{1,2}$ serves as a good generator for $V_\mathrm{bi}^{[1,2]}$ on finite sizes, as it reproduces the perturbation up to a conserved quantity.
A similar argument can be made for $V_\mathrm{bi}^{[1,4]}$.

\subsection{$\beta=1$ and $\gamma$ odd}
Let us see what happens if $\beta = 1$ and $\gamma = 3$.
Now, we still have $J_{1,2;\text{tot}}^{(0)} = 0$, but $J_{3,2;\text{tot}}^{(0)}$ is not a conserved quantity.
Hence, in the final expression
\begin{equation}
\{\widetilde{X}^{1,3}, H_0\} =  V_{\text{bi}}^{[1,3]} - \frac{2}{L}Q_1^{(0)} J_{3,2;\text{tot}}^{(0)}
\end{equation}
the extra term is not a conserved quantity, and this ansatz cannot be viewed as a valid generator of $V_{\text{bi}}^{[1,3]}$.

In general,  we expect that the currents $J_{\gamma,2;\text{tot}}^{(0)}$ for odd $\gamma \geq 3$
are not equal to IoMs, and this ansatz does not work as a valid generator of $V_{\text{bi}}^{[1,\gamma]}$.
This is different from the quantum case, where for free fermions, for example, this was true, and our ansatz was always valid~\cite{Vanovac2024, Pozsgay_notes2023}.

\section{Thermal ensemble estimates for WIBs' AGP norm scaling with system size}
\label{app:analytic_scalings}
In the main text, we numerically estimate the AGP norm for a given perturbation $V$ via long-time trajectory averages and also average over an ensemble of initial conditions.
For WIB perturbations, we expect this to be well captured by a canonical thermal (i.e., equilibrium statistical mechanics) ensemble estimate.
We estimate the system-size scaling of the measured AGP norm---the long-time saturation of the measured AGP variance---for the perturbations used in Fig.~\ref{fig:agp-hoc-a-global} by evaluating the thermal variance of the corresponding generator,
\begin{equation}
\chi_{\text{th}}(V)\ \equiv\ \mathrm{Var_{th}}(X)\ =\ \langle X^2\rangle_{\rm th} - \langle X\rangle_{\rm th}^2 ~,
\label{eq:chi_def_app}
\end{equation}
where $V = \{X,H_0\}$. 
Throughout, we work in the canonical ensemble, $Z = \text{Tr} e^{-\beta_{\text{th}} H_0}$, at inverse temperature
\begin{equation}
\beta_{\rm th} \equiv \frac{1}{k_B T_{\rm th}} ~,
\end{equation}
and denote thermal averages by $\langle \cdots \rangle_{\rm th}$.
Below, we simply write $\beta$ instead of $\beta_{\text{th}}$.

We work on an $L$-site PBC chain with normal-mode variables $\tilde q_k, \tilde p_k$ ($k=\frac{2\pi n}{L},~n \in \mathbb{Z} ~\text{mod}~ L$) satisfying
\begin{equation}
\{\tilde q_k,\tilde q_{k'}\}=0=\{\tilde p_k,\tilde p_{k'}\},\qquad
\{\tilde q_k,\tilde p_{k'}\}=\delta_{k+k',0}.
\end{equation}
With these conventions, the HOC Hamiltonian diagonalizes as
\begin{equation}
H_0=\frac12\sum_k\Big(\tilde p_k\,\tilde p_{-k}+\omega_k^2\,\tilde q_k\,\tilde q_{-k}\Big),
\qquad
\omega_k^2=2(1-\cos k)=4\sin^2\frac{k}{2}.
\label{eq:hoc_diag_app}
\end{equation} 
Assuming a thermal ensemble, from the equipartition theorem we have 
\begin{align}
&\langle \tilde{p}_k \tilde{p}_{k'} \rangle_{\text{th}} = \delta_{k+k'=0} \, \frac{1}{\beta} ~, 
\qquad \langle p_j p_{j'} \rangle_{\text{th}} = \delta_{j,j'} \, \frac{1}{\beta} ~,
\label{eq:ept_p} \\
& \langle \tilde{q}_k \tilde{q}_{k'} \rangle_{\text{th}} = \delta_{k+k'=0} \, \frac{1}{\beta \omega_k^2} ~,
\qquad \langle q_j q_{j'} \rangle_{\text{th}} = \frac{1}{\beta L} \sum_{k \neq 0} \frac{\cos[k(j-j')]}{\omega_k^2} ~,
\label{eq:ept_q}
\end{align}
where we have also shown the corresponding correlation functions in real space, which will be useful in what follows.
Equation~(\ref{eq:ept_p}) holds for all $k$ including $k = 0$ and $k = \pi$.
On the other hand, Eq.~(\ref{eq:ept_q}) does not hold for $k=0$ and $\tilde{q}_{k=0}$, which we treat as follows:
For $P_{\text{tot}}$ ($\sim \tilde{p}_{k=0}$) conserving systems, we have $\omega^2_{k=0} = 0$, and it is natural to exclude the mode $\tilde{q}_{k=0}$ from the integrals specifying the equilibrium statistical mechanics problem, making it mathematically well-defined.

We note that because of the Goldstone modes in our momentum-conserving system, we have $\omega_k^2$ vanishing at small $k$, and the fluctuations of the oscillator position variable $q_j$ grow with the system size: 
\begin{equation}
\langle q_j^2 \rangle_{\text{th}} = \frac{1}{\beta L} \sum_{k = \frac{2\pi n}{L} \neq 0} \frac{1}{4\sin^2(k/2)} \approx \frac{1}{\beta L} \sum_{\substack{n=-\infty \\ n \neq 0}}^\infty \frac{L^2}{4 \pi^2 n^2} = \frac{L}{12 \beta} ~.
\label{eq:HOC_thermal_varq}
\end{equation}
Here we have noted that the sum over $k$ is dominated by small momenta, approximated $4\sin^2(k/2) \approx (2\pi n/L)^2$, extended the sum over $n$ to $\pm\infty$ (since it converges quickly at large $n$), and used $\sum_{n=1}^\infty \frac{1}{n^2} = \frac{\pi^2}{6}$.
On the other hand, the fluctuations of differences in oscillator position remain finite in the thermodynamic limit:
\begin{equation}
\langle (q_j - q_{j+m})^2 \rangle_{\text{th}} = \frac{1}{\beta L} \sum_{k \neq 0} \frac{2[1 - \cos(km)]}{\omega_k^2} \xrightarrow[L\to\infty]{} \frac{1}{\beta} \int_{-\pi}^\pi \frac{dk}{2\pi} \, \frac{\sin^2\left(\frac{km}{2}\right)}{\sin^2\left(\frac{k}{2}\right)} = \frac{m}{\beta} ~,
\label{eq:HOC_thermal_varDq}
\end{equation}
since the integrand is finite at small $k$.

We are now ready to compute $\chi_{\text{th}}(V)$ for the generators used in Fig.~\ref{fig:agp-hoc-a-global}, focusing on the scaling with system size.
Details that follow are meant to support the discussion and findings in Sec.~\ref{sec:numerics}.

\subsection{Scalings of AGP norm for extensive local generators }
\subsubsection{Momentum-conserving $X_{\text{ex}}$}
\label{subsubapp:mcXex}
We begin with a simple illustration where we can carry out all calculations explicitly.
Consider the momentum-conserving generator from Eq.~(\ref{eq:HOC_Xex_for_Vbo3}):
\begin{align}
X_{\text{ex}} &= \frac{1}{8} \sum_j (-2 q_j p_j + q_j p_{j+1} + p_j q_{j+1}) = \frac{1}{4} \sum_k \tilde{p}_k \tilde{q}_{-k} (\cos k - 1) ~, \quad\implies\quad 
\langle X_{\text{ex}} \rangle_{\text{th}} = 0 ~,\\
\text{Var}_{\text{th}}(X_{\text{ex}}) &=  
\frac{1}{16} \sum_{k, k'} \langle \tilde{p}_{k} \tilde{q}_{-k} \tilde{p}_{k'} \tilde{q}_{-k'}  \rangle_{\text{th}} \, (\cos k - 1) \, (\cos k' - 1)
= \frac{1}{16} \sum_{k \neq 0}  \langle |\tilde{p}_{k}|^2 \rangle_{\text{th}} \, \langle |\tilde{q}_{-k}|^2 \rangle_{\text{th}} (\cos k-1)^2 \notag \\
&= \frac{1}{16} \sum_{k \neq 0} \frac{1}{\beta} \frac{1}{\beta \omega_k^2} (\cos k-1)^2
= \frac{1}{16\beta^2} \sum_{k \neq 0} \frac{1}{2(1 - \cos k)} (\cos k - 1)^2 = 
\boxed{\frac{L}{32\beta^2}} ~,
\end{align}

This generator gives perturbation $V_{\text{bo}}^{[3]}$ [up to addition of IoMs as given in Eqs.~(\ref{eq:HOC_Vbo3})-(\ref{eq:HOC_Xex_for_Vbo3})], which we simply referred to as $V_{\text{bo}}$ in our numerical study of the AGP norm in Fig.~\ref{fig:agp-hoc-a-global}, where we observed scaling $\chi^{\text{sat}}(V_{\text{bo}}) \sim L$ in agreement with the above prediction.

In this Appendix in Fig.~\ref{fig:scalings_comp} we show additional analysis where we fit the data presented earlier in Fig.~\ref{fig:agp-hoc-a-global}.
We also attempt to verify our finding above more quantitatively by drawing initial conditions from a canonical distribution closer to the one we assumed in the calculations. 
We see that our numerical results are very close to the theoretical prediction indeed.
(The reason we think that our data from the main text does not give the exact same numerical coefficient when we estimate the temperature from the energy density is because the simple-minded canonical distribution used here is probably $O(1)$ different from that for the initial conditions used in the main text as far as the distribution of all IoMs---i.e., generalized Gibbs ensemble---is concerned.)

We can understand the result also by thinking directly in real space.
Consider first general writing
\begin{equation}
X_{\text{ex}} = \sum_j x_j ~, \qquad
\text{Var}(X_{\text{ex}})_{\text{th}} = \sum_{j,j'} (\langle x_j x_{j'} \rangle_{\text{th}} - \langle x_j \rangle_{\text{th}} \langle x_{j'} \rangle_{\text{th}}) ~.
\label{eqapp:Var_th_Xexgen}
\end{equation}
In the preceding example, a particularly simple grouping of terms is $x_j = \frac{1}{8} p_j (q_{j+1} + q_{j-1} - 2 q_j)$, and using Eq.~(\ref{eq:ept_p}) we obtain
\begin{equation}
\langle x_j x_{j'} \rangle_{\text{th}} =  \delta_{jj'} \frac{1}{64 \beta} \langle (q_{j+1} + q_{j-1} - 2 q_j)^2 \rangle_{\text{th}} = \delta_{jj'} \frac{1}{64 \beta} \frac{1}{L} \sum_k \frac{1}{\beta \omega_k^2} [2(\cos k - 1)]^2 = \delta_{jj'} \frac{1}{32 \beta^2} ~,
\end{equation}
reproducing the previous calculation.
The real-space version of the calculation brings up the most important point, which generalizes to all cases with momentum-conserving generators, such as cubic and quartic extensive local generators listed in the main text and studied in Fig.~\ref{fig:agp-hoc-a-global}, like cubic $V_{\text{ex}}^{(3),(-,-)}$ in Eq.~(\ref{eq:HOC_Vex3_m_m}) [simply referred to as $V_{\text{ex}}$ in Fig.~\ref{fig:agp-hoc-a-global}] with generator in Eq.~(\ref{eqapp:HOC_Vex3_m_m}).
Namely, for such generators $\langle x_j x_{j'} \rangle_{\text{th}} - \langle x_j \rangle_{\text{th}} \langle x_{j'} \rangle_{\text{th}}$ is $O(1)$ (set by the equilibrium temperature) for $j,j'$ that are nearby and decays exponentially with $|j-j'|$.
[In fact, in the cases with cubic or quartic such generators in the HOC problem, the connected correlations are zero beyond few neighbors in the thermodynamic limit; we can see this, e.g., by writing $x_j$ as a function of $\{q_{j+m+1} - q_{j+m}\}$ (containing few $m \in 0, \pm 1, \dots$), changing variables to $\theta_j = q_{j+1} - q_j$, and noting that the $\theta_j$ variables decouple in the equilibrium ensemble.]
Thus, for fixed $j$, the sum over $j'$ converges to a finite value.
Then the sum over $j$ gives the variance scaling with $L$ in all such cases.

As we will discuss next, for $X_{\text{ex}}$ that are not momentum-conserving, $\langle x_j^2 \rangle_{\text{th}}$ can grow with the system sizes, invalidating the above argument.

\begin{figure}
    \centering
    \includegraphics[width=0.95\linewidth]{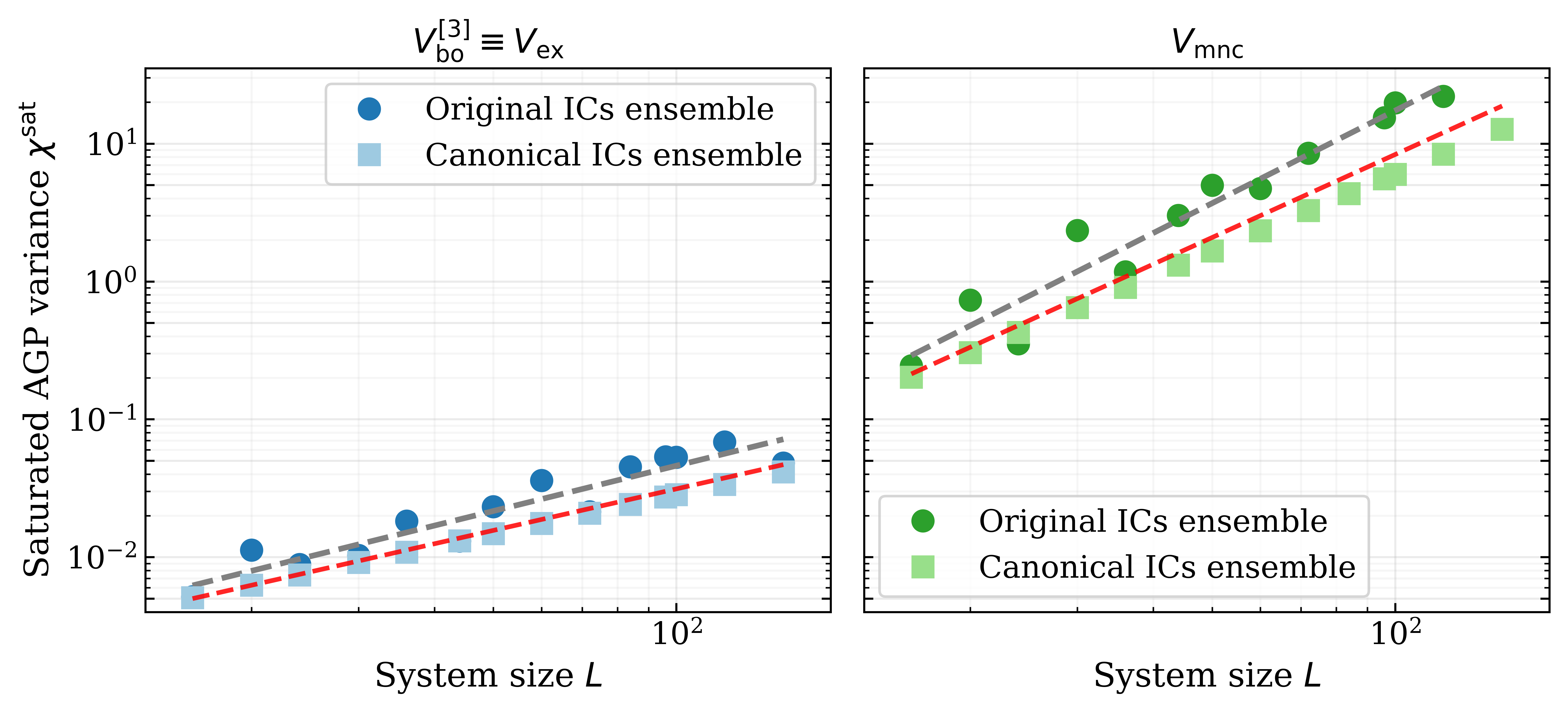}
    \caption{
    \textbf{Left:} Scaling of the saturated AGP variance $\chi^{\mathrm{sat}}$ with system size $L$ for the perturbation $V_{\mathrm{bo}}^{[3]}\equiv V_{\mathrm{ex}}$ that has momentum-conserving extensive local generator given in Eq.~(\ref{eq:HOC_Vbo3}), also discussed in App.~\ref{subsubapp:mcXex}.
    \textbf{Right:} Same analysis for the perturbation $V_{\mathrm{mnc}}$ in Eq.~(\ref{eq:HOC_Vex0}) whose extensive local generator is momentum-nonconserving, also discussed in App.~\ref{subsubapp:Xmnc}.
    In both panels, dark-filled circles show numerical results obtained from the original initial-condition (IC) ensemble used in the main text with energy density $\epsilon = 0.1$, while light-filled squares correspond to values obtained from initial conditions drawn from a canonical thermal distribution as discussed in this Appendix with inverse temperature $\beta=10$.
    The dashed gray lines show power-law fits to the numerical data from the original ensemble, yielding $\chi^{\mathrm{sat}}\sim L^{\gamma_{\rm fit}}$ with $\gamma_{\rm fit}\approx 1.09$ (left panel) and $\gamma_{\rm fit}\approx 2.28$ (right panel).
    Dashed red lines indicate the theoretical predictions obtained from thermal ensemble estimates of the generator variance, $\chi_{\mathrm{th}}\propto L$ for momentum-conserving generators and $\chi_{\mathrm{th}}\propto L^{2}$ for momentum nonconserving generators.
    The canonical ensemble calculations follow the predicted scaling closely, while the original ensemble shows small $O(1)$ deviations in the prefactor. 
    These differences are expected because the analytic calculation assumes a canonical distribution, whereas the ensemble of initial conditions used in the simulations in the main text is likely from an $O(1)$ different generalized Gibbs ensemble over the integrals of motion of the HOC.
} 
\label{fig:scalings_comp}
\end{figure}

\subsubsection{Momentum-nonconserving $X_{\text{mnc}}$}
\label{subsubapp:Xmnc}
Now we turn our attention to momentum-nonconserving generators. 
We consider $X_{\text{mnc}} = \sum_j p_j q_j $ that was used to generate $V_{\text{mnc}} = \sum_j [p_j^2 - (q_j - q_{j+1})^2]$ in Eq.~(\ref{eq:HOC_Vex0}).
Performing the same calculations as above for the generator in terms of Fourier modes, $X_{\text{mnc}} = \sum_k \tilde{p}_{k} \tilde{q}_{-k}$, we have $\langle X_{\text{mnc}}\rangle_{\text{th}} = 0$ and
\begin{align}
\text{Var}_{\text{th}}(X_{\text{mnc}}) = \langle X^2_{\text{mnc}} \rangle_{\text{th}} 
= \sum_k \langle |\tilde{p}_{k}|^2 \, \rangle_{\text{th}} \, \langle |\tilde{q}_{-k}|^2 \rangle_{\text{th}} = \sum_k \frac{1}{\beta} \frac{1}{\beta \omega_k^2} 
= \frac{1}{\beta^2} \sum_k \frac{1}{2(1 - \cos k)} \sim L^2 ~.
\end{align}
Here we come across the subtlety in dealing with PBC and the $\tilde{q}_{k=0}$ mode, and the solution is to exclude $k=0$ from the sum.
We cannot take the usual thermodynamic limit $\sum_k \dots = L \int \frac{dk}{2\pi} \dots$ since the integral diverges near $k=0$.
Instead, we need to consider the discrete sum with the minimal momentum $k_{\text{min}} = \pm 2\pi/L$, and this is the origin of the different scaling with $L$.

Alternatively, we can also perform calculation working in real space and use Eq.~(\ref{eq:HOC_thermal_varq}):
\begin{equation}
\text{Var}_{\text{th}}(X_{\text{mnc}}) = \sum_j \langle p_j^2 \rangle_{\text{th}} \, \langle q_j^2 \rangle_{\text{th}} \approx \frac{L^2}{12 \beta^2} ~.
\label{eqapp:HOC_Var_Xmnc}
\end{equation}
We see that the origin of the different scaling is the divergence (linear growth) with $L$ of the variance of $q_j$.
We report the scaling in Fig.~\ref{fig:scalings_comp} like earlier and see that our predictions match our findings.

We can apply similar reasoning to understand scaling of the AGP norms for the other WIBs in Fig.~\ref{fig:agp-hoc-a-global} obtained using momentum-nonconserving generators.
For example, for the perturbation in Eq.~(\ref{eq:HOC_Vex3_p_m}) referred to as $X_{\text{mnc,2}}$ in Fig.~\ref{fig:agp-hoc-a-global}, we first rewrite the generator from Eq.~(\ref{eqapp:HOC_Vex3_p_m}) as
\begin{equation}
X_{\text{mnc,2}} = \sum_j p_j (q_{j-1}^2 - q_{j+1}^2) = \sum_j p_j (q_{j-1} - q_{j+1}) (q_{j-1} + q_{j+1}) ~.
\end{equation}
We then calculate
\begin{equation}
\text{Var}_{\text{th}}(X_{\text{mnc,2}}) = \sum_j \langle p_j^2 \rangle_{\text{th}} \langle (q_{j-1} - q_{j+1})^2 (q_{j-1} + q_{j+1})^2 \rangle_{\text{th}}  = \sum_j \langle p_j^2 \rangle_{\text{th}} \langle (q_{j-1} - q_{j+1})^2 \rangle_{\text{th}} \langle (q_{j-1} + q_{j+1})^2  \rangle_{\text{th}} \approx \frac{8 L^2}{\beta^3} ~,
\end{equation}
where we have applied Wick's theorem to the quartic term in $q$ and used Eqs.~(\ref{eq:HOC_thermal_varq}) and (\ref{eq:HOC_thermal_varDq}) to obtain
\begin{equation*}
\begin{aligned}
& \langle (q_{j-1} - q_{j+1}) (q_{j-1} + q_{j+1}) \rangle_{\text{th}} = 0 ~, \qquad
\langle (q_{j-1} - q_{j+1})^2  \rangle_{\text{th}} \approx \frac{2}{\beta} ~, \\
& \langle (q_{j-1} + q_{j+1})^2 \rangle_{\text{th}} = \langle 2 q_{j-1}^2 + 2 q_{j+1}^2 - (q_{j-1} - q_{j+1})^2 \rangle_{\text{th}} \approx \frac{4L}{\beta} ~,
\end{aligned}
\end{equation*}
showing only the leading term in each case.
The specific scaling with $L$ arises because in the expression for $X_{\text{mnc},2}$, the quadratic part in $q$ can be rewritten as a product of two factors where only one of them is not momentum-conserving.

Similar considerations apply to generator $X_{\text{mnc},1}$ of $V_{\text{mnc,1}}$ in Fig.~\ref{fig:agp-hoc-a-global}, which refers to generator Eq.~(\ref{eqapp:HOC_Vex3_m_p}) of perturbation in Eq.~(\ref{eq:HOC_Vex3_m_p}) in the main text.
Here again only one of the three factors is momentum-nonconserving $\sim (q_j + q_{j+1})$ with its variance growing $\sim L$, so the total variance of $X_{\text{mnc,1}}$ grows $\sim L^2$.
Thus, our thermal estimates of the AGP norms of $V_{\text{mnc},1}$ and $V_{\text{mnc},2}$ reproduce the numerical findings in Fig.~\ref{fig:agp-hoc-a-global}.

\subsubsection{Quartic $V_{\text{ex}}$ with unusually large AGP norm}
We now also look at an example of the quartic, extensive local perturbation generated from $X_{\text{mnc}^2}= \sum_j p_j q_j^3$. In this case, both the generator and the perturbation are momentum non-conserving. 
We can easily calculate the variance working in real space and obtain
\begin{equation}
\text{Var}_{\text{th}}(X_{\text{mnc}^2}) = \sum_j \langle p_j^2 \rangle_{\text{th}} \, \langle q_j^6 \rangle_{\text{th}} 
= \sum_j \langle p_j^2 \rangle_{\text{th}} \, 15 (\langle q_j^2 \rangle_{\text{th}})^3 
\approx \frac{5 L^4}{576 \beta^4} ~,
\end{equation}
where we have applied Wick's theorem to express $\langle q_j^6 \rangle_{\text{th}}$ in terms of products of $\langle q_j^2 \rangle_{\text{th}}$.
This matches the scaling we found numerically in Fig.~\ref{fig:agp-hoc-a-global} in the main text.

\subsection{Scalings of AGP norm for non-local generators of WIBs: cubic potentials in the HOC and bilocal and trilocal generators}
Apart from dealing with $\tilde{q}_{k=0}$, scalings from extensive local generators are the easiest to understand. 
It is more challenging to obtain the scalings of boosted, bilocal, and tri-local generators, where the generator itself is not necessarily a local object (in fact, in general we do not know the finite-size expression, which in the quantum case to study quantum AGPs as proxies, and such finite-size uncertainty in our knowledge of the generators mattered especially for the boost generated WIBs~\cite{Vanovac2024}).
In the case of the HOC, as we discussed after Eq.~(\ref{eq:HOC_Xex_for_Vbo3}), perturbations obtained using boosted generators can also be represented using extensive local generators, so here we focus on the bilocal and tri-local generators.

We start by calculating the thermal variance of the AGP given by Eq.~(\ref{eq:Xkspace}).
We clearly have $\langle X \rangle_{\text{th}} = 0$.
Since the thermal distribution is Gaussian, we can apply Wick's theorem, and using symmetries of $g(k,k')$ and $f(k,k')$ from App.~\ref{app:AGPkspace} we obtain
\begin{align}
\text{Var}_{\text{th}}(X) &= \sum_{k,k'} g(k,k') g(-k,-k') \frac{1}{\beta^3} \times 6 + \sum_{k_1,k_2} g(k_1,-k_1) g(k_2,-k_2) \frac{1}{\beta^3} \times 9 \\
& + \sum_{k,k'} f(k,k') f(-k,-k') \frac{1}{\beta^3 \omega_k^2 \omega_{k'}^2} \times 2 + \sum_{k_1,k_2} f(k_1,-k_1) f(k_2,-k_2) \frac{1}{\beta^3 \omega_{k_1}^2 \omega_{k_2}^2} \\
&+ 2\sum_{k_1,k_2} g(k_1,-k_1) f(k_2,-k_2) \frac{1}{\beta^3 \omega_{k_2}^2} \times 3 ~.
\end{align}
This expression simplifies for our choice of $g$ and $f$ at special momenta described in App.~\ref{app:AGPkspace}, namely $g(k,k') = 0$ and $f(k,k') = 0$ whenever one of $k$, $k'$, or $-k-k'$ is zero:
\begin{align}
\text{Var}_{\text{th}}(X) &= \frac{6}{\beta^3} 
\underbrace{
\sum_{\substack{k,k'\\ k \neq 0,\, k' \neq 0,\\ -k-k' \neq 0}} g(k,k') g(-k,-k')
}_{W^{gg}}
\;~+~\; \frac{2}{\beta^3} 
\underbrace{
\sum_{\substack{k,k'\\ k \neq 0,\, k' \neq 0,\\ -k-k' \neq 0}} f(k,k') f(-k,-k') \frac{1}{\omega_k^2 \omega_{k'}^2}
}_{W^{ff}} ~.
\end{align}

\subsubsection{$V_{\alpha\text{-FPUT}}$ obtained from $X_{\text{FPUT}}$}
We now specialize to the FPUT case with $g_{\text{FPUT}}$ and $f_{\text{FPUT}}$ given in Eq.~(\ref{eq:gfFPUT}).
We consider the two parts in the above expression for $\text{Var}_{\text{th}}(X)$ separately, starting with the $gg$ part:
\begin{align}
W^{gg}_{\text{FPUT}} = \frac{1}{144\,L} \sum_{\substack{k,k'\\ k \neq 0,\, k' \neq 0,\\ -k-k' \neq 0}} \frac{1}{\sin^2\left(\frac{k}{2}\right) \sin^2\left(\frac{k'}{2}\right) \sin^2\left(\frac{-k-k'}{2}\right)} 
\approx \frac{1}{144\,L} \sum_{\substack{n,n' = -\infty \\ n \neq 0,\, n' \neq 0,\\ n+n' \neq 0}}^\infty \frac{L^6}{\pi^6 n^2 (n')^2 (n+n')^2} = C L^5 ~.
\end{align}
In the above, we noted that the sum over $k,k'$ is dominated by cases where both momenta are small, and for small such $k = 2\pi n/L$ we approximated $\sin(k/2) \approx \pi n/L$, etc., allowing both positive and negative $n, n'$ (which is natural, e.g., starting from the Brillouin zone centered at zero momentum).
Since the resulting sum over $n, n'$ converges quickly at large $n, n'$, we can extend the summation ranges to $\pm \infty$, obtaining a fixed $O(1)$ number, which we lumped together with the other numerical prefactors into a constant $C$.
We expect that such an analysis extracts the leading in $L$ contribution.

We similarly analyze the $ff$ part:
\begin{align}
W^{ff}_{\text{FPUT}} =  \frac{1}{L} \sum_{\substack{k,k'\\ k \neq 0,\, k' \neq 0,\\ -k-k' \neq 0}} \frac{\cos^2\left(\frac{-k-k'}{2}\right)}{\sin^2\left(\frac{-k-k'}{2}\right)} \, \frac{1}{16\sin^2\left(\frac{k}{2}\right) \sin^2\left(\frac{k'}{2}\right)} 
\approx \frac{1}{16 L} \sum_{\substack{n,n' = -\infty \\ n \neq 0,\, n' \neq 0,\\ n+n' \neq 0}}^\infty \frac{L^6}{\pi^6 n^2 (n')^2 (n+n')^2} = 9 C L^5 ~,
\end{align}
where we noted again that the sum over $k,k'$ is dominated by small $k,k'$ and extracted the leading in $L$ contribution as in the $gg$ part, which happens to involve an identical infinite sum.

We hence conclude that both the $W^{gg}$ and $W^{ff}$ parts contribute $\sim L^5 / \beta^3$ to $\text{Var}_{\text{th}}(X_{\text{FPUT}})$.
This reproduces our numerical findings for the AGP norm scaling in this case, as reported in Fig.~\ref{fig:agp-hoc-a-global}.

\subsubsection{ $V_{\text{bi}}$ obtained from $X_{\text{bi}}$}
We now perform similar calculations for the bi-local perturbation with potential in Eq.~(\ref{eq:vbi_from_vfput}); $g_{\text{bi}}$ and $f_{\text{bi}}$ are related to $g_{\text{FPUT}}$ and $f_{\text{FPUT}}$ by the same factor, which we can use to readily obtain initial expressions for $W^{gg}_{\text{bi}}$ and $W^{ff}_{\text{bi}}$.
Thus, for the former, we obtain: 
\begin{equation}
W^{gg}_{\text{bi}} = \frac{1}{144\,L} \sum_{\substack{k,k'\\ k \neq 0,\, k' \neq 0,\\ -k-k' \neq 0}} \frac{\left[1 - \cos\left(\frac{k}{2}\right) \cos\left(\frac{k'}{2}\right) \cos\left(\frac{-k-k'}{2}\right)\right]^2}{\sin^2\left(\frac{k}{2}\right) \sin^2\left(\frac{k'}{2}\right) \sin^2\left(\frac{-k-k'}{2}\right)} ~.
\label{eq:Wgg_bi}
\end{equation}
Analysis at large $L$ here is somewhat different from the $\alpha$-FPUT case, since, e.g., in the $k \to 0$ limit the numerator eliminates the $k' \to 0$ singularity fully, so the scaling is weaker on one hand but also important $k'$ no longer need to reside only at small momentum.
The precise treatment is as follows.
We first rewrite
\begin{equation}
1 - \cos\left(\frac{k}{2}\right) \cos\left(\frac{k'}{2}\right) \cos\left(\frac{-k-k'}{2}\right) = \frac{1}{2} \left[\sin^2\left(\frac{k}{2}\right) + \sin^2\left(\frac{k'}{2}\right) + \sin^2\left(\frac{-k-k'}{2}\right) \right] ~.
\end{equation}
Plugging this into Eq.~(\ref{eq:Wgg_bi}) and using symmetries under exchanges among $k$, $k'$, and $-k-k'$, we obtain
\begin{equation*}
W^{gg}_{\text{bi}} = \frac{1}{144\,L} \sum_{\substack{k,k'\\ k \neq 0,\, k' \neq 0,\\ -k-k' \neq 0}} \left[ 
\frac{3}{2} \frac{1}{\sin^2\left(\frac{k}{2}\right)} + \frac{3}{4} \frac{\sin^2\left(\frac{-k-k'}{2}\right)}{\sin^2\left(\frac{k}{2}\right) \sin^2\left(\frac{k'}{2}\right)}
\right] 
= \frac{1}{144\,L} \sum_{\substack{k,k'\\ k \neq 0,\, k' \neq 0,\\ -k-k' \neq 0}} \left[ 
3 \frac{1}{\sin^2\left(\frac{k}{2}\right)} - \frac{3}{2} +  \frac{3}{2} \frac{\cos\left(\frac{k}{2}\right) \cos\left(\frac{k'}{2}\right)}{\sin\left(\frac{k}{2}\right) \sin\left(\frac{k'}{2}\right)}
\right] ~,
\end{equation*}
where we have used
\begin{equation*}
2\sin^2\left(\frac{-k-k'}{2}\right) = 1 - \cos(k) \cos(k') + \sin(k) \sin(k') = (1 + \cos(k))(1 - \cos(k')) + \cos(k') - \cos(k) + \sin(k) \sin(k') ~.
\end{equation*}
We now see that the dominant part for large $L$ is given by
\begin{equation}
W^{gg}_{\text{bi}} \approx \frac{1}{144\,L} \left[ \sum_{\substack{n=-\infty \\ n \neq 0}}^\infty 3 \frac{L^2}{\pi^2 n^2} \times (L - 2) + O(L^2) \right]\approx \frac{L^2}{144} ~.
\end{equation}

We can similarly perform analysis of $W^{ff}_{\text{bi}}$, obtaining similar scaling $\sim L^2$.
We thus conclude by exact calculation that $\text{Var}_{\text{th}}(X_{\text{bi}}) \sim L^2/\beta^3$, reproducing our numerical findings for the corresponding AGP norm scaling in Fig.~\ref{fig:agp-hoc-a-global}.

\subsection{Toda chain}
We can carry out direct counterparts of the HOC calculations for extensive local generators in the Toda chain.
While the full Toda equilibrium measure is more complicated than that of the HOC, the scaling with system size for extensive local generators is qualitatively similar.
In some cases, we can carry out thermal calculations fully, as in the HOC chain, and in some cases we cannot but can still infer qualitative scaling.
In particular, we expect a similar difference between momentum-conserving and momentum non-conserving extensive local generators.
Below, we illustrate this explicitly for a simple choice of momentum-conserving $X_{\text{ex}}$ which depends only on the momenta, and for a simple example of a momentum-nonconserving generator $X_{\text{mnc}}$.

Consider the extensive local generator
\begin{align}
X_{\text{ex}} =  \sum_j \frac{p_j^3}{3},
\end{align}
which gives perturbation in Eq.~(\ref{eq:Toda_Vex1}) and is labeled $V_{\text{ex,1}}$ in Fig.~\ref{fig:agp-toda-global}. 
Since $X_{\rm ex}$ is odd under time reversal and depends only on $\{p_j\}$, we have $\langle X_{\rm ex}\rangle_{\rm th}=0.$
The thermal variance is therefore
\begin{align}
\chi_{\rm th}(V_{\rm ex}) \equiv \mathrm{Var}_{\rm th}(X_{\rm ex})
= \Big\langle\Big(\sum_{j}\frac{p_j^3}{3}\Big)^2\Big\rangle_{\rm th}
= \frac{1}{9} \sum_{j,j'} \langle p_j^3 p_{j'}^3\rangle_{\rm th} = \frac{1}{9} \sum_j 15 \, \left(\langle p_j^2 \rangle_{\rm th} \right)^3 = \frac{5 L}{3 \beta^3} ~.
\end{align}
This yields the expected extensive scaling $\chi_{\rm th} \sim L$, in agreement with the numerical scaling of the saturated AGP variance reported for the corresponding perturbation in Fig.~\ref{fig:agp-toda-global}.
By general arguments similar to those in Eq.~(\ref{eqapp:Var_th_Xexgen}), we expect similar scaling for extensive local generators whose terms are momentum-conserving.

Consider now the momentum-nonconserving generator producing perturbation in Eq.~(\ref{eq:Toda_Vex0}), which is studied in Fig.~\ref{fig:agp-toda-global} where it is referred to as $X_{\text{mnc}}$.
We have already analyzed this generator for the HOC model in Eq.~(\ref{eqapp:HOC_Var_Xmnc}), and calculations for the Toda chain are essentially the same:
\begin{equation}
\text{Var}_{\text{th}}(X_{\text{mnc}}) = \sum_j \langle p_j^2 \rangle_{\text{th}} \, \langle q_j^2 \rangle_{\text{th}} = L \frac{1}{\beta} \langle q_j^2 \rangle_{\text{th}} \sim L^2 ~.
\label{eqapp:Toda_Var_Xmnc}
\end{equation}
The difference from the HOC model is that we cannot write a simple expression for $\langle q_j^2 \rangle_{\text{th}}$, but we still expect it to scale $\sim L$ because the Toda chain, as a 1d statistical mechanics problem, has a similar Goldstone mode at low energies as the HOC model. 

\end{document}